\documentclass{aa}  

\usepackage{graphicx}
\usepackage{subcaption}
\usepackage{amsmath}
\usepackage{textcomp, gensymb}
\usepackage{tabularx}
\usepackage{booktabs}
\usepackage{lipsum}
\usepackage{comment}
\usepackage{txfonts}
\usepackage{hyperref}
\hypersetup{
    colorlinks=true,
    linkcolor=blue,
    filecolor=magenta,      
    urlcolor=blue,
    citecolor=blue
}
\usepackage{enumitem}
\usepackage[normalem]{ulem}
\usepackage{changepage}
\usepackage[para]{threeparttable}
\usepackage[switch]{lineno}

\usepackage{array}
\usepackage{changepage}
\usepackage{makecell}
\usepackage{orcidlink}
\usepackage{etoolbox}

\RequirePackage{soul}
\RequirePackage{xcolor}

\newcommand{\origin}{\textsc{Paper~I}\,}
\newcommand{\originII}{\textsc{Paper~II}\,}

\newcommand{\seecite}[4]{(\citeauthor{#1} \citeyear{#1}, hereafter \origin; \citeauthor{#3} \citeyear{#3}, hereafer \originII; see also \citeauthor{#2} \citeyear{#2} and \citeauthor{#4} \citeyear{#4})}
\newcommand{\saysay}[1]{“#1”}
\newcommand{\TXS}{TXS~0506$+$056}
\newcommand{\PKS}{PKS~1424$+$240}
\newcommand{\Gsource}{5BZB~J0630-2406}
\newcommand{\RXJ}{5BZB~J0035$+$1515}
\newcommand{\changlook}{5BZQ~J1243$+$4043}
\newcommand{\Lsource}{5BZB~J1150$+$2417}
\newcommand{\accretion}{L_{\rm BLR} / L_{\rm Edd}}
\newcommand{\gratio}{L_{\gamma} / L_{\rm Edd}}

\newcommand{\Pradio}{P_{1.4\,{\rm GHz}}}
\newcommand{\Msun}{M_{\odot}}
\newcommand{\mbh}{M_{\rm BH}}

\newcommand{\Ha}{H$\alpha$\,}
\newcommand{\Hb}{H$\beta$\,}

\newcommand{\MgII}{Mg~\MakeUppercase{\romannumeral2}\,}
\newcommand{\OII}{[O~\MakeUppercase{\romannumeral2}]\,}
\newcommand{\OIII}{[O~\MakeUppercase{\romannumeral3}]\,}

\newcommand{\CIV}{C~\MakeUppercase{\romannumeral4}\,}

\makeatletter
\renewcommand*\aa@pageof{, page \thepage{} of \pageref*{LastPage}}
\makeatother

\begin{document}

   \title{The physical properties of candidate neutrino-emitter blazars}

   \subtitle{}

  \author{Alessandra Azzollini\orcidlink{0000-0002-2515-1353}
         \inst{1}
        \and
           Sara Buson\orcidlink{0000-0002-3308-324X}
         \inst{1,2}
         \and
           Alexis Coleiro\orcidlink{0000-0003-0860-440X}
         \inst{3}
         \and
           Gaëtan Fichet de Clairfontaine\orcidlink{0000-0002-0786-7307}
         \inst{1}\thanks{Now at Departament d’Astronomia i Astrofísica, Universitat de València, C/ Dr. Moliner, 50, 46100, Burjassot, València, Spain}
         \and
           Leonard Pfeiffer\orcidlink{0000-0003-2497-6836}
         \inst{1}
         \and
         Jose Maria Sanchez Zaballa\orcidlink{0009-0001-9486-1252}
         \inst{1}
         \and
         Margot Boughelilba\orcidlink{0000-0003-1046-1647}
         \inst{2}
         \and 
         Massimiliano Lincetto\orcidlink{0000-0002-1460-3369}
         \inst{1}}

  \institute{
       Julius-Maximilians-Universit\"at W\"urzburg, Fakultät f\"ur Physik und Astronomie, Institut f\"ur Theoretische Physik und Astrophysik, Lehrstuhl f\"ur Astronomie,  Emil-Fischer-Str. 31, D-97074 W\"urzburg, Germany\\
             \email{alessandra.azzollini@uni-wuerzburg.de}
       \and 
       Deutsches Elektronen-Synchrotron DESY, Platanenallee 6, 15738 Zeuthen, Germany
        \and
            Université Paris Cité, CNRS, Astroparticule et Cosmologie, F-75013 Paris, France
           }

   \date{Received 01/07/2024; accepted 04/07/2025}

  \abstract
   {The processes governing the production of astrophysical high-energy neutrinos are still a matter of debate, and the sources that originate them remain an open question. Among the putative emitters, active galactic nuclei (AGN) have gained increasing attention in recent years. Blazars, in particular, stand out due to their capability of accelerating particles in environments with external radiation fields. Recent observations suggest that they may play a role in the production of high-energy neutrinos detected by the IceCube observatory.}
   {We studied the physical properties of a subsample of $52$ blazars, that have been proposed as candidate neutrino emitters, based on a positional cross-correlation statistical analysis between IceCube hotspots and the Fifth Edition of the Roma BZCat catalog. We provide a first characterization of their central engines and inner physical nature, which may help to explore the potential link with neutrino production.}
   {This study carries out an analysis of the optical spectroscopic properties of a sample of $52$ candidate neutrino-emitter blazars, to infer their accretion regime. It is complemented by data at the radio and $\gamma$-ray frequencies, which carry the information about the intrinsic power of the relativistic jet. We compared the properties of the sample of candidate neutrino-emitter blazars to other blazar samples from the literature. To this end, we performed statistical tests and also explored, through simulations, the applicability of methods that include limits (censored data) on the quantities of our interest.}
   {Overall, the sample of candidate neutrino-emitter blazars displays properties compatible with those of the reference samples. 
    We observe a mild tendency to prefer objects with intense radiation fields (which are typical of radiatively efficient accretors), and high radio power, such as high-excitation radio galaxies (HERGs).
    Among the blazars in our sample, $24$ are detected in $\gamma$-rays; they cover various ranges of $\gamma$-ray luminosities, compatible with the overall population. 
    Additionally, we show that the statistical tests commonly used in the literature need to be used with caution, as they are highly sensitive to the amount of censored data and the sample size.}
   {}


   \maketitle

\section{Introduction} 
\label{sec:intro}

The IceCube Neutrino Observatory at the South Pole is the most sensitive detector currently operating to investigate the origin of astrophysical high-energy neutrinos. Neutrinos are unique messengers that allow us to study our Universe because they travel long cosmological distances almost unimpeded and carry information about the location of their sources. In 2013, the IceCube Collaboration reported the detection of a diffuse high-energy neutrino flux consistent with an astrophysical origin in the $\gtrsim 100$ TeV to $10$ PeV energy range \citep{icecube2013, IC_north_hard_spectrum:2016}. High-energy neutrinos are created in processes that involve the interaction of cosmic rays with matter or radiation. Therefore, they represent the \saysay{smoking gun} signature of cosmic-ray sources. The astrophysical sources that originate them remain largely uncertain. 

Among the most luminous and persistent sources in the universe are active galactic nuclei (AGN), which are highly powerful astrophysical objects able to emit throughout the whole electromagnetic spectrum, from radio frequencies up to the $\gamma$-ray band. According to theoretical models \citep{Blandford_Payne}, they are powered by accretion onto a central supermassive black hole (SMBH), and have masses that range from millions to billions of solar masses. These can be spinning Kerr BHs that release a fraction of their kinetic energy via a large-scale magnetic field \citep{Blandford_Znajek}.
Blazars are a subclass of AGN, characterized by ultra-relativistic jets that point toward the observer's line of sight. The spectral energy distribution (SED) of blazars typically displays a double-humped shape, with a low-energy peak typically in the infrared (IR)--X-rays, and the high-energy peak in the hard X-rays and/or $\gamma$-rays \citep{Fossati:1998, Fan:2016}. The first peak is explained by synchrotron radiation emitted by the population of accelerated electrons in the relativistic jet, while, in leptonic models, the second peak is attributed to inverse Compton radiation. Although this framework has been effective in describing the multiwavelength emission of the majority of blazars, it may be reasonable to expect that hadrons are present alongside leptons. The possibility of relativistic jets being able to accelerate hadrons requires exploring lepto-hadronic and/or hadronic scenarios. Hadronic frameworks, although less favored compared to the leptonic scenario for many blazars, open up the perspective of exploring the link between blazars and neutrinos. 

In blazar jets within the hadronic scenarios, neutrinos can be the result of either proton-proton (p-p) or proton-photon (p-$\gamma$) interactions \citep[for a review, see][]{Cerruti:2020}. In the former case, a proton of higher energy interacts with a less energetic one, producing pions and neutrinos as a result. In the latter case, protons of the jet interact with the lower-energy radiation fields, either internal or external, through photo-meson (or photo-pion) interaction or Bethe-Heitler pair production \citep{Mannheim:1993, Boettcher:2013, Dermer:2014}. This leads to muons and neutrinos as by-products.

For this work we were mostly interested in p-$\gamma$ interactions as the combination of efficient particle acceleration and external radiation fields can foster neutrino production in blazars \citep{Dermer:2014}. Blazars can host external radiation fields, i.e., from the accretion disk and the re-processed disk emission by the broad and narrow line region, and the dusty torus through external Compton (EC). The photo-pion production efficiency, i.e., the physics governing the production of neutrinos, is tightly linked to the different modes of accretion. These are primarily traced by emission in the optical--IR band.

Studies in the literature suggest that blazars may originate neutrinos, contributing to the observed high-energy diffuse astrophysical neutrino flux \citep{Mannheim:1993, Dermer:2014, IceCube2017_2LAC, IceCube_10y_reprocessed:2022}. In 2017, the $\gamma$-ray blazar \TXS was associated with a candidate muon-neutrino event, with a chance association at the $3\sigma$ level \citep{icecube2018}.
More recent works have provided evidence for a correlation between a subset of 52 blazars and IceCube neutrino data, suggesting that blazars may be the first population of extragalactic high-energy neutrino emitters \seecite{Buson:2022}{Buson_erratum}{Buson:2023}{Bellenghi_2023}. 

While the statistical significance indicates that the correlation between the two samples is unlikely to arise by chance, it should not be mistaken for the probability of individual neutrino--blazar associations. It is improbable that all $52$ candidates are genuinely connected to the neutrinos. Among them, one should expect a non-negligible number of spurious associations by chance (simply by Poisson probability). Assessing the genuineness of the individual associations remains challenging, especially given the proprietary nature of the IceCube data and analysis tools (see also \cite{Barbano_2023}, Lincetto et al., in prep.).

Bearing this in mind, in this paper we explore the physical properties of the full set of 52 candidate neutrino-emitters $-$ which may include a non-negligible number of spurious associations $-$ and compare them to those of the general blazar population. Based on the energy of the IceCube neutrino dataset, these candidate neutrino emitters may be capable of accelerating protons to energies up to the petaelectronvolt (PeV) range. Therefore, we refer to them as candidate \saysay{PeVatron blazars}.

We provide a first multiwavelength characterization of the intrinsic physical properties of the $52$ candidate neutrino-emitter blazars, exploiting optical spectroscopy complemented by radio and $\gamma$-ray information from the literature.

As the production of neutrinos depends on the object's accretion regime and radiation field properties (within lepto-hadronic frameworks), studying the behavior at different frequencies probes various aspects of the underlying physics \citep[see, e.g.,][]{FichetDC:2023, SanchezZaballa:2025, Pfeiffer_BT, Pfeiffer_MT}.
Section \ref{sec:class} introduces the blazar classification scheme based on the intrinsic physical properties used in this work. Section \ref{sec: sample} addresses the blazar sample. In Sections \ref{sec:dataset} and \ref{sec: analysis}, we describe the employed dataset and the procedure that we adopted in the analysis. Then, in Section \ref{sec:comparison} we present our findings about the properties of the sources and discuss the results in the broader context of the overall blazar population. This work assumes a flat $\Lambda$CDM cosmology with $\rm{H}_{\rm 0} = 69.6~\rm{km}\cdot{\rm s}^{-1}\cdot\rm{Mpc}^{-1}$, $\Omega_{\rm 0} = 0.29$ and $\Omega_{\Lambda} = 0.71$.

%%%%%%%%%%%%%%%%%%%%%%%%%%%%%%%%%%%%%%%%%%%%%%
\section{The classification of blazars}
\label{sec:class}
\subsection{The traditional classification scheme}
In the AGN unification model \citep{Urry:1995,Urry:2004}, jetted AGN are traditionally classified into BL Lacertae (BL Lacs, BLLs) and flat spectrum radio quasars (FSRQs) based on the strength of the lines in the optical spectrum. The historical dividing criterion between the two classes is set by the rest-frame equivalent width (EW) of the emission lines. The latter measures the observed line strength, by definition, being the integrated line flux with respect to the underlying continuum emission. EW $< 5$ \AA\ defines BLLs and EW $> 5$ \AA\ FSRQs. Phenomena such as variations in the jet power and the orientation with respect to the observer's line of sight mean that this purely observational classification is not always reliable and/or consistent over time. Smaller or larger EW values could be due to beaming effects and variability in the nonthermal continuum, without necessarily indicating intrinsically lower or higher line luminosity \citep{Ghisellini:2011}. 

Attempts in the literature to follow the observation-driven classification scheme required to coin new nomenclature for subsets of blazars that may show ambiguous properties (e.g., properties of both the FSRQs and BL Lacs classes). This is, for example, the case of \saysay{blue flat spectrum radio quasars} \citep{Ghisellini:2012}, also known as \saysay{high-power high-synchrotron-peak blazars},  also known as \saysay{masquerading BL Lacs} \citep[][]{Padovani:2012, Padovani:2019}. Blue FSRQs (masquerading BLLs) appear as BL Lacs based on the featureless optical spectrum and high synchrotron peak. However, they are proposed to be intrinsically FSRQs with their broad emission lines swamped by the jet's powerful synchrotron emission. 

Similarly, objects that have been observed to transition from BL Lac to FSRQ, and/or vice versa, following the appearance or disappearance of optical lines, have been dubbed as \saysay{changing-look blazars} \citep[][Azzollini et al., in prep.]{Vermeulen:1995, Ghisellini:2011, Ruan:2014, PenaHerazo_changinglook}.
It remains unclear whether the observed changes correspond to intrinsic physical variations.
These transition objects may be sources with highly beamed jets, and radiatively efficient accretion, in which sudden variations in the interplay between the thermal and the nonthermal emission may affect the appearance of the optical spectra, leading to the (apparent) variability of the emission lines over time.

As the nature of AGN is closely related to the accretion of matter onto the central SMBH, a physically driven distinction is based on the different accretion modes. The latter characterizes the AGN activity and the main energy mechanism governing their evolution \citep{BestHeckman:2012, Heckman:2014}. On the one hand, when the dominant energetic output is in the form of electromagnetic radiation, we are dealing with \saysay{radiative-mode} AGN (i.e., high-excitation radio galaxies; hereafter HERGs), which show the traditional internal structure of an SMBH surrounded by a geometrically thin, optically thick accretion disk (AD). A hot corona surrounds the accretion disk, which scatters the photons of the disk to X-ray energies. The SMBH and AD are surrounded by a dusty torus (DT) of obscuring molecular gas on larger scales. In this schematic view, the luminous ultraviolet (UV) radiation from the AD illuminates the broad line region (BLR), which is located close to it, and the narrow line region (NLR) further out. The radiatively efficient accretion occurs through cold gas piling up on the core of the AGN, and this leads to the establishment of a stable disk. Objects of this class are generally associated with star formation in the host galaxy. The bulk of the energy output is emitted through radiation, however, they can often show traces of radio jets that extend for tens or hundreds of kiloparsecs. 
On the other hand, in \saysay{jet-mode} AGN (i.e., low-excitation radio galaxies; hereafter LERGs), the radiative emission is usually less powerful, and the energy output is dominated by bulk kinetic energy transported in powerful jets.
For this second category, it is suggested that in place of the thin accretion disk, there is a geometrically thick advection-dominated accretion flow (ADAF). These radiatively inefficient accretors are thought to be fuelled via hot gas (\saysay{hot-mode} accretion) subject to the Bondi mechanism \citep{Bondi:1952}.
The distribution of observed radio energies of the two classes covers a broad range. However, HERGs are found to dominate at high radio luminosity, while LERG objects dominate at lower values. Adopting the radio power at $1.4\, {\rm GHz}$, the dividing threshold falls at $P_{1.4 {\rm GHz}} \sim 10^{26}\, {\rm W} \cdot {\rm Hz}^{-1}$ \citep{BestHeckman:2012}.

It has long been argued that the different accretion modes could originate a dichotomy in jetted AGN \citep[e.g.,][]{Ghisellini_Celotti:2001,Maraschi:2003,Ghisellini:2009,Giommi:2012,Antonucci:2012}.
\cite{Ghisellini:2011} proposed a classification scheme for blazars based on the accretion efficiency: FSRQ objects, where the presence of broad optical emission lines implies intense radiation fields, are related to efficient radiative-mode sources (i.e., HERGs), while BLL objects, which are characterized by the lack of prominent emission lines, correspond to inefficient hot-mode accretors (LERGs). The proposed discriminating parameter is the luminosity of the BLR in Eddington units with the threshold reference $\accretion \gtrsim 5 \times 10^{-4}$. The $\gamma$-ray luminosity in Eddington units is also a good tracer of both the accretion efficiency (in the case of intense radiation fields) and the jet power (in all cases), and the dividing line has been proposed at $\gratio\gtrsim 0.1$ in this case. The distinction places FSRQs above these thresholds, whereas BL Lacs with the ADAF are found below \citep{Ghisellini:2010, Ghisellini:2011, Sbarrato:2012, Sbarrato:2014}. While there is likely continuity between the two extremes rather than a clear-cut separation between the two classes, this distinction is helpful for depicting the underlying physics.
The $\accretion$ and $\gratio$ are proxies for the accretion power and take into account the differences in black hole mass due to the normalization factor being the Eddington luminosity. Within this physically driven scheme, the LERG--HERG discrimination brings less ambiguity in transitional objects, naturally describing blue FSRQs (masquerading BL Lacs) as HERG objects based on their physical properties.
Blue FSRQs encompass sources which are intrinsically FSRQ whose emission lines are washed out by a bright, Doppler-boosted jet continuum, unlike \saysay{true} BL Lacs, which are intrinsically weak-line objects \citep[see][]{Ghisellini:2011, Padovani:2022}.
 Similarly, for changing-look blazars, the apparent line luminosity variations are due to observational biases, while the underlying physics of the objects remains the same.
 The boundary between the two regimes is only loosely defined, the quantities have rather large errors and the models involved are rather simplistic. As a consequence, we caution that the HERG--LERG distinction is still expected to suffer from observational limitations and the transition between the two classes is expected to be gradual.
In the context of potential neutrino emission, the interesting aspect is not the precise classification of an object, but rather the possible presence or absence of radiation fields external to the jet. 
HERG-like objects benefit from multiple external radiation fields, including the accretion disk, photons reprocessed in the BLR, and emission from the dusty torus. These additional photon targets for protons may foster neutrino production compared to LERGs.

\subsection{This work's taxonomy}
\label{subsec: taxonomy}
%%%%%%%%%%%%
In this work, we departed from the historical observation-driven classifications of blazars and hereafter embraced the physically driven view of HERGs and LERGs, based on the accretion regime and high- and low-excitation properties. In blazars, intense radiation fields are traced by the spectral emission lines produced in the BLR and NLR. The radio power $\Pradio$ and the $\gamma$-ray luminosity $L_{\gamma}$ carry the information about the intrinsic power of the relativistic jet. 
In the simplified view that we adopted, the distinction between the two classes may be primarily traced by the following properties,

\begin{itemize}[label=$\bullet$]
    \item 
    The Eddington-scaled accretion rate to the central black hole mass, which sets the separation between radiatively efficient and radiatively inefficient accretion modes; HERGs are then defined by $\accretion \gtrsim 5 \times 10^{-4}$;
    \item 
    The $\gamma$-ray Eddington ratio, where higher values, $\gratio \gtrsim 0.1$, are typically displayed by HERGs compared to LERGs;
    \item
    The radio power, which informs about the jet power, where HERGs have typically higher power, $\Pradio \gtrsim 10^{26}\, {\rm W} \cdot {\rm Hz}^{-1}$ compared to LERGs.
\end{itemize}

This classification based on the radiative efficiency of accretion, the intensity of radiation fields, and the radio and $\gamma$-ray power, is similar to the one used in previous works \citep{Ghisellini:2010, Giommi:2012, BestHeckman:2012, Heckman:2014, Padovani:2022} for discriminating HERGs--LERGs following physically driven markers.
However, the boundaries between HERGs and LERGs are not sharply defined, and the transition between these categories is likely to be gradual.

%%%%%%%%%%%%%%%%%%%%%%%%%%%%%%%%%%%%%%%%%%%%%%
\section{The blazar sample}
\label{sec: sample}
\subsection{Target sample}

In our previous studies, we explored the potential link between blazars and high-energy neutrinos \citep[][\origin\, and \originII]{Buson:2022, Buson:2023}.
To mitigate biased classifications affecting the selection of our sample, we used the fifth data release of the Roma BZCat catalog \cite[$5$BZCat, see][]{BZCat}. This compilation of $3561$ sources does not rely on selections in specific wavelengths or survey strategies and has already proved to be effective in unraveling unexpected discoveries \citep[e.g., the \saysay{WISE blazar strip},][]{Massaro:2011}. The 5BZCat catalog is not complete as the sky coverage of the surveys used to build it is not uniform, however, it is rather \saysay{pure}. For each object of the 5BZCat, an optical spectrum is available and individually inspected to ensure the blazar nature. In 5BZCat, objects are further taxonomically categorized according to the historical classification based on optical spectroscopy; for example, BL Lacs are indicated as \saysay{BZB}, FSRQs as \saysay{BZQ}, and blazars of uncertain type as \saysay{BZU}. In addition, blazars displaying an optical spectrum dominated by the contribution of the host galaxy are referred to as \saysay{BZG}. 

In \origin\ and \originII, 5BZCat sources were positionally cross-matched with all-sky high-level neutrino data, in the form of the skymaps publicly released by the IceCube Collaboration. For the southern celestial hemisphere ($-85^{\circ} < \delta < -5^{\circ}$), the analysis was based on the $7$ years data recorded between 2008--2015 \citep{IceCube7y:2017, Buson:2022}. For the northern celestial hemisphere ($-3^{\circ} \leq \delta \leq 81^{\circ}$), the analysis exploited the most recent neutrino sky-map released in \citet{IceCube_10y_reprocessed:2022}, that encompasses $9$ years of neutrino observations \citep[2011--2020, ][]{Buson:2023}.

The analysis focused on sky locations displaying the highest discrepancy from background expectations, i.e., \saysay{hotspots}, as the most promising locations of putative neutrino cosmic sources. The hotspots are identified by the highest $L$ values, defined as the negative logarithm of the provided local pvalue in the IceCube maps, $L = -\log \left( {\rm pvalue} \right)$. The positional cross-correlation analysis was performed separately for the southern and northern celestial hemispheres. The analysis resulted in a post-trial chance probability $2\times10^{-6} \left( 4.5\sigma \right)$ for the southern dataset and $6.79\times10^{-3} \left( 2.47\sigma \right)$ for the northern dataset, suggesting that the correlation between the blazar and neutrino samples is unlikely to arise by chance. The investigative approach pinpointed a subset of $52$ blazars, ten hosted in the southern and $42$ in the northern celestial hemisphere, as promising candidates for the production of IceCube neutrinos. As anticipated in Section \ref{sec:intro}, the sample includes a non-negligible number of spurious associations. The reliability of the individual blazar-neutrino associations is difficult to access and will be addressed in a forthcoming work.

Table \ref{tab: associations} lists the 5BZCat sources proposed as associated with neutrino hotspots. The first four columns report the name, coordinates, and L value of the neutrino hotspots (for more details see \origin, \originII). The following two columns list the associated $5$BZCat blazars with the corresponding redshift information. For the redshift, we report the values collected by inspecting the most recent literature and cross-checking with the optical spectrum available from this work. The seventh column reports the blazar \emph{Fermi}-LAT counterpart from the Third Data Release of the Fourth \emph{Fermi} Large Area Telescope (LAT) AGN Catalog \citep[$4$LAC-DR$3$, ][]{4LAC-DR3}. Finally, the last column lists the reference to the optical spectrum used for the object. Eight blazars of our sample of interest have already been included in other works as promising candidates for the production of high-energy IceCube neutrinos \citep[][see Appendix \ref{sec:append dataset} for more details on the individual objects]{icecube2018, Krauss:2018, Plavin:2020, Hovatta:2021, Padovani_PKS:2022, Stathopoulos:2022, Plavin:2023}. They are highlighted with the $\diamond$ symbol in Table \ref{tab: associations}.

\subsection{Comparison samples}
\label{subsec: comparison samples}

We were interested in the properties of the candidate PeVatron blazars in the radio, optical, and $\gamma$-ray ranges to investigate the thermal and nonthermal components of their central engine, the accretion properties, and the power of the relativistic jet.
First, we characterized the properties of the sources in our sample. Then, we compared them with those of the overall population of blazars. For the latter, we referred to previous literature studies that tackled the physical properties of relatively large sets of blazars. The primary selection criteria were the public availability of optical spectra or line spectroscopy measurements, provided they were obtained using a methodology consistent with our study. Three samples were found to be suitable, i.e., 
\citet{Sbarrato:2012} and \citet{Paliya:2017, Paliya:2021}. By estimating the quantities of interest using a consistent approach, these samples ensured the absence of biases in our analysis while still encompassing a wide range of properties. More details are provided in Section \ref{sec: analysis}, in the following we introduce the comparison samples.

In \citet[][hereafter S$12$]{Sbarrato:2012}, the sample of blazars was selected from the Seventh Data Release of the Sloan Digital Sky Survey \citep[SDSS-DR$7$, analyzed in][]{Shen:2011} and with a \emph{Fermi}-LAT detection in the First Catalog of Active Galactic Nuclei Detected by the Fermi Large Area Telescope \citep[$1$LAC, ][]{1LAC}. The authors also included the optically selected BL Lac candidates with redshift measurement of \citet{Plotkin:2011}, among which $71$ have $1$LAC detection and $61$ do not. Finally, they considered additional intermediate objects between BLLs and FSRQs selected from SDSS-DR$6$ \citep{SDSS6} and with $\gamma$-ray $1$LAC counterpart. The S$12$ sample includes a total of $163$ sources, with a measurement or an upper limit (UL) on either $L_{\gamma}$ and/or $L_{\rm BLR}$, and enables us to explore the whole space of radiative efficiency spanned by LERG and HERG objects (see Section \ref{sec:comparison}). The paper provides the values of the redshift, the mass of the black hole ($\mbh \left[ \Msun \right]$), $L_{\rm BLR} \left[ {\rm erg} \cdot {\rm s}^{-1} \right]$ and $L_{\gamma} \left[ {\rm erg} \cdot {\rm s}^{-1} \right]$, either measured or with UL, for each object.

In \citet[][hereafter P$17$]{Paliya:2017}, the Candidate Gamma-Ray Blazar Survey Catalog \citep[CGRaBS, ][]{CGRaBS} was cross-matched with the \emph{Fermi}-LAT catalogs \citep[3FGL, 2FGL, 1FGL, and 0FGL, sequentially; ][]{3FGL, 2FGL, 1LAC, 0FGL} considering all the data up to 2016 April 1. This sample was further reduced based on the availability of multiwavelength data, particularly X-rays, in the HEASARC archive \citep{Paliya:2017}. This results in two subsamples of objects with and without LAT detection (324 \saysay{$\gamma$-loud} and 191 \saysay{$\gamma$-quiet} sources, respectively). Several techniques were used to estimate the physical quantities, %among which 
including also optical spectroscopy (P$17$ optical). To ensure consistency with our analysis strategy (see Section \ref{sec: analysis}), for the accretion properties, we narrowed down to the $54$ objects that were analyzed by optical spectroscopy ($50$ $\gamma$-loud, and four $\gamma$-quiet). The redshift, $\mbh$, $L_{\rm disk}$ and size of the BLR ($r_{\rm BLR} \left[ {\rm cm} \right]$) are available from the study. The advantage of this sample is that it allows us to explore both $\gamma$-ray-detected and nondetected sources using properties derived in a consistent way (optical spectroscopy). Although the $\gamma$-quiet blazars we consider are only four, they complement the $\gamma$-quiet sources overview along with the sample of S$12$.

The sample of \citet[][hereafter P$21$]{Paliya:2021} includes all the $\gamma$-ray emitting blazars and blazar candidates listed in the LAT $10$-years Source Catalog \citep[$4$FGL-DR$2$, ][]{4FGL-DR2}, that have been cross-matched with SDSS-DR$16$ \citep{SDSS16} and the optical spectroscopy literature. This sample of 1016 objects includes both measurements (for $674$ objects) and upper limits (for $342$ sources) on the optical lines' luminosities. The paper provides the redshift and the lines' parameters (luminosity $L_{\rm line}$ and full width at half maximum ${\rm FWHM}_{\rm line}$) for each object. Using these, we estimated the physical properties of interest using the approach described in Section \ref{sec: analysis}. By selection, this sample offers a view of $\gamma$-ray-detected blazars.

The main characteristics of these three comparison samples are listed in Table \ref{tab: comparison samples}, which reports each sample with the corresponding total number of sources and of the subsample for which the physical properties were analyzed via optical spectroscopy. The fourth, fifth and sixth column reports the number of objects for which either the luminosity of the broad line region ($L_{\rm BLR}$), the $\gamma$-ray luminosity ($L_{\gamma}$) or both are estimated through upper limits. The last column lists the quantities that are provided in the corresponding papers.

\bgroup
\def\arraystretch{0.93}%
\begin{table*}[!h]
\caption{Summary of the main characteristics of the chosen comparison samples.}
\label{tab: comparison samples}
\centering
\begin{threeparttable}[t]
\begin{tabular}{ccccccc}        
\hline
Sample & $\#$ sources & Via optical spectroscopy & UL on $L_{\rm BLR}$ & UL on $L_{\gamma}$ & UL on $L_{\gamma}, L_{\rm BLR}$ & Quantities \\
\hline
S$12$ & $163$   & $163$     & $22$  & $-$   & $62$  & z, $\mbh$, $L_{\rm BLR}$, $L_{\gamma}$   \\
P$17$ & $515$   & $54$      & $-$   & $4$   & $-$   & z, $\mbh$, $L_{\rm disk}$, $r_{\rm BLR}$ \\ 
P$21$ & $1016$  & $1016$    & $342$ & $-$   & $-$   & z, ${\rm FWHM}_{\rm line}$, $L_{\rm line}$     \\
\hline
\end{tabular}
\tablefoot{The first column reports the sample: S$12$ \citep{Sbarrato:2012}, P$17$ \citep{Paliya:2017} and P$21$ \citep{Paliya:2021}, respectively. The second and third columns contain the total number of sources and the corresponding subsample for which the properties were estimated via optical spectroscopy. The fourth, fifth and sixth columns report the number of objects for which limits were placed on either the $\gamma$-ray luminosity, the BLR luminosity or both. The last column lists the properties that are provided in the corresponding papers (see Section \ref{subsec: comparison samples} for further details).}
\end{threeparttable}
\end{table*}

As visualized in the Venn diagram in Fig. \ref{fig: Venn diagram}, among the blazars in the three comparison samples, there is a certain amount of overlap:
\begin{itemize}
    \item There are $41$ S$12$ sources that are also included in P$17$. They are all $\gamma$-loud.
    \item There are $99$ common sources between S$12$ and P$21$. Among them, $53$ blazars have no upper limits in either sample, while P$21$ provides limits for $46$. Of these $46$, S$12$ includes limits on both $L_{\rm BLR}$, $L_{\gamma}$ for $31$, on $L_{\rm BLR}$ for $11$ and measurements for four objects.
    \item There are $230$ common sources between P$17$ and P$21$.
    \item Among our PeVatron blazar candidates, one is also included in S$12$, seven in P$17$ and seven in P$21$.
\end{itemize}

\begin{figure}
    \centering
    \includegraphics[width = 0.9\linewidth]{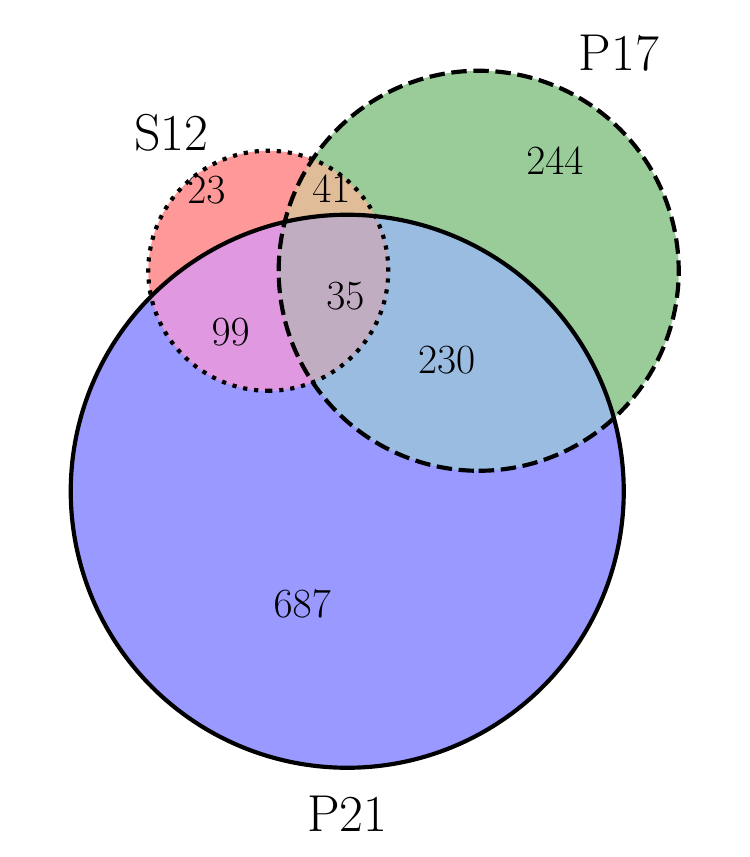}
    \caption{Venn diagram showing the overlap between the three comparison samples S$12$ \citep{Sbarrato:2012}, P$17$ \citep{Paliya:2017} and P$21$ \citep{Paliya:2021}.}
    \label{fig: Venn diagram}
\end{figure}

%%%%%%%%%%%%%%%%%%%%%%%%%%%%%%%%%%%%%%%%%%%%%%
\section{The multiwavelength dataset}
\label{sec:dataset}

In this section, we describe the collection of multiwavelength data in the different bands, available for the $52$ candidate PeVatron blazars. Optical spectroscopy constitutes the basis of our data sample. At these energies, we can delve into the mass accretion process on the central SMBH. Broad emission lines may appear in the spectra as a result of the gas of the BLR photoionized by the radiation produced in the accretion disk. 
The optical dataset is complemented with ancillary properties at radio and $\gamma$-ray energies. 

%%%%%%%%%%%%%%%%%%%%%%%%%%%%%%%%%%%%%%%%%%%%%%
\subsection{Optical spectroscopy}
\label{subsec:optical}
We collected the data from public archives and the literature when available. We observed $12$ objects for which neither a public spectrum was available in the archives, nor information in the literature (see also Table \ref{tab: associations} and Appendix \ref{sec:spectra}). For two of them, we acquired spectra with the European Southern Observatory (ESO) Very Large Telescope at Paranal Observatory (VLT) X-Shooter spectrograph \citep{XShooter}. The observations employed all three arms (NIR, VIS, and UVB) of the instrument. We observed three other targets using the R$150$ grism of the Gemini Multi-Object Spectrographs of Gemini South \citep[GMOS-S, ][]{GMOS:1998, GMOS:2016}. We observed seven additional targets with the R$1000$B grism of the Optical System for Imaging and low-Intermediate-Resolution Integrated Spectroscopy of the Gran Telescopio Canarias \citep[GTC OSIRIS, ][]{osiris}. Further details on the complete spectroscopic dataset are provided in Appendix \ref{sec:append dataset}.

%%%%%%%%%%%%%%%%%%%%%%%%%%%%%%%%%%%%%%%%%%%%%%
\subsection{Gamma-ray properties}
\label{subsec:gamma}

When measured, the $\gamma$-ray luminosity of blazars is a good tracer of both the luminosity of the accretion disk and the power of the relativistic jet \citep{Ghisellini:2010, Sbarrato:2014}. The $L_{\gamma}$ provides a lower limit on the power of the jet as some of the energy will inevitably go toward accelerating the plasma and, thus, it is proportional to the fraction of the energy that goes into radiation  \citep{Sbarrato:2012}. 
Among the candidate PeVatron blazars, $24$ are detected in $\gamma$-rays. We collected the information for these sources from the Third Data Release of the Fourth \emph{Fermi} Large Area Telescope (LAT) AGN Catalog \citep[$4$LAC-DR$3$, ][]{4LAC-DR3}, which encompasses frequencies between $\nu_{1} = 1.21\times 10^{22} \, {\rm Hz} \left( 50 \, 
{\rm MeV} \right)$ and $\nu_{2} = 2.41\times 10^{27} \, {\rm Hz} \left( 1 \, {\rm TeV} \right)$. To estimate the luminosity $L_{\gamma}$, we adopted the following method,

\begin{equation}
    \label{eqn:Lgamma}
    L_{\gamma} = 4 \pi d^{2}_{\rm L} \cdot \frac{S_{\gamma} \left( \nu_{1}, \nu_{2} \right)}{\left( 1 + {\rm z} \right)^{1 - \alpha_{\gamma}}}\, .
\end{equation}

Here $d_{\rm L}$ is the luminosity distance of the source\footnote{The estimation employs the value of the redshift and the \texttt{Astropy} cosmology package. For more details, see \url{https://docs.astropy.org/en/stable/cosmology/index.html}}, $\alpha_{\gamma} = \Gamma_{\gamma} - 1$ was evaluated from the photon spectral index $\Gamma_{\gamma}$ in the LAT energy band \citep{Ghisellini:2009}. $S_{\gamma} \left( \nu_{1}, \nu_{2} \right)$ is the $\gamma$-ray energy flux between the frequencies $\nu_{1}$ and $\nu_{2}$, calculated as integration of a power-law spectrum over the \emph{Fermi}-LAT band as,

\begin{equation}
S_{\gamma} \left( \nu_{1}, \nu_{2} \right) =
\begin{cases}
\frac{\alpha_{\gamma} h \nu_{1} F_{\gamma}}{1 - \alpha_{\gamma}} \cdot \left[ \left( \frac{\nu_{2}}{\nu_{1}} \right)^{1 - \alpha_{\gamma}} -1 \right] & \text{for } \alpha_{\gamma} \neq 1 \,, \\
h \nu_{1} F_{\gamma} \ln \left( \frac{\nu_{2}}{\nu_{1}} \right) & \text{for } \alpha_{\gamma} = 1 \,
\end{cases}
\end{equation}

where $F_{\gamma} \left[ {\rm ph} \cdot {\rm cm}^{-2} \cdot {\rm s}^{-1} \right]$ is the photon flux.
For the remaining $28$ blazars with no LAT detection, we placed an upper limit on the $\gamma$-ray luminosity considering the sensitivity of the instrument for a power-law-spectrum point-source and 10 yr exposure\footnote{See \url{https://www.slac.stanford.edu/exp/glast/groups/canda/lat_Performance.htm} for further details.} ($F_{\gamma} = 5 \times 10^{-10} \, {\rm ph} \cdot {\rm cm}^{-2} \cdot {\rm s}^{-1}$, $\Gamma_{\gamma} = 2.0$). % \citep[a similar approach was used in][]{Sbarrato:2012}. 
While these limits provide straightforward initial information about the jet's power, they should be interpreted with caution. This is especially true for objects whose high-energy emission may elude detection in the LAT band (e.g., those expected to peak at $\gamma$-ray energies $\lesssim$ MeV range).

%%%%%%%%%%%%%%%%%%%%%%%%%%%%%%%%%%%%%%%%%%%%%%
\subsection{Radio properties}
\label{subsec:radio}
 
The radio power offers a complementary proxy for the intrinsic jet power \citep{Sbarrato:2014, Heckman:2014}, especially useful for objects with no detection at $\gamma$ rays. Besides, the radio power $\Pradio$ yields further insights regarding the HERG--LERG nature (see Sections \ref{sec:class} and \ref{subsec: taxonomy}). 
We used the radio information provided by $5$BZCat. For each blazar, the catalog lists the radio flux density at $1.4 \; {\rm GHz}$ from NRAO VLA Sky Survey \citep[NVSS, ][]{NVSS} or Very Large Array (VLA) Faint Images of the Radio Sky at Twenty-cm \citep[FIRST, ][]{FIRST}.

%%%%%%%%%%%%%%%%%%%%%%%%%%%%%%%%%%%%%%%%%%%%%%
\section{Optical spectroscopy data analysis}
\label{sec: analysis}

%%%%%%%%%%%%%%%%%%%%%%%%%%%%%%%%%%%%%%%%%%%%%%
\subsection{Data reduction procedure}
\label{subsec:reduc}

The first step was the reduction of the optical spectra for the sources that we observed. For the X-Shooter spectra, we used the \texttt{EsoReflex} version \texttt{2.11.5} pipeline \citep{EsoReflex} and, for GMOS-S, the Data Reduction for Astronomy from Gemini Observatory North and South (\texttt{DRAGONS}) software v\texttt{3.1.0} \citep{DRAGONS}. For the OSIRIS spectra, we used the PypeIt v\texttt{1.16.0} pipeline \citep{Pypeit_1}. The reduction was performed through the standard steps of bias and dark current subtraction, flat field correction, and cosmic ray removal. The arc lamp exposures were reduced separately and used for the wavelength calibration. Then, sky subtraction was applied to remove the background contribution from the science spectra. The spectra of the correspondent standard star were also reduced separately and used for the estimation of the sensitivity curve. Finally, flux calibration was applied to the science data. After that, for the X-Shooter dataset, the \texttt{Molecfit} version \texttt{4.2.3} \citep{Molecfit1, Molecfit2} software tool was used to correct the science spectra for atmospheric contamination features, in particular the contribution of telluric absorption lines. 

The remaining blazars in the sample were investigated using various pipelines to extract and examine spectra from different archives. The data taken from the Six-degree Field Galaxy Survey \citep[$6$dFGS, ][]{6dFGS:2004, 6dFGS:2009}, and the 2dF QSO Redshift Survey \citep[$2$QZ, ][]{2QZ:1998, 2QZ:2004} catalogs are not flux calibrated, and in this case, we used the conversion factor ${\rm ct} = 7.69 \times 10^{-17} \,{\rm erg} \cdot {\rm s}^{-1} \cdot {\rm cm}^{-2} \cdot$ \AA$^{-1}$\ from photon counts to flux density units \citep{Koss:2017}. For low-resolution spectra of the Large Sky Area Multi-Object Fiber Spectroscopic Telescope \citep[LAMOST, ][]{LAMOST:2012}, the flux calibration is relative. This affects the accuracy level of the absolute flux, which is in units of $10^{-17} \,{\rm erg}\cdot{\rm cm}^{-2}\cdot{\rm s}^{-1}\cdot{\rm \AA}^{-1}$. More details are provided in Appendix \ref{sec:append dataset}.

%%%%%%%%%%%%%%%%%%%%%%%%%%%%%%%%%%%%%%%%%%%%%%
\subsection{Estimation of physical quantities}
\label{subsec:quantities}

%To perform the analysis, 
We extracted and visually inspected the optical spectra of interest to identify possible emission line profiles. The hydrogen recombination lines of the Lyman or Balmer series Ly$\alpha$, \Ha and \Hb are among the strongest lines detected in the environment of an AGN. They are produced in the BLR along with the more highly ionized permitted \MgII, \CIV lines. Instead, \OII and \OIII spectral lines are emitted in the NLR and produced by forbidden transitions. Permitted transitions (e.g., those occurring in the BLR) are electric dipole transitions with high probabilities ($\sim 10^{5}-10^{9}\,{\rm s}^{-1}$) and originate in regions of higher-density plasma \citep[on average, $n_{\rm e}\gtrsim 10^{9}\,{\rm cm}^{-3}$ in observed BLRs, ][]{Osterbrock}. The observed broad line profiles are the result of permitted transitions, while forbidden transitions give rise to narrow profiles. For broad lines, the density of the forming region is so high that all levels of the ions that would be otherwise responsible for forbidden transitions are collisionally de-excited \citep{Osterbrock}. In forbidden transitions, electrons have much lower probabilities ($\sim10^{-6}-10^{2}\,{\rm s}^{-1}$) of being excited to (or de-excited from) these levels and require very low electron densities $\lesssim 10^{6}\,{\rm cm}^{-3}$ to populate the levels above the ground state via electron collisions \citep{Longair:book}. 

We applied a rebinning of the spectra when needed, identified the lines, and then cross-checked the redshift available from the literature with our estimate.
We used the \texttt{SpectRes} \texttt{PYTHON} tool \citep{SpectRes} for the rebinning. Then, we used the Image Reduction and Analysis Facility \citep[\texttt{IRAF}, ][]{Tody:1986, Tody:1993} software to fit the detected lines with either one or multiple Gaussian functions according to the morphological shape of the profile. The results from the fit\footnote{splot function for profile deblending and Gaussian fitting: \url{https://astro.uni-bonn.de/~sysstw/lfa_html/iraf/noao.onedspec.splot.html}} were extracted, compared to independent fits performed with \texttt{SciPy}\footnote{More details at \url{https://docs.scipy.org/doc/scipy/reference/generated/scipy.optimize.curve_fit.html}} and used for further analysis. We used the flux, EW, and full width at half maximum (FWHM) to estimate the central engine properties of the blazars in our sample \citep[similarly to ][]{Celotti:1997, Ghisellini:2008, Shen:2011, Shen:2012, Paliya:2017, Padovani:2019, Paliya:2021}. The total luminosity of the spectral line was estimated as $L_{\rm line} = 4 \pi  F_{\rm line} \rm{d_{\rm L}}^{2}$. The corresponding errors with the fitted quantities are usually $\sim 5\%$ for the line flux and $10\%$ for the FWHM \citep{Decarli:2008, Decarli:2011}.

Thirteen targets lack BLR lines. They display either a featureless optical spectrum or only NLR lines. 
Therefore, for these, we estimated upper limits on the luminosity of the nondetected emission lines: a power-law fit was applied to a region of $500$ \AA\ at the position of the expected line, depending on the redshift \citep[following the approach of ][]{Sbarrato:2012}. Then, we simulated an emission line of Gaussian profile with fixed FWHM $v_{\rm FWHM} = 4000 \, {\rm km} \cdot {\rm s}^{-1}$ \citep{Decarli:2011} and flux $F_{\rm line}$. We assumed an uncertainty of $10 \%$ on the flux $F_{\rm line}$, and let $F_{\rm line}$ vary up to a maximum value of $1 \%$ of the fitted continuum. To determine the upper limit, we performed a $\chi^{2}$ test and accepted as flux limit, the value $F_{\rm line}$ for which $\chi^{2} < \chi^{2} \left( 99\% \right)$.

Thereafter, we employed the luminosity (or the upper limit) of the emission lines to infer the total luminosity of the broad line region,

\begin{equation} 
\label{eqn:lumblr}
    L_{\rm BLR} = L_{\rm line} \cdot  \frac{\left< L_{\rm BLR} \right>}{L_{\rm rel.\,frac.}} \,,
\end{equation}

with $L_{\rm rel.\,frac.} = 77, 22, 34, 63$ for \Ha, \Hb, \MgII and \CIV, respectively, being the contribution of each line to the total BLR luminosity as estimated based on the composite spectrum of \citet{Francis:1991}, assuming a relative flux of $100$ for Ly$\alpha$. The total broad line flux is $\left< L_{\rm BLR} \right> = 555.76$ \citep{Francis:1991}. When more than one emission line was present in the spectrum, we estimated the BLR luminosity by averaging the measured line profiles \citep{Celotti:1997, Sbarrato:2012}. 

To estimate the BLR luminosity, which is then used to infer the accretion regime, we used only the BLR line fluxes or limits. Although for some objects NLR lines are detected, and works suggest to use them to derive estimates for the BLR luminosity \citep{Padovani:2019, Padovani_PKS:2022}, based on our tests, the procedure leads to not fully consistent values of the accretion regime in our sample (Azzollini et al., in prep.).
The luminosity of the disk is estimated by assuming a thin Shakura-Sunyaev accretion disk \citep{Shakura:1973} with BLR covering factor of $0.1$ to it, so that $L_{\rm BLR} \simeq 10 \%\,L_{\rm disk}$ \citep{Sbarrato:2012, Ghisellini:2014}. It directly links the accretion disk luminosity with the observed broad emission lines and provides an estimation independent of the viewing angle (the lines are assumed to be isotropically emitted) and it avoids contamination from the nonthermal continuum. The expected average uncertainty on the resulting value is a factor of $2$ \citep{Sbarrato:2012, Ghisellini:2014}. 

Optical spectral emission lines inform about the bolometric thermal luminosity of the accretion flow in AGN. To this end, we used the broad \Hb, \MgII lines and the narrow \OII, \OIII lines, when available \citep{Punsly:2011, Padovani:2019},

\begin{equation} \label{eqn:lumbol}
    \log_{10} \left( L_{\rm bol} \right) = a + b \cdot  \log_{10} \left( L_{\rm line} \right) \,,
\end{equation} 

where the coefficients are $\left( {\rm a}, {\rm b} \right) =$  $(12.32\pm0.20$, $ 0.78\pm0.01)$, $\left(16.76\pm0.26, 0.68\pm0.01\right),\left(26.50\pm0.32, 0.46\pm0.01\right),$ $ \left(33.96\pm0.33, 0.29\pm0.01\right)$ for \Hb, \MgII, \OII and \OIII, respectively.

The luminosity and FWHM of the lines, and the empirical scaling laws of \citet{Shen:2011} and \citet{Shen:2012} were employed in the estimation of the virial mass of the central black hole. The approach assumes that the BLR is gravitationally bounded to the central BH potential, in such a way that we can estimate the mass of the central object by evaluating the orbital radius and Doppler velocity of the region through an inspection of the emitted lines. The evaluation was made through the following equation:

\begin{equation} \label{eqn:bhmass}
    \log_{10} \left( \frac{\mbh}{\Msun} \right) = a + b \cdot  \log_{10} \left( \frac{\lambda \cdot L_{\rm line}}{10^{44} \;  {\rm erg} \cdot  {\rm s}^{-1}} \right) + c \cdot \log_{10} \left( \frac{\rm FWHM}{{\rm km} \cdot  {\rm s}^{-1}} \right) \, .
\end{equation} 

Here the coefficients are \citep{Shen:2011, McLure:2004, Vestergaard:2009},

\begin{equation}
\left( {\rm a}, {\rm b}, {\rm c} \right) =
\begin{cases}
\left(0.379, 0.43, 2.1\right) & \text{for } {\rm H}\alpha \,, \\
\left(0.672, 0.61, 2.0\right) & \text{for } {\rm H}\beta \,, \\
\left(0.740, 0.62, 2.0\right) & \text{for } {\rm \MgII} \,, \\
\left(0.660, 0.53, 2.0\right) & \text{for } {\rm \CIV} \,. \\
\end{cases}
\end{equation}

The typical uncertainty on the value of $\mbh$ is a factor of $4$ \citep{Vestergaard:2006}.
From the black hole mass, we derived the Eddington luminosity of the sources as $L_{\rm Edd} \simeq 3 \times 10^{4} \left( \frac{M}{\Msun} \right) L_{\odot}$ \citep{Eddington:1926, Lang:1974}. The resulting value only depends on the mass of the central accreting object. By definition, it sets an upper limit on the accretion luminosity of the central compact core. Therefore, the disk luminosity in Eddington units traces the normalized accretion rate \citep{Shen:2011, Ghisellini:2014} $\dot{m} \equiv \frac{\dot{M}_{\rm acc}}{\dot{M}_{\rm Edd}} = \frac{L_{\rm disk}}{\eta L_{\rm Edd}}$, where $\eta$ is the radiative efficiency of accretion. For geometrically thin accretion disks, it depends on the location of the innermost stable orbit of the disk and the spin of the central black hole. 

At $L_{\rm disk} \lesssim 10^{-2}\, L_{\rm Edd}$, the accretion structure becomes radiatively inefficient, with the standard accretion disk turning into a hot accretion flow (ADAF). The dependence of the disk luminosity on the accretion rate changes ($L_{\rm disk} \propto \dot{M}^{2}$). This leads to an analogous change in the dependence of $L_{\rm BLR}$ on $\dot{M}$ \citep{Ghisellini:2014, Sbarrato:2014}. However, the response of the jet to these different inner physical properties of the black hole is not known and remains an open debate. 

In this model, the radiation component takes into account the contributions of the emission disk, the broad line region, the dusty torus surrounding the disk, and the part of the radiation that is intercepted in the IR and re-emitted. This structure allows us to estimate the size of the broad line region and the dusty torus from the luminosity of the disk. Following the works of \citet{Ghisellini:2008, Ghisellini:2014, Ghisellini:2017}, we derive the radius of the BLR and the torus as

\begin{equation} 
\label{eqn:rblr}
   r_{\rm BLR} = 10^{17} \cdot \left( \frac{L_{\rm disk}}{10 ^{45}\,{\rm erg} \cdot {\rm s}^{-1}} \right)^{1/2} \, {\rm cm} \,,
\end{equation}

\begin{equation} 
\label{eqn:rdt}
    r_{\rm DT} = 2.5 \times 10^{18} \cdot \left( \frac{L_{\rm disk}}{10 ^{45}\,{\rm erg} \cdot {\rm s}^{-1}} \right)^{1/2} \, {\rm cm} \,.
\end{equation}

The resulting quantities are listed in Table \ref{table:results}.

%%%%%%%%%%%%%%%%%%%%%%%%%%%
\section{Comparison with reference blazar samples}
\label{sec:comparison}

In this section, we discuss the adopted approach and present the outcomes of our study. We investigated the physical properties of the sample of candidate neutrino-emitter blazars and the reference samples presented in Section \ref{subsec: comparison samples} in light of the more physically driven approach we adopted. To this end, we focused on the accretion regime traced by the $\accretion$ and $\gratio$, and the jet power traced by the radio power $\Pradio$ alongside the $L_{\gamma}$.
The results for the accretion regime properties are presented in  Fig. \ref{fig:accretion regime no highlights}. Our sample (in black, with \TXS, \PKS, and \Gsource\ highlighted in cyan, lime, and magenta, respectively; see Section \ref{subsec: masquerading} for further details) is compared to the $\gamma$-loud and $\gamma$-quiet blazars studied in P$17$ and P$21$ and the sources in S$12$ (see Section \ref{subsec: comparison samples}). The arrows indicate the estimated upper limits on either the optical and/or $\gamma$-ray luminosity (see also Appendix \ref{sec:append upper lim}). The dotted lines represent the distinction for the accretion efficiency regime, respectively $\accretion\sim5 \times 10^{-4}$ and $\gratio\sim0.1$ \citep{Ghisellini:2011, Sbarrato:2012, Sbarrato:2014}. On the sides, the histograms show the distribution of the two ratios for all the samples, excluding estimated limits. The black solid lines indicate the median value for the candidate PeVatron blazars.
In Fig. \ref{fig:accretion vs radio no highlight}, the luminosity of the broad line region as a function of the radio power allowed us to study the disk-jet connection even in the sources that are not-detected at $\gamma$-rays \citep[see a similar approach in][]{Sbarrato:2014}. We compared our sample to the other samples (S$12$, P$17$, P$21$), cross-matching them with NVSS \citep{NVSS} and FIRST \citep{FIRST} to infer the $\Pradio$. 

Fig. \ref{fig: histograms} shows the distribution of the other estimated quantities, i.e., the histograms of the redshift, black hole mass, disk luminosity, size of the BLR and DT, and $\gamma$-ray luminosity, aside from the accretion regime and the radio power, shown in Figs.\ref{fig:accretion regime no highlights}-\ref{fig:accretion vs radio no highlight}. The vertical black solid line represents the median for our sample.

\begin{figure}
    \centering
    \includegraphics[width = \linewidth]{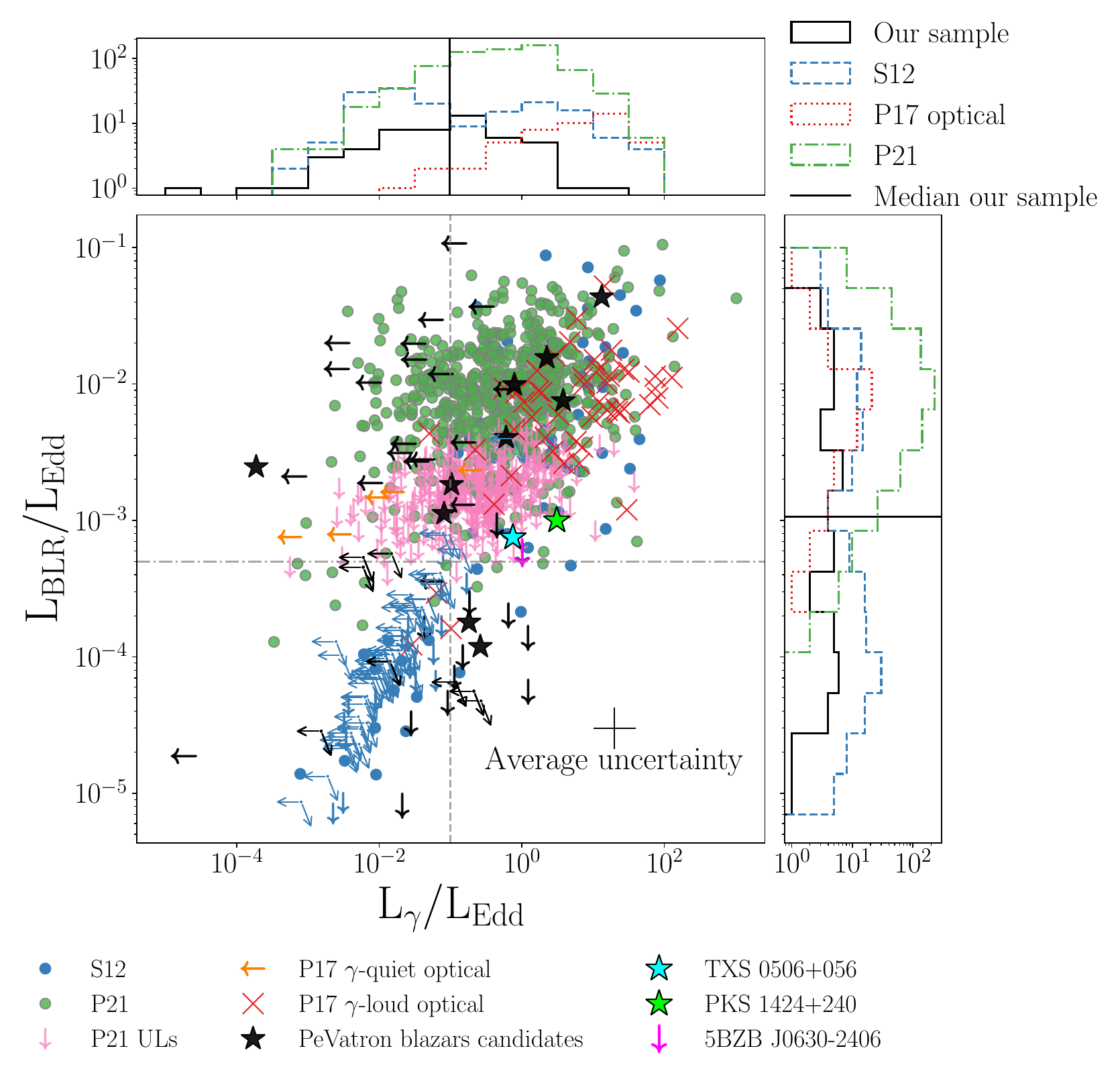}
    \caption{$\accretion$  (accretion regime proxy) vs. $\gratio$ (power of the jet proxy) of the candidate PeVatron blazars (black). The three blue FSRQs (masquerading BL Lacs) \TXS, \PKS, and \Gsource\ that were previously identified as promising neutrino-emitter blazar candidates are highlighted in cyan, lime, and magenta, respectively (see Section \ref{subsec: masquerading}). As a comparison, the plot also shows the samples of S$12$, P$17$, and P$21$ (see Section \ref{subsec: comparison samples}). For P$17$, we plot the subsample with properties analyzed through optical spectroscopy. The arrows represent upper limits on either one or both of the shown quantities (see also Appendix \ref{sec:append upper lim}). The dotted gray lines represent the boundaries for the different accretion efficiencies, respectively $\accretion\sim5\times10^{-4}$ and $\gratio\sim0.1$ \citep{Ghisellini:2011, Sbarrato:2012}. On the sides are the histograms of the corresponding quantities for all the samples, excluding all the upper limits. Here the black solid lines represent the median values for the candidate PeVatron blazars.
    }
    \label{fig:accretion regime no highlights}
\end{figure}
\begin{figure}
    \centering
    \includegraphics[width = 0.93\linewidth]{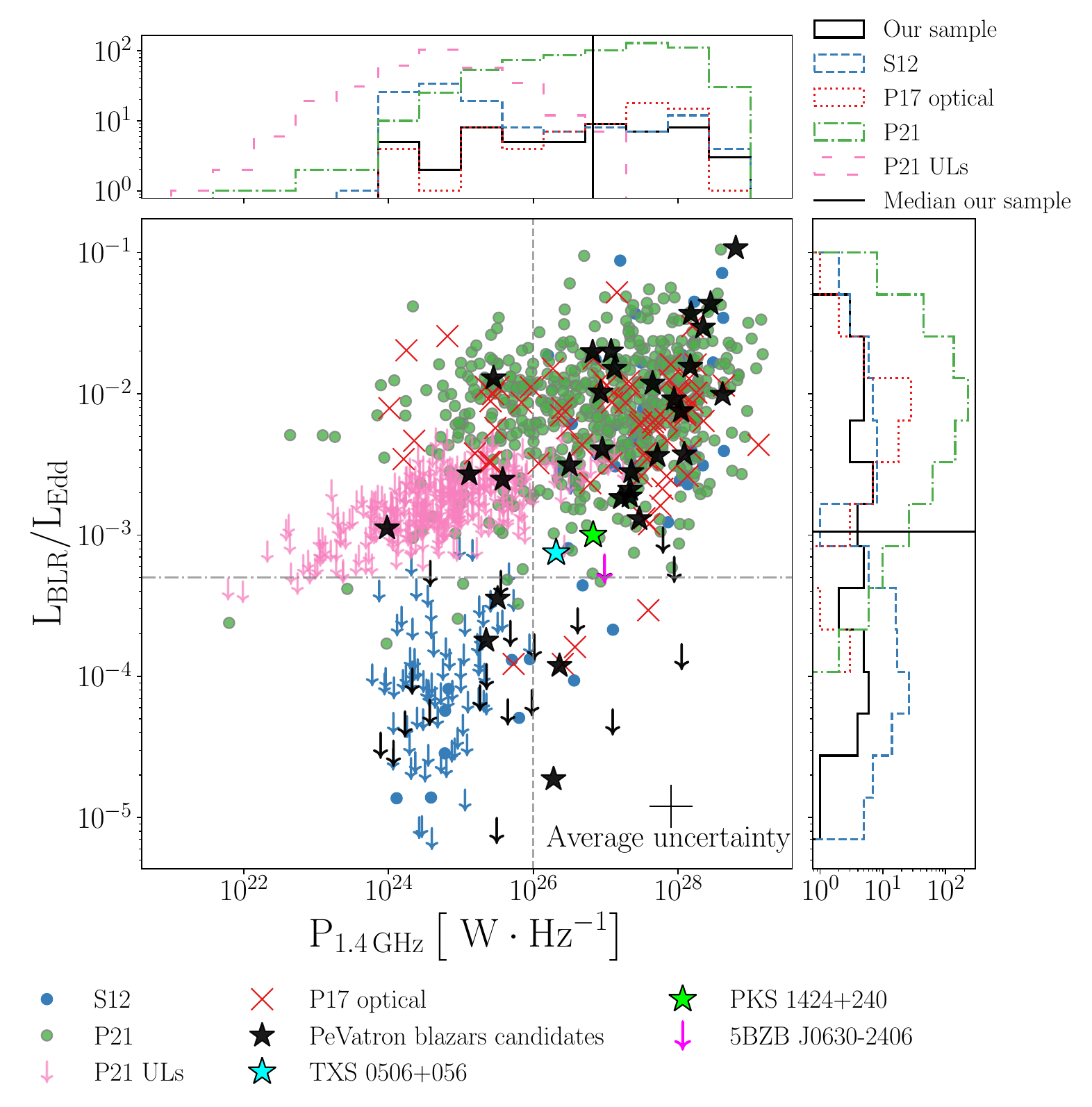}
    \caption{\label{fig:accretion vs radio no highlight} Accretion regime $\accretion$ as a function of the radio power at $1.4\,{\rm GHz}$, compared to the samples of \citep[S$12$, P$17$, and P$21$ in the plot, ][see Section \ref{subsec: comparison samples} for further details]{Sbarrato:2012, Paliya:2017, Paliya:2021}. For P$17$ we plot the subsample whose properties are analyzed through optical spectroscopy. The arrows indicate the upper limits on the optical luminosity. The three blue FSRQs (masquerading BL Lacs) \TXS, \PKS, and \Gsource\ that were already identified as promising neutrino-emitter blazar candidates are highlighted in cyan, lime, and magenta, respectively (see Section \ref{subsec: masquerading}). The dashed horizontal gray line represents the separation limits for the accretion efficiency $\accretion\sim5\times10^{-4}$ \citep{Ghisellini:2011, Sbarrato:2012}. Instead, the vertical dotted line is the threshold $\Pradio \sim 10^{26} \, {\rm W}\cdot {\rm Hz}^{-1}$ above which HERGs dominate over LERGs \citep{BestHeckman:2012, Padovani:2019}. On the sides are the normalized histograms of the corresponding quantities for the samples, excluding all the upper limits. Here, the black solid lines represent the median value for the candidate PeVatron blazars.
    }
\end{figure}

\subsection{Statistical analysis procedure}
\label{subsec: statistical analysis}

We addressed to which extent the properties of the target sample, i.e., the candidate neutrino-emitter objects, are compatible with those of the other reference samples, by means of statistical tests.
We quantified this first by using the Anderson-Darling (A-D) statistical test \citep{Anderson_Darling} on the physical properties of different populations, considering only the measurements of the interested quantity (i.e., excluding limits). As a check, we then included the upper limits, as if they were actual measurements, and by repeating the same analysis we verified that the overall findings of the A-D test remained consistent.
\\
However, this approach is not ideal and has limited statistical power for samples which contain both measured values and a significant fraction of upper limits. To address this limitation, we explored the applicability of alternative methods that incorporate upper limits, as utilized in previous studies, i.e., the Kaplan-Meier estimator used in combination with the logrank and the Peto logrank statistical tests \citep[e.g., ][]{statistical_methods_I, Padovani:1992, Padovani:2022, Paiano:2023}.
\\
To this end, the Kaplan-Meier estimator was used to estimate the survival function of a given physical property, in a stepwise manner, considering both measured values and including censored data \citep[limits; similar to the approach in][]{ASURV, statistical_methods_I, statistical_methods_II, statistical_methods_III, statistical_methods_IV}.
We, then, assessed the performance of the logrank and the Peto logrank tests for comparing survival curves across different samples through simulations, focusing in particular on Type I errors, i.e., false positives.  
As discussed in  Appendix \ref{sec: append survival analysis}, these methods are highly sensitive to the number and distribution of censored data points, as well as variations in sample sizes, requiring strong caution when interpreting the outcomes. Tailored investigations can assess the robustness of each statistically significant result individually (see Appendix \ref{sec: append survival analysis} for further details).
In this study, we present the results only for the Peto logrank test, which we found to be less affected by differences in the censoring patterns between the two samples \citep[as previously discussed in][]{statistical_methods_I}. For reference, we performed the analysis by using the \texttt{PYTHON} package \texttt{lifelines}\footnote{See \url{https://lifelines.readthedocs.io/en/latest/Survival\%20analysis\%20with\%20lifelines.html} for further details}, and the \texttt{ASURV} routine \citep{ASURV}, finding consistent results between the two.
\\
The main findings are summarized in Table \ref{table: AD test}. The first column reports the analyzed quantity, while the second, third, and fourth columns detail the tested samples, including the number of measurements and the number of upper limits either excluded or included in the analysis. The fifth and sixth columns present the resulting pre-trial pvalues from the A-D and Peto logrank tests, respectively.
Our focus is on comparisons that yield significant results in the pre-trial Peto logrank test, noting that the corresponding A-D test outcomes generally follow those of the Peto logrank test. Before discussing the implications of the observed findings, there are a few caveats to note.
%%%
%%%%%%%
\begin{itemize}
    \item 
    The tests we performed are strongly correlated. 
    A statistically significant result in one test can increase the likelihood of significant findings in related tests, due to the inherent correlations among the variables. This interdependence can affect the interpretation of pvalues and the overall significance of the findings.
    \item 
        When conducting multiple statistical tests, the likelihood of observing at least one significant result due to chance increases, that is known as the look-elsewhere effect. To account for this, a correction is needed to limit the overall Type I error rate.
\end{itemize}
%%%%%%
%%%
In this exploratory study, where identifying potential leads for further investigation is a priority, we used the Benjamini-Hochberg procedure to correct for trials \citep{Benjamini_Hochberg_1995}. This is suitable for independent and positively correlated tests, and offers a more balanced approach by controlling the false discovery rate (FDR), allowing for a higher rate of true discoveries while maintaining control over false positives.

To correct for trials, first we ranked the pre-trial pvalues in ascending order. For each pvalue, we computed the Benjamini-Hochberg critical value using the formula,

\begin{center}
Critical Value = $\frac{i}{m} \times$ Q     
\end{center}
\noindent

where $i$ is the rank, $m$ the total number of tests (i.e., $32$ in our case), and Q the chosen FDR level (i.e., $0.003$ in our case).

In this case, the Benjamini-Hochberg critical values are,

    \begin{align}
    \text{rank }\, \text{i} &= 1: \quad 1 \times 0.003/32 \sim 9.4 \times 10^{-5}, \nonumber \\
    \text{rank }\, \text{i} &= 2: \quad 2 \times 0.003/32 \sim 1.9 \times 10^{-4}, \nonumber \\
    \text{rank }\, \text{i} &= 3: \quad 3 \times 0.003/32 \sim 2.8 \times 10^{-4}, \nonumber \\
    \text{rank }\, \text{i} &= 4: \quad 4 \times 0.003/32 \sim 3.7 \times 10^{-4}, \nonumber \\
    \text{etc.}  \nonumber
\end{align}

We identified that the third-ranked pre-trial pvalue,  $1.53 \times 10^{-4}$, obtained from comparing the radio power of the target sample to the S$12$ one, is the largest pvalue that does not exceed its corresponding critical value. Following the Benjamini-Hochberg procedure, all pvalues ranked up to and including this rank are considered statistically significant. They are highlighted with a $\ast$ symbol in Table \ref{table: AD test} (see Appendix \ref{subsec: app applicability}).

%%%%%%%%%%%%%%%%%%%%%%%%%%
%%%%%%%%%%%%%%%%%%%%%%%%%%
\subsection{Discussion}

The three highest post-trial significances $(\gtrsim3\sigma)$ are observed for the comparison of $\mbh$  between the target sample and P$17$, while the other two are found when comparing the redshift and $\Pradio$ with S$12$. All of the other quantities are found statistically compatible with the reference samples, according to the  Benjamini$-$Hochberg test.
\\
Comparing the $\mbh$ distributions between the target sample and the P$17$ sample, we observed a significance $(\gtrsim3\sigma)$ and noted that the former sample includes $38\%$ censored data. To assess the robustness of this result, we conducted simulations to evaluate potential biases introduced by the high rate of censored data. These simulations, presented in  Appendix \ref{sec: append survival analysis}, suggest that the observed discrepancy may arise from an intrinsic difference between the two samples. The mass distribution of the candidate PeVatron blazars extends to larger values, with two objects above $\gtrsim10^{10}\Msun$ and with $\sim35\%$ of the objects (18 out of 52) above $\gtrsim 10^{9}\,\Msun$, compared to the P17 sample, which instead comprises objects with $\mbh$ values up to $2.40\times10^{9}\,\Msun$, six of which ($\sim13\%$) above $\gtrsim 10^{9}\,\Msun$. 
This suggests that our target sample tends to host more massive black holes than those in P17. However, no significant discrepancy is observed when compared to other reference samples, indicating that these properties remain consistent with the overall population of $\gamma$-loud blazars (P21) and sources with diverse accretion properties (S12).

Referring to Fig. \ref{fig:accretion vs radio no highlight}, most candidate PeVatron blazars ($\sim63\%$) show powerful relativistic jets and occupy the region of the $\Pradio$ distribution dominated by HERGs \citep{Padovani:2022, Kalfountzou:2012, BestHeckman:2012}. In this case, the median value is $\tilde{\mu}\left(\Pradio\right)= 6.67\times10^{26}\,{\rm W} \cdot {\rm Hz}^{-1}$, above the $\Pradio\sim10^{26}\,{\rm W}\cdot{\rm Hz}^{-1}$ HERG--LERG threshold. As displayed in Table \ref{table: AD test}, the $\Pradio$ distribution of our sample is compatible with the other samples except in the case of S$12$, for which a difference at $\gtrsim3\sigma$ level with the Peto logrank is found. A similar difference is found with the redshift distribution of S$12$. The distributions of the S$12$ sample are skewed toward lower redshifts and reduced radio powers relative to our sample. Among the S$12$ sample, only the $29\%$ of the objects ($40$ out of $138$) lie above $\Pradio\gtrsim10^{26}\,{\rm W}\cdot{\rm Hz}^{-1}$. As mentioned in Section \ref{subsec: comparison samples}, the S12 reference sample allows for an exploration of the full parameter space of HERG and LERG populations. These findings reinforce, from both the $\Pradio$ and redshift distribution perspectives, the mild tendency shown by the candidate neutrino-emitting blazars to preferably exhibiting HERG-like characteristics.
\\
A consistent picture to the $\Pradio$ distribution emerges looking at the $\accretion$  properties: candidate PeVatron blazars show a mild preference ($\sim 61\%$) for populating the region of the $\accretion$ occupied by sources with efficient accretion properties and, therefore, by HERG-like objects. Although from a statistical standpoint their accretion properties remain consistent with those of the other samples, the effectiveness of these tests is limited by the presence of a large number of censored data.
As shown in Fig.~\ref{fig:accretion regime no highlights}, our sample displays a median value $\tilde{\mu} \left( \accretion \right) = 10^{-3}$ and $\tilde{\mu} \left( \gratio \right) = 9.73\times10^{-2}$. When excluding objects with upper limits, as expected, $\sim 87\%$ occupy the HERG region, with $\tilde{\mu} \left( \accretion \right) = 3.76\times10^{-3}$ and $\tilde{\mu} \left( \gratio \right) = 6.77\times10^{-1}$.
\\
For what concerns the $\gamma$-ray luminosity, the blazars in our sample ($24\,\gamma$-loud and $28\,\gamma$-quiet) are distributed among a broad range of values, from $1.43\times10^{42}\,{\rm erg}\cdot{\rm s}^{-1}$ to $1.21\times10^{48}\,{\rm erg}\cdot{\rm s}^{-1}$. The median value of the detections is $\tilde{\mu}\left(L_{\gamma}\right)= 2.35\times10^{46}\,{\rm erg} \cdot {\rm s}^{-1}$, and the $79\%$ lies between $10^{44}\lesssim L_{\gamma} \lesssim 10^{47}\,{\rm erg}\cdot {\rm s}^{-1}$.
In terms of $\gamma$-ray luminosity, the candidate PeVatron blazars are compatible with the overall population, as suggested by the results of the Peto logrank test with the $5$BZCat and $4$LAC-DR$3$.

\begin{figure*}
    \centering
    \includegraphics[width = \textwidth]{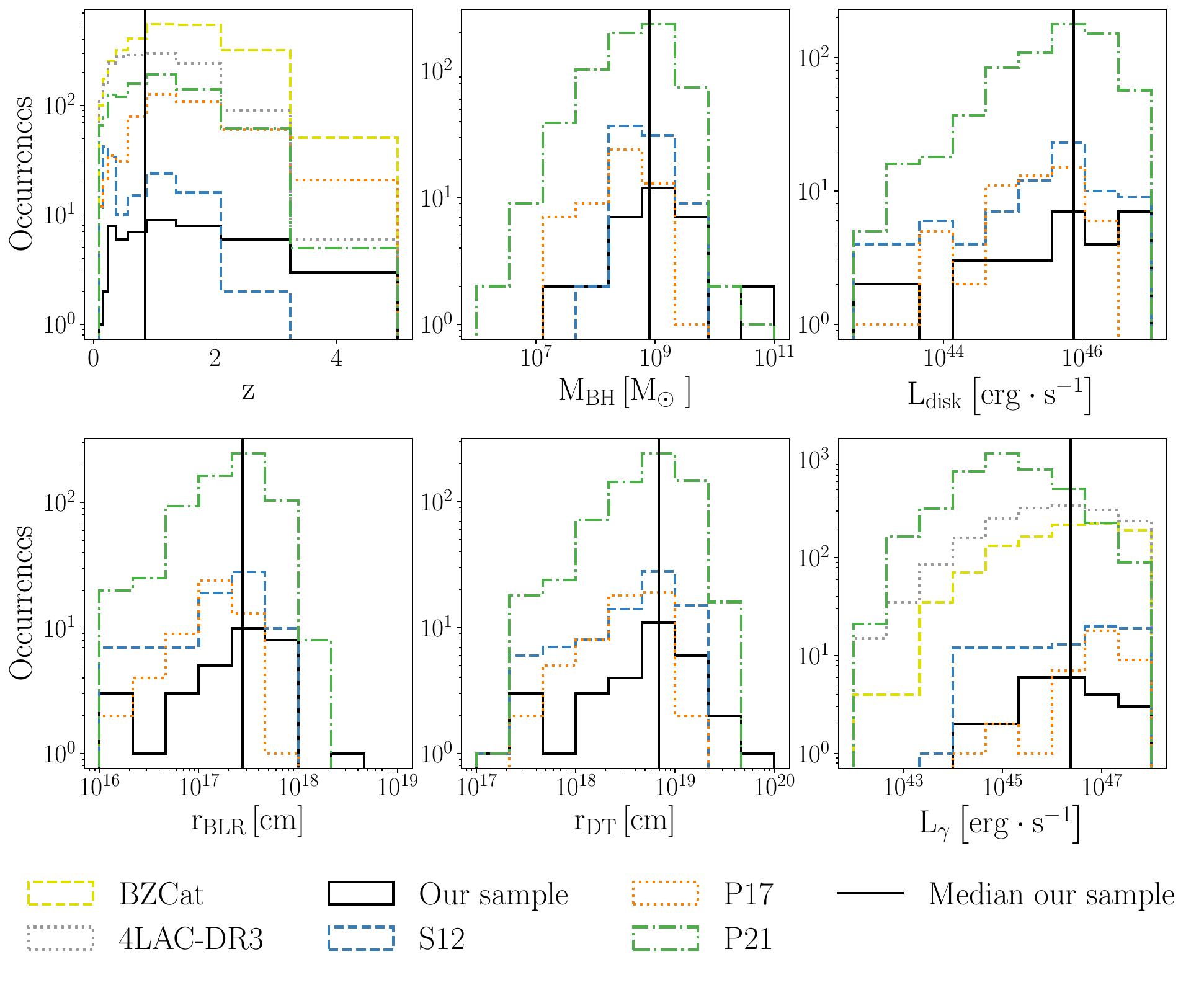}
    \caption{Distribution of the redshift, black hole mass, disk luminosity, size of the broad line region and dusty torus, and $\gamma$-ray luminosity. The comparison samples are \citet{Sbarrato:2012, Paliya:2017, Paliya:2021} (S$12$, P$17$, P$21$, respectively), BZCat and $4$LAC-DR$3$, with the exclusion of upper limits on the plotted quantity. The black solid lines correspond to the median value for the candidate PeVatron blazars.}
    \label{fig: histograms}
\end{figure*}

\subsection{Blue FSRQs -- masquerading BL Lacs among the candidate PeVatron blazars}
\label{subsec: masquerading}

Among the sample of candidate PeVatron blazars, four objects had been proposed as blue FSRQs (masquerading BL Lacs) in the literature:
\TXS, \PKS, \Gsource, and \RXJ. In this work, we conclude that the first three of them share HERG-like characteristics. Based on our newly acquired optical data of the fourth object, we establish the LERG nature of \RXJ.

The candidate IceCube neutrino source \TXS\ (5BZB~J0509$+$0541), was identified as belonging to the subclass of masquerading BL Lacs following the detection of the \OII\ and \OIII\ lines in an optical spectrum taken with the Gran Telescopio Canarias (GTC) between the end of 2017 and the beginning of 2018 \citep{Paiano:2018, Padovani:2019}.
Historically classified as a BL Lac, the HERG nature of this source is supported by the high radio power and \OII, \OIII\ luminosity ($\Pradio\sim 1.8\times10^{26}\,{\rm W}\cdot {\rm Hz}^{-1}$ and $L_{{\rm \OII},\,{\rm \OIII}}\sim 2\times 10^{41} \,{\rm erg}\cdot {\rm s}^{-1}$, respectively) and the accretion regimes $\accretion\sim 10^{-3}$, $\gratio\sim 0.7$ \citep{Padovani:2019}.

Similarly, the PeVatron blazar candidate \PKS\ (5BZB~J1427$+$2348) has been previously identified as a potential neutrino source and blue FSRQ (masquerading BL Lac). 
\PKS\ displayed \OII\ and \OIII\ emission lines in a Gran Telescopio Canarias (GTC) spectrum taken in 2015 \citep{Paiano:2017}. It was proposed as a HERG object based on the high power $\Pradio\sim 5\times10^{26}\,{\rm W}\cdot {\rm Hz}^{-1}$, $L_{\rm \OII}\sim 4\times 10^{41} \,{\rm erg}\cdot {\rm s}^{-1}$, and efficient accretion regimes $\accretion \gtrsim 1.5\times 10^{-3}$, $\gratio>2$ \citep{Padovani_PKS:2022}.

\Gsource\ has been traditionally classified as a BL Lac object. \citet{Ghisellini:2012} highlighted its high power, unusual for BL Lacs, and proposed that it belongs to the subclass of blue FSRQ (masquerading BL Lacs).
We analyzed the VLT spectrum obtained by \citet{Shaw:2013} in $2012$. We confirmed the absence of emission lines in the spectrum, and the lower limit estimate ${\rm z} \gtrsim 1.239$ for the redshift, which was placed by the authors based on the absorption features. 
A recent study estimated an upper limit $z < 1.33$ based on the effect of the extragalactic background light attenuation \citep[][]{Lainez_2024}.

Assuming the value of ${\rm z} \gtrsim 1.239$, we placed an upper limit on the luminosities of the \MgII\ $\lambda 2798$ \AA\ and ${\rm \OII} \lambda 3727$ \AA\ lines as explained in Section \ref{subsec:quantities}. 
 We found $L_{\rm disk}  < 2.53 \times 10^{45}~{\rm erg} \cdot {\rm s}^{-1}$, $\accretion < 5.79\times 10^{-4}$ and $\gratio\sim1.02$. 
In the plot of Figure \ref{fig:accretion regime no highlights}  it is located next to \TXS\ and \PKS, within the HERG regime. 
The inferred black hole mass is $<3.79 \times 10^9$, which is toward the upper hand of the values usually observed. We note that a lower BH mass estimate than this limit would lead to a higher accretion efficiency, keeping the validity of the HERG-nature hypothesis. 

The nature of this blazar appears consistent with the HERG-like behavior, as further supported by its radio power $\Pradio = 9.64\times10^{26}\,{\rm W}\cdot{\rm Hz}^{-1}$ and theoretical modeling. In a previous study, we modeled the quasi-simultaneous broadband spectral energy distribution of \Gsource, revealing that the source hosts a luminous ($L_{\rm disk} \sim 4 \times 10^{45}~{\rm erg} \cdot {\rm s}^{-1}$) accretion disk with accretion rates $\accretion\sim2\times10^{-4}$, $\gratio\sim0.15$ and dissipation radius on the outer edge of the BLR  \citep{FichetDC:2023}.
The conclusions are in agreement with the results of \citet{Ghisellini:2012}, in which a limit $L_{\rm disk} \sim 5.5 \times 10^{45}~{\rm erg} \cdot {\rm s}^{-1}$ was obtained using an empirical relation between $L_{\rm BLR}-L_{\gamma}$\footnote{The empirical relation links the BLR and $\gamma$-ray luminosity through $L_{\rm BLR}\propto 4\cdot L_{\gamma}^{0.93}$ (see \citet{Sbarrato:2012, Ghisellini:2012} for further details).}, and are in agreement with this work. The overall findings suggest the likely presence of external radiation fields (accretion disk, narrow and broad line regions) whose emission is overcome by the jet's synchrotron emission. \\

The nature of \RXJ\ as blue FSRQ was discussed in previous works \citep{Ghisellini:2012, Padovani:2012}. Starting from the featureless SDSS spectrum available for this source, the works assumed an average mass $\mbh = 5\times 10^{8}\,\Msun$ and adopted the photometric redshift $z = 1.28$ to derive the values $L_{\gamma} \sim 10^{45} {\rm erg} \cdot {\rm s}^{-1}$, $L_{\rm BLR} < 3\times 10^{44} {\rm erg} \cdot {\rm s}^{-1}$ and $\Pradio = 1.58\times 10^{26}\,{\rm W} \cdot {\rm Hz}^{-1}$, leading to the blue FSRQ scenario. However, the lack of a spectroscopic redshift determination at the time prevented the authors to settle the debate. As noted, a smaller value of the redshift would result in lower luminosity estimations and different conclusions, i.e., the source is intrinsically a LERG \citep{Ghisellini:2012}. 
Our new GTC OSIRIS observation resulted in the detection of \MgII\ and \OII\ emission lines. We estimated a redshift of ${\rm z} = 0.57$, which is consistent with the lower limit of ${\rm z} > 0.55$ reported \citep{Paiano:2017}.
We derived $\accretion \sim 1.79\times 10^{-4}$ and $\gratio\sim 0.182$, while for the radio power, we found a value below the $\sim 10^{26} \,{\rm W}\cdot{\rm Hz}^{-1}$. Therefore, we conclude that \RXJ\ belongs to the LERG class, based on the accretion regime and radio power properties. This is consistent with what was already foreseen in case of a redshift lower than the photometric estimation \citep{Ghisellini:2012}.

Adopting the physically motivated classification describes the three known blue FSRQs (masquerading BL Lacs) \TXS, \PKS, and \Gsource\ included in our sample with less ambiguity as high emitting power sources with their broad emission lines swamped by the jet synchrotron emission \citep{Giommi:2013, Padovani:2022}. The implications of their intrinsic nature are explored in further detail in the next Section \ref{subsec: multimessenger implications} in the context of neutrino stacking studies. \TXS, \PKS, and \Gsource\  are highlighted in cyan, lime, and magenta in Figs.\ref{fig:accretion regime no highlights}-\ref{fig:accretion vs radio no highlight}. Their location beside the other HERG-like sources is consistent with the mild tendency shown by the candidate PeVatron blazars to preferentially exhibit intense radiation fields, radiatively efficient accretion and powerful jets.

\subsection{Implications for neutrino multimessenger studies}
\label{subsec: multimessenger implications}
Several studies have been hampered by blindly adhering to the historical categorization of objects, sometimes relying solely on catalog classifications or attempting to fit them in observation-based schemes, leading to classifications that change over time \citep[][]{Antonucci:2012}. A notorious example in the field of multimessenger neutrino astrophysics is \TXS, the first probable high-energy neutrino source. Due to its featureless optical spectrum, historically it was classified as a BLL and therefore considered a LERG-like object with no prominent radiation fields. Later, it was proposed to be an FSRQ-like object, therefore being intrinsically a HERG. 
As addressed in the previous section, this is not a standalone case in the context of candidate neutrino-emitter blazars. The impact on previous neutrino studies focused on blazars may be not negligible.

For instance, several works investigated the contribution of $\gamma$-ray blazars to the IceCube neutrino diffuse flux  \citep{IceCube2017_2LAC}.
The studies performed a neutrino stacking analysis of the whole blazar sample, along with an analysis of the separate BLL and FSRQ subsamples. They derived limits on the maximal contributions to the diffuse neutrino flux: $19\%-27\%$ for the whole blazar sample; $3\%-13\%$ for BL Lacs, and $5\%-17\%$ for FSRQs \citep{IceCube2017_2LAC}. 

The rationale behind examining different blazar subpopulations was due to their distinct physical environments and potential neutrino production mechanisms. In FSRQs, characterized by strong radiation fields, the leading hypothesis for neutrino production is proton-photon interactions. In contrast, BL Lacs, with no or weaker radiation fields, are hypothesized to produce neutrinos primarily through proton-synchrotron mechanisms \citep{Mucke:2003}.
In light of their observational-driven classification, the aforementioned stacking study listed \TXS, \PKS, and \Gsource\  among the BLL subsample. However, as discussed in Section \ref{subsec: masquerading}, they belong to the HERG class and display quasar-like properties, hence should be accounted for within the FSRQ subsample.
As pointed out by our earlier works (\origin, \originII), these promising neutrino-emitters are associated with fairly prominent anisotropies in the time-integrated IceCube skymap, therefore their contribution to a stacking analysis would unlikely be negligible. Similar objects may be present in the considered sample of $\gamma$-ray blazars, and erroneously included in the BL Lacs subpopulation.
Our findings advocate for the need to embrace the physical-driven selection of blazars and put into question the limits (possibly also the nondetection) derived from the individual subsamples.

\section{Summary and conclusions}
\label{sec:conclusions}

In this work we studied a selected sample of candidate neutrino-emitter blazars proposed as counterparts of IceCube hotspots (i.e., the $52$ candidate PeVatron blazars), using a multiwavelength approach, collecting information in optical, radio, and $\gamma$-rays. The sample includes sources classified both as BL Lacs and FSRQs according to the traditional scheme based on the EW of the spectral lines. To infer the nature of these objects, we adopted a physically driven approach, which allowed us to pinpoint with less ambiguity the HERGs or LERGs properties \citep{Ghisellini:2011, Sbarrato:2012, Sbarrato:2014, BestHeckman:2012, Padovani:2022}. 
To this end, we investigated the mass accretion properties of the central engine and the relativistic jet power. We performed statistical tests to compare the properties of our target sample of candidate neutrino-emitter blazars with reference samples from the literature. We explored the applicability of methods that include limits in the analysis, showing that these are highly sensitive to the number and distributions of limits, and sample sizes, requiring caution when interpreting the results.

Our work suggests discrepancies at the $\gtrsim3\sigma$ level post-trial in three cases, where the black hole mass, redshift and radio power of the candidate PeVatron blazars extend to larger values compared to two of the reference samples. The findings suggest that these blazars are more HERG-like in terms of jet power, as evidenced by a slightly higher representation of HERG sources. This may indicate a preference for objects with higher accretion efficiency, stronger radiation fields, and significant radio power.
In a multimessenger view, this would be consistent with theoretical models that predict enhanced neutrino production in such environments \citep{Dermer:2014}, and with evidence supporting radio-bright objects as neutrino candidate emitters \citep{Plavin:2020, Plavin_2021, Zhou:2021, Kun:2022, Plavin:2023, Kovalev_2023, Kouch:2024}.

The presence of blue FSRQs (also known as masquerading BLLs) among candidate PeVatron blazars warrants caution in interpreting previous stacking studies that constrain the diffuse flux contribution from different blazar subpopulations. Our findings challenge the limit, and potentially even the nondetection, derived from individual subsamples \citep{IceCube2017_2LAC}, and emphasize the need for a physically driven selection of blazars.
While our statistical analysis did not provide conclusive evidence for a difference in the overall behavior of the candidate neutrino-emitter blazars, this initial investigation of the candidate PeVatron blazar sample revealed a broad range of properties.
The challenge in confirming individual neutrino -- blazar associations currently limits definitive conclusions about the properties of candidate neutrino-emitter blazars. Future dedicated studies will focus on assessing the genuineness of these associations to shed light on the connection between blazars and neutrinos.

\begin{acknowledgements}
The authors thank the anonymous referee for the insightful comments that helped improve the manuscript.
We are thankful to Andrea Tramacere, Stefano Marchesi, Marco Ajello, Paola Marziani, Nicola Masetti, Roger Romani, and Karl Mannheim for the constructive discussions. AA thanks German Gimeno and Kathleen Labrie for their support during the phases of observation and reduction of the Gemini data, and Tapio Pursimo for the help with the NOT spectrum.
This work was supported by the European Research Council, ERC Starting grant \emph{MessMapp}, S.B. Principal Investigator, under contract no. $949555$. 
Based on observations collected at the European Southern Observatory under ESO programs $109.22{\rm WZ}.001$, $109.22{\rm WZ}.001$, $084.{\rm B}-0711({\rm A})$, $087.{\rm B}-0614({\rm A})$, $077.{\rm B}-0045$. 
Based on observations obtained at the international Gemini Observatory, a program of NSF’s NOIRLab [Program ID \texttt{GS-2022B-DD-201}], which is managed by the Association of Universities for Research in Astronomy (AURA) under a cooperative agreement with the National Science Foundation on behalf of the Gemini Observatory partnership: the National Science Foundation (United States), National Research Council (Canada), Agencia Nacional de Investigaci\'{o}n y Desarrollo (Chile), Ministerio de Ciencia, Tecnolog\'{i}a e Innovaci\'{o}n (Argentina), Minist\'{e}rio da Ci\^{e}ncia, Tecnologia, Inova\c{c}\~{o}es e Comunica\c{c}\~{o}es (Brazil), and Korea Astronomy and Space Science Institute (Republic of Korea).
This work is (partly) based on data obtained with the instrument OSIRIS, built by a Consortium led by the Instituto de Astrofísica de Canarias in collaboration with the Instituto de Astronomía of the Universidad Autónoma de México. OSIRIS was funded by GRANTECAN and the National Plan of Astronomy and Astrophysics of the Spanish Government. This work is (partly) based on data from the GTC Public Archive at CAB (INTA-CSIC), developed in the framework of the Spanish Virtual Observatory project supported by the Spanish MINECO through grants AYA 2011-24052 and AYA 2014-55216. The system is maintained by the Data Archive Unit of the CAB (INTA-CSIC).
This research has made use of the NASA/IPAC Extragalactic Database, which is funded by the National Aeronautics and Space Administration and operated by the California Institute of Technology.
Guoshoujing Telescope (the Large Sky Area Multi-Object Fiber Spectroscopic Telescope LAMOST) is a National Major Scientific Project built by the Chinese Academy of Sciences. Funding for the project has been provided by the National Development and Reform Commission. LAMOST is operated and managed by the National Astronomical Observatories, Chinese Academy of Sciences.
The 2dF QSO Redshift Survey (2QZ) was compiled by the 2QZ survey team from observations made with the $2$-degree Field on the Anglo-Australian Telescope. 
Funding for the Sloan Digital Sky Survey IV has been provided by the Alfred P. Sloan Foundation, the U.S. Department of Energy Office of Science, and the Participating Institutions. SDSS-IV acknowledges support and resources from the Center for High-Performance Computing at the University of Utah. The SDSS website is www.sdss4.org. SDSS-IV is managed by the Astrophysical Research Consortium for the Participating Institutions of the SDSS Collaboration including the Brazilian Participation Group, the Carnegie Institution for Science, Carnegie Mellon University, Center for Astrophysics | Harvard \& Smithsonian, the Chilean Participation Group, the French Participation Group, Instituto de Astrof\'isica de Canarias, The Johns Hopkins University, Kavli Institute for the Physics and Mathematics of the Universe (IPMU) / University of Tokyo, the Korean Participation Group, Lawrence Berkeley National Laboratory, Leibniz Institut f\"ur Astrophysik Potsdam (AIP),  Max-Planck-Institut f\"ur Astronomie (MPIA Heidelberg), Max-Planck-Institut f\"ur Astrophysik (MPA Garching), Max-Planck-Institut f\"ur Extraterrestrische Physik (MPE), National Astronomical Observatories of China, New Mexico State University, New York University, University of Notre Dame, Observat\'ario Nacional / MCTI, The Ohio State University, Pennsylvania State University, Shanghai Astronomical Observatory, United Kingdom Participation Group, Universidad Nacional Aut\'onoma de M\'exico, University of Arizona, University of Colorado Boulder, University of Oxford, University of Portsmouth, University of Utah, University of Virginia, University of Washington, University of Wisconsin, Vanderbilt University, and Yale University.
This research has made use of the SIMBAD database, operated at CDS, Strasbourg, France. 
Part of this work is based on archival data, software, or online services provided by the Space Science Data Center - ASI.
\end{acknowledgements}

\bibliographystyle{aa}
\bibliography{aa51328-24}

\begin{thebibliography}{164}
\expandafter\ifx\csname natexlab\endcsname\relax\def\natexlab#1{#1}\fi

\bibitem[{{Aartsen} {et~al.}(2017{\natexlab{a}}){Aartsen}, {Abraham},
  {Ackermann}, {Adams}, {Aguilar}, {Ahlers}, {Ahrens}, {Altmann}, {Andeen},
  {Anderson}, {Ansseau}, {Anton}, {Archinger}, {Arguelles}, {Arlen},
  {Auffenberg}, {Axani}, {Bai}, {Barwick}, {Baum}, {Bay}, {Beatty}, {Becker
  Tjus}, {Becker}, {BenZvi}, {Berghaus}, {Berley}, {Bernardini}, {Bernhard},
  {Besson}, {Binder}, {Bindig}, {Bissok}, {Blaufuss}, {Blot}, {Boersma},
  {Bohm}, {B{\"o}rner}, {Bos}, {Bose}, {B{\"o}ser}, {Botner}, {Braun},
  {Brayeur}, {Bretz}, {Burgman}, {Casey}, {Casier}, {Cheung}, {Chirkin},
  {Christov}, {Clark}, {Classen}, {Coenders}, {Collin}, {Conrad}, {Cowen},
  {Cruz Silva}, {Daughhetee}, {Davis}, {Day}, {de Andr{\'e}}, {De Clercq}, {del
  Pino Rosendo}, {Dembinski}, {De Ridder}, {Desiati}, {de Vries}, {de
  Wasseige}, {de With}, {DeYoung}, {D{\'\i}az-V{\'e}lez}, {di Lorenzo},
  {Dujmovic}, {Dumm}, {Dunkman}, {Eberhardt}, {Ehrhardt}, {Eichmann}, {Euler},
  {Evenson}, {Fahey}, {Fazely}, {Feintzeig}, {Felde}, {Filimonov}, {Finley},
  {Flis}, {F{\"o}sig}, {Franckowiak}, {Fuchs}, {Gaisser}, {Gaior}, {Gallagher},
  {Gerhardt}, {Ghorbani}, {Giang}, {Gladstone}, {Glagla}, {Gl{\"u}senkamp},
  {Goldschmidt}, {Golup}, {Gonzalez}, {G{\'o}ra}, {Grant}, {Griffith}, {Haack},
  {Haj Ismail}, {Hallgren}, {Halzen}, {Hansen}, {Hansmann}, {Hansmann},
  {Hanson}, {Hebecker}, {Heereman}, {Helbing}, {Hellauer}, {Hickford},
  {Hignight}, {Hill}, {Hoffman}, {Hoffmann}, {Holzapfel}, {Homeier}, {Hoshina},
  {Huang}, {Huber}, {Huelsnitz}, {Hultqvist}, {In}, {Ishihara}, {Jacobi},
  {Japaridze}, {Jeong}, {Jero}, {Jones}, {Jurkovic}, {Kappes}, {Karg}, {Karle},
  {Katz}, {Kauer}, {Keivani}, {Kelley}, {Kemp}, {Kheirandish}, {Kim},
  {Kintscher}, {Kiryluk}, {Kittler}, {Klein}, {Kohnen}, {Koirala}, {Kolanoski},
  {Konietz}, {K{\"o}pke}, {Kopper}, {Kopper}, {Koskinen}, {Kowalski}, {Krings},
  {Kroll}, {Kr{\"u}ckl}, {Kr{\"u}ger}, {Kunnen}, {Kunwar}, {Kurahashi},
  {Kuwabara}, {Labare}, {Lanfranchi}, {Larson}, {Lennarz}, {Lesiak-Bzdak},
  {Leuermann}, {Leuner}, {Lu}, {L{\"u}nemann}, {Madsen}, {Maggi}, {Mahn},
  {Mancina}, {Mandelartz}, {Maruyama}, {Mase}, {Maunu}, {McNally}, {Meagher},
  {Medici}, {Meier}, {Meli}, {Menne}, {Merino}, {Meures}, {Miarecki},
  {Middell}, {Mohrmann}, {Montaruli}, {Moulai}, {Nahnhauer}, {Naumann}, {Neer},
  {Niederhausen}, {Nowicki}, {Nygren}, {Obertacke Pollmann}, {Olivas},
  {Omairat}, {O'Murchadha}, {Palczewski}, {Pandya}, {Pankova}, {Penek},
  {Pepper}, {P{\'e}rez de los Heros}, {Pfendner}, {Pieloth}, {Pinat},
  {Posselt}, {Price}, {Przybylski}, {Quinnan}, {Raab}, {R{\"a}del}, {Rameez},
  {Rawlins}, {Reimann}, {Relich}, {Resconi}, {Rhode}, {Richman}, {Riedel},
  {Robertson}, {Rongen}, {Rott}, {Ruhe}, {Ryckbosch}, {Rysewyk}, {Sabbatini},
  {Sanchez Herrera}, {Sandrock}, {Sandroos}, {Sarkar}, {Satalecka}, {Schimp},
  {Schlunder}, {Schmidt}, {Schoenen}, {Sch{\"o}neberg}, {Sch{\"o}nwald},
  {Schumacher}, {Seckel}, {Seunarine}, {Soldin}, {Song}, {Spiczak}, {Spiering},
  {Stahlberg}, {Stamatikos}, {Stanev}, {Stasik}, {Steuer}, {Stezelberger},
  {Stokstad}, {St{\"o}{\ss}l}, {Str{\"o}m}, {Strotjohann}, {Sullivan},
  {Sutherland}, {Taavola}, {Taboada}, {Tatar}, {Ter-Antonyan}, {Terliuk},
  {Te{\v{s}}i{\'c}}, {Tilav}, {Toale}, {Tobin}, {Toscano}, {Tosi},
  {Tselengidou}, {Turcati}, {Unger}, {Usner}, {Vallecorsa}, {Vandenbroucke},
  {van Eijndhoven}, {Vanheule}, {van Rossem}, {van Santen}, {Veenkamp},
  {Vehring}, {Voge}, {Vraeghe}, {Walck}, {Wallace}, {Wallraff}, {Wandkowsky},
  {Weaver}, {Wendt}, {Westerhoff}, {Whelan}, {Wickmann}, {Wiebe}, {Wiebusch},
  {Wille}, {Williams}, {Wills}, {Wissing}, {Wolf}, {Wood}, {Woolsey},
  {Woschnagg}, {Xu}, {Xu}, {Xu}, {Yanez}, {Yodh}, {Yoshida}, {Zoll}, \&
  {IceCube Collaboration}}]{IceCube2017_2LAC}
{Aartsen}, M.~G., {Abraham}, K., {Ackermann}, M., {et~al.} 2017{\natexlab{a}},
  \apj, 835, 45

\bibitem[{{Aartsen} {et~al.}(2016){Aartsen}, {Abraham}, {Ackermann}, {Adams},
  {Aguilar}, {Ahlers}, {Ahrens}, {Altmann}, {Andeen}, {Anderson}, {Ansseau},
  {Anton}, {Archinger}, {Arg{\"u}elles}, {Auffenberg}, {Axani}, {Bai},
  {Barwick}, {Baum}, {Bay}, {Beatty}, {Becker Tjus}, {Becker}, {BenZvi},
  {Berghaus}, {Berley}, {Bernardini}, {Bernhard}, {Besson}, {Binder}, {Bindig},
  {Bissok}, {Blaufuss}, {Blot}, {Bohm}, {B{\"o}rner}, {Bos}, {Bose},
  {B{\"o}ser}, {Botner}, {Braun}, {Brayeur}, {Bretz}, {Burgman}, {Carver},
  {Casier}, {Cheung}, {Chirkin}, {Christov}, {Clark}, {Classen}, {Coenders},
  {Collin}, {Conrad}, {Cowen}, {Cross}, {Day}, {de Andr{\'e}}, {De Clercq},
  {del Pino Rosendo}, {Dembinski}, {De Ridder}, {Desiati}, {de Vries}, {de
  Wasseige}, {de With}, {DeYoung}, {D{\'\i}az-V{\'e}lez}, {di Lorenzo},
  {Dujmovic}, {Dumm}, {Dunkman}, {Eberhardt}, {Ehrhardt}, {Eichmann}, {Eller},
  {Euler}, {Evenson}, {Fahey}, {Fazely}, {Feintzeig}, {Felde}, {Filimonov},
  {Finley}, {Flis}, {F{\"o}sig}, {Franckowiak}, {Friedman}, {Fuchs}, {Gaisser},
  {Gallagher}, {Gerhardt}, {Ghorbani}, {Giang}, {Gladstone}, {Glagla},
  {Gl{\"u}senkamp}, {Goldschmidt}, {Golup}, {Gonzalez}, {Grant}, {Griffith},
  {Haack}, {Haj Ismail}, {Hallgren}, {Halzen}, {Hansen}, {Hansmann},
  {Hansmann}, {Hanson}, {Hebecker}, {Heereman}, {Helbing}, {Hellauer},
  {Hickford}, {Hignight}, {Hill}, {Hoffman}, {Hoffmann}, {Holzapfel},
  {Hoshina}, {Huang}, {Huber}, {Hultqvist}, {In}, {Ishihara}, {Jacobi},
  {Japaridze}, {Jeong}, {Jero}, {Jones}, {Jurkovic}, {Kappes}, {Karg}, {Karle},
  {Katz}, {Kauer}, {Keivani}, {Kelley}, {Kemp}, {Kheirandish}, {Kim},
  {Kintscher}, {Kiryluk}, {Kittler}, {Klein}, {Kohnen}, {Koirala}, {Kolanoski},
  {Konietz}, {K{\"o}pke}, {Kopper}, {Kopper}, {Koskinen}, {Kowalski}, {Krings},
  {Kroll}, {Kr{\"u}ckl}, {Kr{\"u}ger}, {Kunnen}, {Kunwar}, {Kurahashi},
  {Kuwabara}, {Labare}, {Lanfranchi}, {Larson}, {Lauber}, {Lennarz},
  {Lesiak-Bzdak}, {Leuermann}, {Leuner}, {Lu}, {L{\"u}nemann}, {Madsen},
  {Maggi}, {Mahn}, {Mancina}, {Mandelartz}, {Maruyama}, {Mase}, {Maunu},
  {McNally}, {Meagher}, {Medici}, {Meier}, {Meli}, {Menne}, {Merino}, {Meures},
  {Miarecki}, {Mohrmann}, {Montaruli}, {Moulai}, {Nahnhauer}, {Naumann},
  {Neer}, {Niederhausen}, {Nowicki}, {Nygren}, {Obertacke Pollmann}, {Olivas},
  {O'Murchadha}, {Palczewski}, {Pandya}, {Pankova}, {Peiffer}, {Penek},
  {Pepper}, {P{\'e}rez de los Heros}, {Pieloth}, {Pinat}, {Price},
  {Przybylski}, {Quinnan}, {Raab}, {R{\"a}del}, {Rameez}, {Rawlins}, {Reimann},
  {Relethford}, {Relich}, {Resconi}, {Rhode}, {Richman}, {Riedel}, {Robertson},
  {Rongen}, {Rott}, {Ruhe}, {Ryckbosch}, {Rysewyk}, {Sabbatini}, {Sanchez
  Herrera}, {Sandrock}, {Sandroos}, {Sarkar}, {Satalecka}, {Schimp},
  {Schlunder}, {Schmidt}, {Schoenen}, {Sch{\"o}neberg}, {Schumacher}, {Seckel},
  {Seunarine}, {Soldin}, {Song}, {Spiczak}, {Spiering}, {Stahlberg}, {Stanev},
  {Stasik}, {Steuer}, {Stezelberger}, {Stokstad}, {St{\"o}{\ss}l}, {Str{\"o}m},
  {Strotjohann}, {Sullivan}, {Sutherland}, {Taavola}, {Taboada}, {Tatar},
  {Tenholt}, {Ter-Antonyan}, {Terliuk}, {Te{\v{s}}i{\'c}}, {Tilav}, {Toale},
  {Tobin}, {Toscano}, {Tosi}, {Tselengidou}, {Turcati}, {Unger}, {Usner},
  {Vandenbroucke}, {van Eijndhoven}, {Vanheule}, {van Rossem}, {van Santen},
  {Veenkamp}, {Vehring}, {Voge}, {Vraeghe}, {Walck}, {Wallace}, {Wallraff},
  {Wandkowsky}, {Weaver}, {Weiss}, {Wendt}, {Westerhoff}, {Whelan}, {Wickmann},
  {Wiebe}, {Wiebusch}, {Wille}, {Williams}, {Wills}, {Wolf}, {Wood}, {Woolsey},
  {Woschnagg}, {Xu}, {Xu}, {Xu}, {Yanez}, {Yodh}, {Yoshida}, {Zoll}, \&
  {Icecube Collaboration}}]{IC_north_hard_spectrum:2016}
{Aartsen}, M.~G., {Abraham}, K., {Ackermann}, M., {et~al.} 2016, \apj, 833, 3

\bibitem[{{Aartsen} {et~al.}(2017{\natexlab{b}}){Aartsen}, {Abraham},
  {Ackermann}, {Adams}, {Aguilar}, {Ahlers}, {Ahrens}, {Altmann}, {Andeen},
  {Anderson}, {Ansseau}, {Anton}, {Archinger}, {Arg{\"u}elles}, {Auffenberg},
  {Axani}, {Bai}, {Barwick}, {Baum}, {Bay}, {Beatty}, {Becker Tjus}, {Becker},
  {BenZvi}, {Berley}, {Bernardini}, {Bernhard}, {Besson}, {Binder}, {Bindig},
  {Bissok}, {Blaufuss}, {Blot}, {Bohm}, {B{\"o}rner}, {Bos}, {Bose},
  {B{\"o}ser}, {Botner}, {Braun}, {Brayeur}, {Bretz}, {Bron}, {Burgman},
  {Carver}, {Casier}, {Cheung}, {Chirkin}, {Christov}, {Clark}, {Classen},
  {Coenders}, {Collin}, {Conrad}, {Cowen}, {Cross}, {Day}, {de Andr{\'e}}, {De
  Clercq}, {del Pino Rosendo}, {Dembinski}, {De Ridder}, {Desiati}, {de Vries},
  {de Wasseige}, {de With}, {DeYoung}, {D{\'\i}az-V{\'e}lez}, {di Lorenzo},
  {Dujmovic}, {Dumm}, {Dunkman}, {Eberhardt}, {Ehrhardt}, {Eichmann}, {Eller},
  {Euler}, {Evenson}, {Fahey}, {Fazely}, {Feintzeig}, {Felde}, {Filimonov},
  {Finley}, {Flis}, {F{\"o}sig}, {Franckowiak}, {Friedman}, {Fuchs}, {Gaisser},
  {Gallagher}, {Gerhardt}, {Ghorbani}, {Giang}, {Gladstone}, {Glauch},
  {Gl{\"u}senkamp}, {Goldschmidt}, {Golup}, {Gonzalez}, {Grant}, {Griffith},
  {Haack}, {Haj Ismail}, {Hallgren}, {Halzen}, {Hansen}, {Hansmann}, {Hanson},
  {Hebecker}, {Heereman}, {Helbing}, {Hellauer}, {Hickford}, {Hignight},
  {Hill}, {Hoffman}, {Hoffmann}, {Holzapfel}, {Hoshina}, {Huang}, {Huber},
  {Hultqvist}, {In}, {Ishihara}, {Jacobi}, {Japaridze}, {Jeong}, {Jero},
  {Jones}, {Jurkovic}, {Kappes}, {Karg}, {Karle}, {Katz}, {Kauer}, {Keivani},
  {Kelley}, {Kheirandish}, {Kim}, {Kintscher}, {Kiryluk}, {Kittler}, {Klein},
  {Kohnen}, {Koirala}, {Kolanoski}, {Konietz}, {K{\"o}pke}, {Kopper}, {Kopper},
  {Koskinen}, {Kowalski}, {Krings}, {Kroll}, {Kr{\"u}ckl}, {Kr{\"u}ger},
  {Kunnen}, {Kunwar}, {Kurahashi}, {Kuwabara}, {Labare}, {Lanfranchi},
  {Larson}, {Lauber}, {Lennarz}, {Lesiak-Bzdak}, {Leuermann}, {Lu},
  {L{\"u}nemann}, {Madsen}, {Maggi}, {Mahn}, {Mancina}, {Mandelartz},
  {Maruyama}, {Mase}, {Maunu}, {McNally}, {Meagher}, {Medici}, {Meier}, {Meli},
  {Menne}, {Merino}, {Meures}, {Miarecki}, {Mohrmann}, {Montaruli}, {Moulai},
  {Nahnhauer}, {Naumann}, {Neer}, {Niederhausen}, {Nowicki}, {Nygren},
  {Obertacke Pollmann}, {Olivas}, {O'Murchadha}, {Palczewski}, {Pandya},
  {Pankova}, {Peiffer}, {Penek}, {Pepper}, {P{\'e}rez de los Heros}, {Pieloth},
  {Pinat}, {Price}, {Przybylski}, {Quinnan}, {Raab}, {R{\"a}del}, {Rameez},
  {Rawlins}, {Reimann}, {Relethford}, {Relich}, {Resconi}, {Rhode}, {Richman},
  {Riedel}, {Robertson}, {Rongen}, {Rott}, {Ruhe}, {Ryckbosch}, {Rysewyk},
  {Sabbatini}, {Sanchez Herrera}, {Sandrock}, {Sandroos}, {Sarkar},
  {Satalecka}, {Schlunder}, {Schmidt}, {Schoenen}, {Sch{\"o}neberg},
  {Schumacher}, {Seckel}, {Seunarine}, {Soldin}, {Song}, {Spiczak}, {Spiering},
  {Stanev}, {Stasik}, {Stettner}, {Steuer}, {Stezelberger}, {Stokstad},
  {St{\"o}ssl}, {Str{\"o}m}, {Strotjohann}, {Sullivan}, {Sutherland},
  {Taavola}, {Taboada}, {Tatar}, {Tenholt}, {Ter-Antonyan}, {Terliuk},
  {Te{\v{s}}i{\'c}}, {Tilav}, {Toale}, {Tobin}, {Toscano}, {Tosi},
  {Tselengidou}, {Turcati}, {Unger}, {Usner}, {Vandenbroucke}, {van
  Eijndhoven}, {Vanheule}, {van Rossem}, {van Santen}, {Veenkamp}, {Vehring},
  {Voge}, {Vogel}, {Vraeghe}, {Walck}, {Wallace}, {Wallraff}, {Wandkowsky},
  {Weaver}, {Weiss}, {Wendt}, {Westerhoff}, {Whelan}, {Wickmann}, {Wiebe},
  {Wiebusch}, {Wille}, {Williams}, {Wills}, {Wolf}, {Wood}, {Woolsey},
  {Woschnagg}, {Xu}, {Xu}, {Xu}, {Yanez}, {Yodh}, {Yoshida}, {Zoll}, \&
  {IceCube Collaboration}}]{IceCube7y:2017}
{Aartsen}, M.~G., {Abraham}, K., {Ackermann}, M., {et~al.} 2017{\natexlab{b}},
  \apj, 835, 151

\bibitem[{{Aartsen} {et~al.}(2020){Aartsen}, {Ackermann}, {Adams}, {Aguilar},
  {Ahlers}, {Ahrens}, {Alispach}, {Andeen}, {Anderson}, {Ansseau}, {Anton},
  {Arg{\"u}elles}, {Auffenberg}, {Axani}, {Backes}, {Bagherpour}, {Bai},
  {Balagopal}, {Barbano}, {Barwick}, {Bastian}, {Baum}, {Baur}, {Bay},
  {Beatty}, {Becker}, {Becker Tjus}, {BenZvi}, {Berley}, {Bernardini},
  {Besson}, {Binder}, {Bindig}, {Blaufuss}, {Blot}, {Bohm}, {B{\"o}rner},
  {B{\"o}ser}, {Botner}, {B{\"o}ttcher}, {Bourbeau}, {Bourbeau}, {Bradascio},
  {Braun}, {Bron}, {Brostean-Kaiser}, {Burgman}, {Buscher}, {Busse}, {Carver},
  {Chen}, {Cheung}, {Chirkin}, {Choi}, {Clark}, {Classen}, {Coleman}, {Collin},
  {Conrad}, {Coppin}, {Correa}, {Cowen}, {Cross}, {Dave}, {De Clercq},
  {DeLaunay}, {Dembinski}, {Deoskar}, {De Ridder}, {Desiati}, {de Vries}, {de
  Wasseige}, {de With}, {DeYoung}, {Diaz}, {D{\'\i}az-V{\'e}lez}, {Dujmovic},
  {Dunkman}, {Dvorak}, {Eberhardt}, {Ehrhardt}, {Eller}, {Engel}, {Evenson},
  {Fahey}, {Fazely}, {Felde}, {Filimonov}, {Finley}, {Fox}, {Franckowiak},
  {Friedman}, {Fritz}, {Gaisser}, {Gallagher}, {Ganster}, {Garrappa},
  {Gerhardt}, {Ghorbani}, {Glauch}, {Gl{\"u}senkamp}, {Goldschmidt},
  {Gonzalez}, {Grant}, {Griffith}, {Griswold}, {G{\"u}nder}, {G{\"u}nd{\"u}z},
  {Haack}, {Hallgren}, {Halliday}, {Halve}, {Halzen}, {Hanson}, {Haungs},
  {Hebecker}, {Heereman}, {Heix}, {Helbing}, {Hellauer}, {Henningsen},
  {Hickford}, {Hignight}, {Hill}, {Hoffman}, {Hoffmann}, {Hoinka},
  {Hokanson-Fasig}, {Hoshina}, {Huang}, {Huber}, {Huber}, {Hultqvist},
  {H{\"u}nnefeld}, {Hussain}, {In}, {Iovine}, {Ishihara}, {Japaridze}, {Jeong},
  {Jero}, {Jones}, {Jonske}, {Joppe}, {Kang}, {Kang}, {Kappes}, {Kappesser},
  {Karg}, {Karl}, {Karle}, {Katz}, {Kauer}, {Kelley}, {Kheirandish}, {Kim},
  {Kintscher}, {Kiryluk}, {Kittler}, {Klein}, {Koirala}, {Kolanoski},
  {K{\"o}pke}, {Kopper}, {Kopper}, {Koskinen}, {Kowalski}, {Krings},
  {Kr{\"u}ckl}, {Kulacz}, {Kurahashi}, {Kyriacou}, {Labare}, {Lanfranchi},
  {Larson}, {Lauber}, {Lazar}, {Leonard}, {Leszczy{\'n}ska}, {Leuermann},
  {Liu}, {Lohfink}, {Lozano Mariscal}, {Lu}, {Lucarelli}, {L{\"u}nemann},
  {Luszczak}, {Lyu}, {Ma}, {Madsen}, {Maggi}, {Mahn}, {Makino}, {Mallik},
  {Mallot}, {Mancina}, {Mari{\c{s}}}, {Maruyama}, {Mase}, {Matis}, {Maunu},
  {McNally}, {Meagher}, {Medici}, {Medina}, {Meier}, {Meighen-Berger}, {Menne},
  {Merino}, {Meures}, {Micallef}, {Mockler}, {Moment{\'e}}, {Montaruli},
  {Moore}, {Morse}, {Moulai}, {Muth}, {Nagai}, {Naumann}, {Neer},
  {Niederhausen}, {Nisa}, {Nowicki}, {Nygren}, {Obertacke Pollmann}, {Oehler},
  {Olivas}, {O'Murchadha}, {O'Sullivan}, {Palczewski}, {Pandya}, {Pankova},
  {Park}, {Peiffer}, {P{\'e}rez de los Heros}, {Philippen}, {Pieloth}, {Pinat},
  {Pizzuto}, {Plum}, {Porcelli}, {Price}, {Przybylski}, {Raab}, {Raissi},
  {Rameez}, {Rauch}, {Rawlins}, {Rea}, {Reimann}, {Relethford}, {Renschler},
  {Renzi}, {Resconi}, {Rhode}, {Richman}, {Robertson}, {Rongen}, {Rott},
  {Ruhe}, {Ryckbosch}, {Rysewyk}, {Safa}, {Sanchez Herrera}, {Sandrock},
  {Sandroos}, {Santander}, {Sarkar}, {Sarkar}, {Satalecka}, {Schaufel},
  {Schieler}, {Schlunder}, {Schmidt}, {Schneider}, {Schneider}, {Schr{\"o}der},
  {Schumacher}, {Sclafani}, {Seckel}, {Seunarine}, {Shefali}, {Silva},
  {Snihur}, {Soedingrekso}, {Soldin}, {Song}, {Spiczak}, {Spiering},
  {Stachurska}, {Stamatikos}, {Stanev}, {Stein}, {Steinm{\"u}ller}, {Stettner},
  {Steuer}, {Stezelberger}, {Stokstad}, {St{\"o}{\ss}l}, {Strotjohann},
  {St{\"u}rwald}, {Stuttard}, {Sullivan}, {Taboada}, {Tenholt}, {Ter-Antonyan},
  {Terliuk}, {Tilav}, {Tollefson}, {Tomankova}, {T{\"o}nnis}, {Toscano},
  {Tosi}, {Trettin}, {Tselengidou}, {Tung}, {Turcati}, {Turcotte}, {Turley},
  {Ty}, {Unger}, {Unland Elorrieta}, {Usner}, {Vandenbroucke}, {Van Driessche},
  {van Eijk}, {van Eijndhoven}, {Vanheule}, {van Santen}, {Vraeghe}, {Walck},
  {Wallace}, {Wallraff}, {Wandkowsky}, {Watson}, {Weaver}, {Weindl}, {Weiss},
  {Weldert}, {Wendt}, {Werthebach}, {Whelan}, {Whitehorn}, {Wiebe}, {Wiebusch},
  {Wille}, {Williams}, {Wills}, {Wolf}, {Wood}, {Wood}, {Woschnagg}, {Wrede},
  {Xu}, {Xu}, {Xu}, {Yanez}, {Yodh}, {Yoshida}, {Yuan}, \&
  {Z{\"o}cklein}}]{IceCube10y:2020}
{Aartsen}, M.~G., {Ackermann}, M., {Adams}, J., {et~al.} 2020, \prl, 124,
  051103

\bibitem[{{Abbasi} {et~al.}(2022){Abbasi}, {Ackermann}, {Adams}, {Aguilar},
  {Ahlers}, {Ahrens}, {Alameddine}, {Alispach}, {Alves}, {Amin}, {Andeen},
  {Anderson}, {Anton}, {Arg{\"u}elles}, {Ashida}, {Axani}, {Bai}, {Balagopal},
  {Barbano}, {Barwick}, {Bastian}, {Basu}, {Baur}, {Bay}, {Beatty}, {Becker},
  {Becker Tjus}, {Bellenghi}, {Benzvi}, {Berley}, {Bernardini}, {Besson},
  {Binder}, {Bindig}, {Blaufuss}, {Blot}, {Boddenberg}, {Bontempo}, {Borowka},
  {B{\"o}ser}, {Botner}, {B{\"o}ttcher}, {Bourbeau}, {Bradascio}, {Braun},
  {Brinson}, {Bron}, {Brostean-Kaiser}, {Browne}, {Burgman}, {Burley}, {Busse},
  {Campana}, {Carnie-Bronca}, {Chen}, {Chen}, {Chirkin}, {Choi}, {Clark},
  {Clark}, {Classen}, {Coleman}, {Collin}, {Conrad}, {Coppin}, {Correa},
  {Cowen}, {Cross}, {Dappen}, {Dave}, {de Clercq}, {Delaunay}, {Delgado
  L{\'o}pez}, {Dembinski}, {Deoskar}, {Desai}, {Desiati}, {de Vries}, {de
  Wasseige}, {de With}, {Deyoung}, {Diaz}, {D{\'\i}az-V{\'e}lez}, {Dittmer},
  {Dujmovic}, {Dunkman}, {Duvernois}, {Dvorak}, {Ehrhardt}, {Eller}, {Engel},
  {Erpenbeck}, {Evans}, {Evenson}, {Fan}, {Fazely}, {Fedynitch}, {Feigl},
  {Fiedlschuster}, {Fienberg}, {Filimonov}, {Finley}, {Fischer}, {Fox},
  {Franckowiak}, {Friedman}, {Fritz}, {F{\"u}rst}, {Gaisser}, {Gallagher},
  {Ganster}, {Garcia}, {Garrappa}, {Gerhardt}, {Ghadimi}, {Glaser}, {Glauch},
  {Gl{\"u}senkamp}, {Goldschmidt}, {Gonzalez}, {Goswami}, {Grant},
  {Gr{\'e}goire}, {Griswold}, {G{\"u}nther}, {Gutjahr}, {Haack}, {Hallgren},
  {Halliday}, {Halve}, {Halzen}, {Hanson}, {Hardin}, {Harnisch}, {Haungs},
  {Hebecker}, {Helbing}, {Henningsen}, {Hettinger}, {Hickford}, {Hignight},
  {Hill}, {Hill}, {Hoffman}, {Hoffmann}, {Hokanson-Fasig}, {Hoshina}, {Huang},
  {Huber}, {Huber}, {Hultqvist}, {H{\"u}nnefeld}, {Hussain}, {Hymon}, {in},
  {Iovine}, {Ishihara}, {Jansson}, {Japaridze}, {Jeong}, {Jin}, {Jones},
  {Kang}, {Kang}, {Kang}, {Kappes}, {Kappesser}, {Kardum}, {Karg}, {Karl},
  {Karle}, {Katz}, {Kauer}, {Kellermann}, {Kelley}, {Kheirandish}, {Kin},
  {Kintscher}, {Kiryluk}, {Klein}, {Koirala}, {Kolanoski}, {Kontrimas},
  {K{\"o}pke}, {Kopper}, {Kopper}, {Koskinen}, {Koundal}, {Kovacevich},
  {Kowalski}, {Kozynets}, {Kun}, {Kurahashi}, {Lad}, {Lagunas Gualda},
  {Lanfranchi}, {Larson}, {Lauber}, {Lazar}, {Lee}, {Leonard},
  {Leszczy{\'n}ska}, {Li}, {Lincetto}, {Liu}, {Liubarska}, {Lohfink}, {Lozano
  Mariscal}, {Lu}, {Lucarelli}, {Ludwig}, {Luszczak}, {Lyu}, {Ma}, {Madsen},
  {Mahn}, {Makino}, {Mancina}, {Mari{\c{s}}}, {Martinez-Soler}, {Maruyama},
  {Mase}, {McElroy}, {McNally}, {Mead}, {Meagher}, {Mechbal}, {Medina},
  {Meier}, {Meighen-Berger}, {Micallef}, {Mockler}, {Montaruli}, {Moore},
  {Morse}, {Moulai}, {Naab}, {Nagai}, {Nahnhauer}, {Naumann}, {Necker},
  {Nguyen}, {Niederhausen}, {Nisa}, {Nowicki}, {Nygren}, {Obertack},
  {Pollmann}, {Oehler}, {Oeyen}, {Olivas}, {O'Sullivan}, {Pandya}, {Pankova},
  {Park}, {Parker}, {Paudel}, {Paul}, {P{\'e}rez de Los Heros}, {Peters},
  {Peterson}, {Philippen}, {Pieper}, {Pittermann}, {Pizzuto}, {Plum},
  {Popovych}, {Porcelli}, {Prado Rodriguez}, {Price}, {Pries}, {Przybylski},
  {Rack-Helleis}, {Raissi}, {Rameez}, {Rawlins}, {Rea}, {Rehman},
  {Reichherzer}, {Reimann}, {Renzi}, {Resconi}, {Reusch}, {Rhode}, {Richman},
  {Riedel}, {Roberts}, {Robertson}, {Roellinghoff}, {Rongen}, {Rott}, {Ruhe},
  {Ryckbosch}, {Rysewyk Cantu}, {Safa}, {Saffer}, {Sanchez Herrera},
  {Sandrock}, {Sandroos}, {Santander}, {Sarkar}, {Sarkar}, {Satalecka},
  {Schaufel}, {Schieler}, {Schindler}, {Schmidt}, {Schneider}, {Schneider},
  {Schr{\"o}der}, {Schumacher}, {Schwefer}, {Sclafani}, {Seckel}, {Seunarine},
  {Sharma}, {Shefali}, {Silva}, {Skrzypek}, {Smithers}, {Snihur},
  {Soedingrekso}, {Soldin}, {Spannfellner}, {Spiczak}, {Spiering},
  {Stachurska}, {Stamatikos}, {Stanev}, {Stein}, {Stettner}, {Steuer},
  {Stezelberger}, {Stokstad}, {St{\"u}rwald}, {Stuttard}, {Sullivan},
  {Taboada}, {Ter-Antonyan}, {Tilav}, {Tischbein}, {Tollefson}, {T{\"o}nnis},
  {Toscano}, {Tosi}, {Trettin}, {Tselengidou}, {Tung}, {Turcati}, {Turcotte},
  {Turley}, {Twagirayezu}, {Ty}, {Unland Elorrieta}, {Valtonen-Mattila},
  {Vandenbroucke}, {van Eijndhoven}, {Vannerom}, {van Santen}, {Verpoest},
  {Walck}, {Watson}, {Weaver}, {Weigel}, {Weindl}, {Weiss}, {Weldert}, {Wendt},
  {Werthebach}, {Weyrauch}, {Whitehorn}, {Wiebusch}, {Williams}, {Wolf},
  {Woschnagg}, {Wrede}, {Wulff}, {Xu}, {Yanez}, {Yoshida}, {Yu}, {Yuan},
  {Zhangan}, \& {Zhelnin}}]{IceCube_10y_reprocessed:2022}
{Abbasi}, R., {Ackermann}, M., {Adams}, J., {et~al.} 2022, Science, 378, 538

\bibitem[{{Abdo} {et~al.}(2010){Abdo}, {Ackermann}, {Ajello}, {Allafort},
  {Antolini}, {Atwood}, {Axelsson}, {Baldini}, {Ballet}, {Barbiellini},
  {Bastieri}, {Baughman}, {Bechtol}, {Bellazzini}, {Berenji}, {Blandford},
  {Bloom}, {Bogart}, {Bonamente}, {Borgland}, {Bouvier}, {Bregeon}, {Brez},
  {Brigida}, {Bruel}, {Buehler}, {Burnett}, {Buson}, {Caliandro}, {Cameron},
  {Cannon}, {Caraveo}, {Carrigan}, {Casandjian}, {Cavazzuti}, {Cecchi},
  {{\c{C}}elik}, {Celotti}, {Charles}, {Chekhtman}, {Chen}, {Cheung}, {Chiang},
  {Ciprini}, {Claus}, {Cohen-Tanugi}, {Conrad}, {Costamante}, {Cotter},
  {Cutini}, {D'Elia}, {Dermer}, {de Angelis}, {de Palma}, {De Rosa}, {Digel},
  {Silva}, {Drell}, {Dubois}, {Dumora}, {Escande}, {Farnier}, {Favuzzi},
  {Fegan}, {Ferrara}, {Focke}, {Fortin}, {Frailis}, {Fukazawa}, {Funk},
  {Fusco}, {Gargano}, {Gasparrini}, {Gehrels}, {Germani}, {Giebels},
  {Giglietto}, {Giommi}, {Giordano}, {Giroletti}, {Glanzman}, {Godfrey},
  {Grandi}, {Grenier}, {Grondin}, {Grove}, {Guiriec}, {Hadasch}, {Harding},
  {Hayashida}, {Hays}, {Healey}, {Hill}, {Horan}, {Hughes}, {Iafrate}, {Itoh},
  {J{\'o}hannesson}, {Johnson}, {Johnson}, {Johnson}, {Johnson}, {Kamae},
  {Katagiri}, {Kataoka}, {Kawai}, {Kerr}, {Kn{\"o}dlseder}, {Kuss}, {Lande},
  {Latronico}, {Lavalley}, {Lemoine-Goumard}, {Llena Garde}, {Longo},
  {Loparco}, {Lott}, {Lovellette}, {Lubrano}, {Madejski}, {Makeev}, {Malaguti},
  {Massaro}, {Mazziotta}, {McConville}, {McEnery}, {McGlynn}, {Michelson},
  {Mitthumsiri}, {Mizuno}, {Moiseev}, {Monte}, {Monzani}, {Morselli},
  {Moskalenko}, {Murgia}, {Nolan}, {Norris}, {Nuss}, {Ohno}, {Ohsugi},
  {Omodei}, {Orlando}, {Ormes}, {Ozaki}, {Paneque}, {Panetta}, {Parent},
  {Pelassa}, {Pepe}, {Pesce-Rollins}, {Piranomonte}, {Piron}, {Porter},
  {Rain{\`o}}, {Rando}, {Razzano}, {Reimer}, {Reimer}, {Reposeur}, {Ripken},
  {Ritz}, {Rodriguez}, {Romani}, {Roth}, {Ryde}, {Sadrozinski}, {Sanchez},
  {Sander}, {Saz Parkinson}, {Scargle}, {Sgr{\`o}}, {Shaw}, {Siskind}, {Smith},
  {Spandre}, {Spinelli}, {Starck}, {Stawarz}, {Strickman}, {Suson}, {Tajima},
  {Takahashi}, {Takahashi}, {Tanaka}, {Taylor}, {Thayer}, {Thayer}, {Thompson},
  {Tibaldo}, {Torres}, {Tosti}, {Tramacere}, {Ubertini}, {Uchiyama}, {Usher},
  {Vasileiou}, {Vilchez}, {Villata}, {Vitale}, {Waite}, {Wallace}, {Wang},
  {Winer}, {Wood}, {Yang}, {Ylinen}, \& {Ziegler}}]{1LAC}
{Abdo}, A.~A., {Ackermann}, M., {Ajello}, M., {et~al.} 2010, \apj, 715, 429

\bibitem[{{Abdo} {et~al.}(2009){Abdo}, {Ackermann}, {Ajello}, {Atwood},
  {Axelsson}, {Baldini}, {Ballet}, {Band}, {Barbiellini}, {Bastieri},
  {Battelino}, {Baughman}, {Bechtol}, {Bellazzini}, {Berenji}, {Bignami},
  {Blandford}, {Bloom}, {Bonamente}, {Borgland}, {Bouvier}, {Bregeon}, {Brez},
  {Brigida}, {Bruel}, {Burnett}, {Caliandro}, {Cameron}, {Caraveo},
  {Casandjian}, {Cavazzuti}, {Cecchi}, {Charles}, {Chekhtman}, {Cheung},
  {Chiang}, {Ciprini}, {Claus}, {Cohen-Tanugi}, {Cominsky}, {Conrad}, {Corbet},
  {Costamante}, {Cutini}, {Davis}, {Dermer}, {de Angelis}, {de Luca}, {de
  Palma}, {Digel}, {Dormody}, {do Couto e Silva}, {Drell}, {Dubois}, {Dumora},
  {Farnier}, {Favuzzi}, {Fegan}, {Ferrara}, {Focke}, {Frailis}, {Fukazawa},
  {Funk}, {Fusco}, {Gargano}, {Gasparrini}, {Gehrels}, {Germani}, {Giebels},
  {Giglietto}, {Giommi}, {Giordano}, {Glanzman}, {Godfrey}, {Grenier},
  {Grondin}, {Grove}, {Guillemot}, {Guiriec}, {Hanabata}, {Harding}, {Hartman},
  {Hayashida}, {Hays}, {Healey}, {Horan}, {Hughes}, {J{\'o}hannesson},
  {Johnson}, {Johnson}, {Johnson}, {Johnson}, {Kamae}, {Katagiri}, {Kataoka},
  {Kawai}, {Kerr}, {Kn{\"o}dlseder}, {Kocevski}, {Kocian}, {Komin}, {Kuehn},
  {Kuss}, {Lande}, {Latronico}, {Lee}, {Lemoine-Goumard}, {Longo}, {Loparco},
  {Lott}, {Lovellette}, {Lubrano}, {Madejski}, {Makeev}, {Marelli},
  {Mazziotta}, {McConville}, {McEnery}, {McGlynn}, {Meurer}, {Michelson},
  {Mitthumsiri}, {Mizuno}, {Moiseev}, {Monte}, {Monzani}, {Moretti},
  {Morselli}, {Moskalenko}, {Murgia}, {Nakamori}, {Nolan}, {Norris}, {Nuss},
  {Ohno}, {Ohsugi}, {Omodei}, {Orlando}, {Ormes}, {Ozaki}, {Paneque},
  {Panetta}, {Parent}, {Pelassa}, {Pepe}, {Pesce-Rollins}, {Piron}, {Porter},
  {Poupard}, {Rain{\`o}}, {Rando}, {Ray}, {Razzano}, {Rea}, {Reimer}, {Reimer},
  {Reposeur}, {Ritz}, {Rochester}, {Rodriguez}, {Romani}, {Roth}, {Ryde},
  {Sadrozinski}, {Sanchez}, {Sander}, {Saz Parkinson}, {Scargle}, {Schalk},
  {Sellerholm}, {Sgr{\`o}}, {Shaw}, {Shrader}, {Sierpowska-Bartosik},
  {Siskind}, {Smith}, {Smith}, {Spandre}, {Spinelli}, {Starck}, {Stephens},
  {Strickman}, {Strong}, {Suson}, {Tajima}, {Takahashi}, {Takahashi}, {Tanaka},
  {Thayer}, {Thayer}, {Thompson}, {Tibaldo}, {Tibolla}, {Torres}, {Tosti},
  {Tramacere}, {Uchiyama}, {Usher}, {Van Etten}, {Vilchez}, {Vitale}, {Waite},
  {Wallace}, {Wang}, {Watters}, {Winer}, {Wood}, {Ylinen}, {Ziegler}, \&
  {Fermi/LAT Collaboration}}]{0FGL}
{Abdo}, A.~A., {Ackermann}, M., {Ajello}, M., {et~al.} 2009, \apjs, 183, 46

\bibitem[{{Abdurro'uf} {et~al.}(2022){Abdurro'uf}, {Accetta}, {Aerts}, {Silva
  Aguirre}, {Ahumada}, {Ajgaonkar}, {Filiz Ak}, {Alam}, {Allende Prieto},
  {Almeida}, {Anders}, {Anderson}, {Andrews}, {Anguiano}, {Aquino-Ort{\'\i}z},
  {Arag{\'o}n-Salamanca}, {Argudo-Fern{\'a}ndez}, {Ata}, {Aubert},
  {Avila-Reese}, {Badenes}, {Barb{\'a}}, {Barger}, {Barrera-Ballesteros},
  {Beaton}, {Beers}, {Belfiore}, {Bender}, {Bernardi}, {Bershady}, {Beutler},
  {Bidin}, {Bird}, {Bizyaev}, {Blanc}, {Blanton}, {Boardman}, {Bolton},
  {Boquien}, {Borissova}, {Bovy}, {Brandt}, {Brown}, {Brownstein}, {Brusa},
  {Buchner}, {Bundy}, {Burchett}, {Bureau}, {Burgasser}, {Cabang}, {Campbell},
  {Cappellari}, {Carlberg}, {Wanderley}, {Carrera}, {Cash}, {Chen}, {Chen},
  {Cherinka}, {Chiappini}, {Choi}, {Chojnowski}, {Chung}, {Clerc}, {Cohen},
  {Comerford}, {Comparat}, {da Costa}, {Covey}, {Crane}, {Cruz-Gonzalez},
  {Culhane}, {Cunha}, {Dai}, {Damke}, {Darling}, {Davidson}, {Davies},
  {Dawson}, {De Lee}, {Diamond-Stanic}, {Cano-D{\'\i}az}, {S{\'a}nchez},
  {Donor}, {Duckworth}, {Dwelly}, {Eisenstein}, {Elsworth}, {Emsellem},
  {Eracleous}, {Escoffier}, {Fan}, {Farr}, {Feng}, {Fern{\'a}ndez-Trincado},
  {Feuillet}, {Filipp}, {Fillingham}, {Frinchaboy}, {Fromenteau}, {Galbany},
  {Garc{\'\i}a}, {Garc{\'\i}a-Hern{\'a}ndez}, {Ge}, {Geisler}, {Gelfand},
  {G{\'e}ron}, {Gibson}, {Goddy}, {Godoy-Rivera}, {Grabowski}, {Green},
  {Greener}, {Grier}, {Griffith}, {Guo}, {Guy}, {Hadjara}, {Harding},
  {Hasselquist}, {Hayes}, {Hearty}, {Hern{\'a}ndez}, {Hill}, {Hogg},
  {Holtzman}, {Horta}, {Hsieh}, {Hsu}, {Hsu}, {Huber}, {Huertas-Company},
  {Hutchinson}, {Hwang}, {Ibarra-Medel}, {Chitham}, {Ilha}, {Imig}, {Jaekle},
  {Jayasinghe}, {Ji}, {Johnson}, {Jones}, {J{\"o}nsson}, {Katkov}, {Khalatyan},
  {Kinemuchi}, {Kisku}, {Knapen}, {Kneib}, {Kollmeier}, {Kong}, {Kounkel},
  {Kreckel}, {Krishnarao}, {Lacerna}, {Lane}, {Langgin}, {Lavender}, {Law},
  {Lazarz}, {Leung}, {Leung}, {Lewis}, {Li}, {Li}, {Lian}, {Liang}, {Lin},
  {Lin}, {Lin}, {Lintott}, {Long}, {Longa-Pe{\~n}a}, {L{\'o}pez-Cob{\'a}},
  {Lu}, {Lundgren}, {Luo}, {Mackereth}, {de la Macorra}, {Mahadevan},
  {Majewski}, {Manchado}, {Mandeville}, {Maraston}, {Margalef-Bentabol},
  {Masseron}, {Masters}, {Mathur}, {McDermid}, {Mckay}, {Merloni},
  {Merrifield}, {Meszaros}, {Miglio}, {Di Mille}, {Minniti}, {Minsley},
  {Monachesi}, {Moon}, {Mosser}, {Mulchaey}, {Muna}, {Mu{\~n}oz}, {Myers},
  {Myers}, {Nadathur}, {Nair}, {Nandra}, {Neumann}, {Newman}, {Nidever},
  {Nikakhtar}, {Nitschelm}, {O'Connell}, {Garma-Oehmichen}, {Luan Souza de
  Oliveira}, {Olney}, {Oravetz}, {Ortigoza-Urdaneta}, {Osorio}, {Otter},
  {Pace}, {Padilla}, {Pan}, {Pan}, {Parikh}, {Parker}, {Peirani}, {Pe{\~n}a
  Ram{\'\i}rez}, {Penny}, {Percival}, {Perez-Fournon}, {Pinsonneault},
  {Poidevin}, {Poovelil}, {Price-Whelan}, {B{\'a}rbara de Andrade Queiroz},
  {Raddick}, {Ray}, {Rembold}, {Riddle}, {Riffel}, {Riffel}, {Rix}, {Robin},
  {Rodr{\'\i}guez-Puebla}, {Roman-Lopes}, {Rom{\'a}n-Z{\'u}{\~n}iga}, {Rose},
  {Ross}, {Rossi}, {Rubin}, {Salvato}, {S{\'a}nchez}, {S{\'a}nchez-Gallego},
  {Sanderson}, {Santana Rojas}, {Sarceno}, {Sarmiento}, {Sayres}, {Sazonova},
  {Schaefer}, {Schiavon}, {Schlegel}, {Schneider}, {Schultheis}, {Schwope},
  {Serenelli}, {Serna}, {Shao}, {Shapiro}, {Sharma}, {Shen}, {Shetrone}, {Shu},
  {Simon}, {Skrutskie}, {Smethurst}, {Smith}, {Sobeck}, {Spoo}, {Sprague},
  {Stark}, {Stassun}, {Steinmetz}, {Stello}, {Stone-Martinez},
  {Storchi-Bergmann}, {Stringfellow}, {Stutz}, {Su}, {Taghizadeh-Popp},
  {Talbot}, {Tayar}, {Telles}, {Teske}, {Thakar}, {Theissen}, {Tkachenko},
  {Thomas}, {Tojeiro}, {Hernandez Toledo}, {Troup}, {Trump}, {Trussler},
  {Turner}, {Tuttle}, {Unda-Sanzana}, {V{\'a}zquez-Mata}, {Valentini},
  {Valenzuela}, {Vargas-Gonz{\'a}lez}, {Vargas-Maga{\~n}a}, {Alfaro},
  {Villanova}, {Vincenzo}, {Wake}, {Warfield}, {Washington}, {Weaver},
  {Weijmans}, {Weinberg}, {Weiss}, {Westfall}, {Wild}, {Wilde}, {Wilson},
  {Wilson}, {Wilson}, {Wolf}, {Wood-Vasey}, {Yan}, {Zamora}, {Zasowski},
  {Zhang}, {Zhao}, {Zheng}, {Zheng}, \& {Zhu}}]{SDSS17}
{Abdurro'uf}, {Accetta}, K., {Aerts}, C., {et~al.} 2022, \apjs, 259, 35

\bibitem[{{Acero} {et~al.}(2015){Acero}, {Ackermann}, {Ajello}, {Albert},
  {Atwood}, {Axelsson}, {Baldini}, {Ballet}, {Barbiellini}, {Bastieri},
  {Belfiore}, {Bellazzini}, {Bissaldi}, {Blandford}, {Bloom}, {Bogart},
  {Bonino}, {Bottacini}, {Bregeon}, {Britto}, {Bruel}, {Buehler}, {Burnett},
  {Buson}, {Caliandro}, {Cameron}, {Caputo}, {Caragiulo}, {Caraveo},
  {Casandjian}, {Cavazzuti}, {Charles}, {Chaves}, {Chekhtman}, {Cheung},
  {Chiang}, {Chiaro}, {Ciprini}, {Claus}, {Cohen-Tanugi}, {Cominsky}, {Conrad},
  {Cutini}, {D'Ammando}, {de Angelis}, {DeKlotz}, {de Palma}, {Desiante},
  {Digel}, {Di Venere}, {Drell}, {Dubois}, {Dumora}, {Favuzzi}, {Fegan},
  {Ferrara}, {Finke}, {Franckowiak}, {Fukazawa}, {Funk}, {Fusco}, {Gargano},
  {Gasparrini}, {Giebels}, {Giglietto}, {Giommi}, {Giordano}, {Giroletti},
  {Glanzman}, {Godfrey}, {Grenier}, {Grondin}, {Grove}, {Guillemot}, {Guiriec},
  {Hadasch}, {Harding}, {Hays}, {Hewitt}, {Hill}, {Horan}, {Iafrate}, {Jogler},
  {J{\'o}hannesson}, {Johnson}, {Johnson}, {Johnson}, {Johnson}, {Kamae},
  {Kataoka}, {Katsuta}, {Kuss}, {La Mura}, {Landriu}, {Larsson}, {Latronico},
  {Lemoine-Goumard}, {Li}, {Li}, {Longo}, {Loparco}, {Lott}, {Lovellette},
  {Lubrano}, {Madejski}, {Massaro}, {Mayer}, {Mazziotta}, {McEnery},
  {Michelson}, {Mirabal}, {Mizuno}, {Moiseev}, {Mongelli}, {Monzani},
  {Morselli}, {Moskalenko}, {Murgia}, {Nuss}, {Ohno}, {Ohsugi}, {Omodei},
  {Orienti}, {Orlando}, {Ormes}, {Paneque}, {Panetta}, {Perkins},
  {Pesce-Rollins}, {Piron}, {Pivato}, {Porter}, {Racusin}, {Rando}, {Razzano},
  {Razzaque}, {Reimer}, {Reimer}, {Reposeur}, {Rochester}, {Romani},
  {Salvetti}, {S{\'a}nchez-Conde}, {Saz Parkinson}, {Schulz}, {Siskind},
  {Smith}, {Spada}, {Spandre}, {Spinelli}, {Stephens}, {Strong}, {Suson},
  {Takahashi}, {Takahashi}, {Tanaka}, {Thayer}, {Thayer}, {Thompson},
  {Tibaldo}, {Tibolla}, {Torres}, {Torresi}, {Tosti}, {Troja}, {Van Klaveren},
  {Vianello}, {Winer}, {Wood}, {Wood}, {Zimmer}, \& {Fermi-LAT
  Collaboration}}]{3FGL}
{Acero}, F., {Ackermann}, M., {Ajello}, M., {et~al.} 2015, \apjs, 218, 23

\bibitem[{{Adelman-McCarthy} {et~al.}(2008){Adelman-McCarthy}, {Ag{\"u}eros},
  {Allam}, {Allende Prieto}, {Anderson}, {Anderson}, {Annis}, {Bahcall},
  {Bailer-Jones}, {Baldry}, {Barentine}, {Bassett}, {Becker}, {Beers}, {Bell},
  {Berlind}, {Bernardi}, {Blanton}, {Bochanski}, {Boroski}, {Brinchmann},
  {Brinkmann}, {Brunner}, {Budav{\'a}ri}, {Carliles}, {Carr}, {Castander},
  {Cinabro}, {Cool}, {Covey}, {Csabai}, {Cunha}, {Davenport}, {Dilday}, {Doi},
  {Eisenstein}, {Evans}, {Fan}, {Finkbeiner}, {Friedman}, {Frieman},
  {Fukugita}, {G{\"a}nsicke}, {Gates}, {Gillespie}, {Glazebrook}, {Gray},
  {Grebel}, {Gunn}, {Gurbani}, {Hall}, {Harding}, {Harvanek}, {Hawley},
  {Hayes}, {Heckman}, {Hendry}, {Hindsley}, {Hirata}, {Hogan}, {Hogg}, {Hyde},
  {Ichikawa}, {Ivezi{\'c}}, {Jester}, {Johnson}, {Jorgensen}, {Juri{\'c}},
  {Kent}, {Kessler}, {Kleinman}, {Knapp}, {Kron}, {Krzesinski}, {Kuropatkin},
  {Lamb}, {Lampeitl}, {Lebedeva}, {Lee}, {French Leger}, {L{\'e}pine}, {Lima},
  {Lin}, {Long}, {Loomis}, {Loveday}, {Lupton}, {Malanushenko}, {Malanushenko},
  {Mandelbaum}, {Margon}, {Marriner}, {Mart{\'\i}nez-Delgado}, {Matsubara},
  {McGehee}, {McKay}, {Meiksin}, {Morrison}, {Munn}, {Nakajima}, {Neilsen},
  {Newberg}, {Nichol}, {Nicinski}, {Nieto-Santisteban}, {Nitta}, {Okamura},
  {Owen}, {Oyaizu}, {Padmanabhan}, {Pan}, {Park}, {Peoples}, {Pier}, {Pope},
  {Purger}, {Raddick}, {Re Fiorentin}, {Richards}, {Richmond}, {Riess}, {Rix},
  {Rockosi}, {Sako}, {Schlegel}, {Schneider}, {Schreiber}, {Schwope}, {Seljak},
  {Sesar}, {Sheldon}, {Shimasaku}, {Sivarani}, {Allyn Smith}, {Snedden},
  {Steinmetz}, {Strauss}, {SubbaRao}, {Suto}, {Szalay}, {Szapudi}, {Szkody},
  {Tegmark}, {Thakar}, {Tremonti}, {Tucker}, {Uomoto}, {Vanden Berk},
  {Vandenberg}, {Vidrih}, {Vogeley}, {Voges}, {Vogt}, {Wadadekar}, {Weinberg},
  {West}, {White}, {Wilhite}, {Yanny}, {Yocum}, {York}, {Zehavi}, \&
  {Zucker}}]{SDSS6}
{Adelman-McCarthy}, J.~K., {Ag{\"u}eros}, M.~A., {Allam}, S.~S., {et~al.} 2008,
  \apjs, 175, 297

\bibitem[{{Ahumada} {et~al.}(2020){Ahumada}, {Allende Prieto}, {Almeida},
  {Anders}, {Anderson}, {Andrews}, {Anguiano}, {Arcodia}, {Armengaud},
  {Aubert}, {Avila}, {Avila-Reese}, {Badenes}, {Balland}, {Barger},
  {Barrera-Ballesteros}, {Basu}, {Bautista}, {Beaton}, {Beers}, {Benavides},
  {Bender}, {Bernardi}, {Bershady}, {Beutler}, {Bidin}, {Bird}, {Bizyaev},
  {Blanc}, {Blanton}, {Boquien}, {Borissova}, {Bovy}, {Brandt}, {Brinkmann},
  {Brownstein}, {Bundy}, {Bureau}, {Burgasser}, {Burtin}, {Cano-D{\'\i}az},
  {Capasso}, {Cappellari}, {Carrera}, {Chabanier}, {Chaplin}, {Chapman},
  {Cherinka}, {Chiappini}, {Doohyun Choi}, {Chojnowski}, {Chung}, {Clerc},
  {Coffey}, {Comerford}, {Comparat}, {da Costa}, {Cousinou}, {Covey}, {Crane},
  {Cunha}, {Ilha}, {Dai}, {Damsted}, {Darling}, {Davidson}, {Davies}, {Dawson},
  {De}, {de la Macorra}, {De Lee}, {Queiroz}, {Deconto Machado}, {de la Torre},
  {Dell'Agli}, {du Mas des Bourboux}, {Diamond-Stanic}, {Dillon}, {Donor},
  {Drory}, {Duckworth}, {Dwelly}, {Ebelke}, {Eftekharzadeh}, {Davis Eigenbrot},
  {Elsworth}, {Eracleous}, {Erfanianfar}, {Escoffier}, {Fan}, {Farr},
  {Fern{\'a}ndez-Trincado}, {Feuillet}, {Finoguenov}, {Fofie},
  {Fraser-McKelvie}, {Frinchaboy}, {Fromenteau}, {Fu}, {Galbany}, {Garcia},
  {Garc{\'\i}a-Hern{\'a}ndez}, {Garma Oehmichen}, {Ge}, {Geimba Maia},
  {Geisler}, {Gelfand}, {Goddy}, {Gonzalez-Perez}, {Grabowski}, {Green},
  {Grier}, {Guo}, {Guy}, {Harding}, {Hasselquist}, {Hawken}, {Hayes}, {Hearty},
  {Hekker}, {Hogg}, {Holtzman}, {Horta}, {Hou}, {Hsieh}, {Huber}, {Hunt}, {Ider
  Chitham}, {Imig}, {Jaber}, {Jimenez Angel}, {Johnson}, {Jones},
  {J{\"o}nsson}, {Jullo}, {Kim}, {Kinemuchi}, {Kirkpatrick}, {Kite}, {Klaene},
  {Kneib}, {Kollmeier}, {Kong}, {Kounkel}, {Krishnarao}, {Lacerna}, {Lan},
  {Lane}, {Law}, {Le Goff}, {Leung}, {Lewis}, {Li}, {Lian}, {Lin}, {Long},
  {Longa-Pe{\~n}a}, {Lundgren}, {Lyke}, {Mackereth}, {MacLeod}, {Majewski},
  {Manchado}, {Maraston}, {Martini}, {Masseron}, {Masters}, {Mathur},
  {McDermid}, {Merloni}, {Merrifield}, {M{\'e}sz{\'a}ros}, {Miglio}, {Minniti},
  {Minsley}, {Miyaji}, {Mohammad}, {Mosser}, {Mueller}, {Muna},
  {Mu{\~n}oz-Guti{\'e}rrez}, {Myers}, {Nadathur}, {Nair}, {Nandra}, {Correa do
  Nascimento}, {Nevin}, {Newman}, {Nidever}, {Nitschelm}, {Noterdaeme},
  {O'Connell}, {Olmstead}, {Oravetz}, {Oravetz}, {Osorio}, {Pace}, {Padilla},
  {Palanque-Delabrouille}, {Palicio}, {Pan}, {Pan}, {Parker}, {Paviot},
  {Peirani}, {Ram{\'r}ez}, {Penny}, {Percival}, {Perez-Fournon},
  {P{\'e}rez-R{\`a}fols}, {Petitjean}, {Pieri}, {Pinsonneault}, {Poovelil},
  {Povick}, {Prakash}, {Price-Whelan}, {Raddick}, {Raichoor}, {Ray}, {Rembold},
  {Rezaie}, {Riffel}, {Riffel}, {Rix}, {Robin}, {Roman-Lopes},
  {Rom{\'a}n-Z{\'u}{\~n}iga}, {Rose}, {Ross}, {Rossi}, {Rowlands}, {Rubin},
  {Salvato}, {S{\'a}nchez}, {S{\'a}nchez-Menguiano}, {S{\'a}nchez-Gallego},
  {Sayres}, {Schaefer}, {Schiavon}, {Schimoia}, {Schlafly}, {Schlegel},
  {Schneider}, {Schultheis}, {Schwope}, {Seo}, {Serenelli}, {Shafieloo},
  {Shamsi}, {Shao}, {Shen}, {Shetrone}, {Shirley}, {Silva Aguirre}, {Simon},
  {Skrutskie}, {Slosar}, {Smethurst}, {Sobeck}, {Sodi}, {Souto}, {Stark},
  {Stassun}, {Steinmetz}, {Stello}, {Stermer}, {Storchi-Bergmann},
  {Streblyanska}, {Stringfellow}, {Stutz}, {Su{\'a}rez}, {Sun},
  {Taghizadeh-Popp}, {Talbot}, {Tayar}, {Thakar}, {Theriault}, {Thomas},
  {Thomas}, {Tinker}, {Tojeiro}, {Toledo}, {Tremonti}, {Troup}, {Tuttle},
  {Unda-Sanzana}, {Valentini}, {Vargas-Gonz{\'a}lez}, {Vargas-Maga{\~n}a},
  {V{\'a}zquez-Mata}, {Vivek}, {Wake}, {Wang}, {Weaver}, {Weijmans}, {Wild},
  {Wilson}, {Wilson}, {Wolthuis}, {Wood-Vasey}, {Yan}, {Yang}, {Y{\`e}che},
  {Zamora}, {Zarrouk}, {Zasowski}, {Zhang}, {Zhao}, {Zhao}, {Zheng}, {Zheng},
  {Zhu}, \& {Zou}}]{SDSS16}
{Ahumada}, R., {Allende Prieto}, C., {Almeida}, A., {et~al.} 2020, \apjs, 249,
  3

\bibitem[{{Ajello} {et~al.}(2022){Ajello}, {Baldini}, {Ballet}, {Bastieri},
  {Becerra Gonzalez}, {Bellazzini}, {Berretta}, {Bissaldi}, {Bonino}, {Brill},
  {Bruel}, {Buson}, {Caputo}, {Caraveo}, {Cheung}, {Chiaro}, {Cibrario},
  {Ciprini}, {Crnogorcevic}, {Cutini}, {D'Ammando}, {De Gaetano}, {Di Lalla},
  {Di Venere}, {Dom{\'\i}nguez}, {Ramazani}, {Ferrara}, {Fiori}, {Fukazawa},
  {Funk}, {Fusco}, {Gammaldi}, {Gargano}, {Garrappa}, {Gasparrini},
  {Giglietto}, {Giordano}, {Giroletti}, {Green}, {Grenier}, {Guiriec}, {Horan},
  {Hou}, {Kayanoki}, {Kuss}, {Larsson}, {Latronico}, {Lewis}, {Li}, {Liodakis},
  {Longo}, {Loparco}, {Lott}, {Lovellette}, {Lubrano}, {Madejski}, {Maldera},
  {Manfreda}, {Mart{\'\i}-Devesa}, {Mazziotta}, {Mereu}, {Michelson},
  {Mirabal}, {Mitthumsiri}, {Mizuno}, {Monzani}, {Morselli}, {Moskalenko},
  {Negro}, {Ojha}, {Orienti}, {Orlando}, {Ormes}, {Pei}, {Pe{\~n}a-Herazo},
  {Persic}, {Pesce-Rollins}, {Petrosian}, {Pillera}, {Poon}, {Porter},
  {Principe}, {Rain{\`o}}, {Rando}, {Rani}, {Razzano}, {Razzaque}, {Reimer},
  {Reimer}, {Scotton}, {Serini}, {Sgr{\`o}}, {Siskind}, {Spandre}, {Spinelli},
  {Suson}, {Tajima}, {Torres}, {Valverde}, {Yassin}, \& {Zaharijas}}]{4LAC-DR3}
{Ajello}, M., {Baldini}, L., {Ballet}, J., {et~al.} 2022, \apjs, 263, 24

\bibitem[{Anderson \& Darling(1952)}]{Anderson_Darling}
Anderson, T. \& Darling, D. 1952, The Annals of Mathematical Statistics, 23,
  193

\bibitem[{{Antonucci}(2012)}]{Antonucci:2012}
{Antonucci}, R. 2012, Astronomical and Astrophysical Transactions, 27, 557

\bibitem[{{Appenzeller} {et~al.}(1998){Appenzeller}, {Fricke}, {F{\"u}rtig},
  {G{\"a}ssler}, {H{\"a}fner}, {Harke}, {Hess}, {Hummel}, {J{\"u}rgens},
  {Kudritzki}, {Mantel}, {Meisl}, {Muschielok}, {Nicklas}, {Rupprecht},
  {Seifert}, {Stahl}, {Szeifert}, \& {Tarantik}}]{FORS}
{Appenzeller}, I., {Fricke}, K., {F{\"u}rtig}, W., {et~al.} 1998, The
  Messenger, 94, 1

\bibitem[{{Avni} {et~al.}(1980){Avni}, {Soltan}, {Tananbaum}, \&
  {Zamorani}}]{statistical_methods_III}
{Avni}, Y., {Soltan}, A., {Tananbaum}, H., \& {Zamorani}, G. 1980, \apj, 238,
  800

\bibitem[{{Baldi} {et~al.}(2022){Baldi}, {Laor}, {Behar}, {Horesh}, {Panessa},
  {McHardy}, \& {Kimball}}]{Baldi_2022}
{Baldi}, R.~D., {Laor}, A., {Behar}, E., {et~al.} 2022, \mnras, 510, 1043

\bibitem[{{Ballet} {et~al.}(2020){Ballet}, {Burnett}, {Digel}, \&
  {Lott}}]{4FGL-DR2}
{Ballet}, J., {Burnett}, T.~H., {Digel}, S.~W., \& {Lott}, B. 2020, arXiv
  e-prints, arXiv:2005.11208

\bibitem[{Barbano {et~al.}(2023)Barbano, Buson, de~Clairfontaine, Oswald,
  Tramacere, Illuminati, Azzollini, Baghmanyan, \& Pfeiffer}]{Barbano_2023}
Barbano, E., Buson, S., de~Clairfontaine, G.~F., {et~al.} 2023, PoS, ICRC2023,
  1540

\bibitem[{{Bellenghi} {et~al.}(2023){Bellenghi}, {Padovani}, {Resconi}, \&
  {Giommi}}]{Bellenghi_2023}
{Bellenghi}, C., {Padovani}, P., {Resconi}, E., \& {Giommi}, P. 2023, arXiv
  e-prints, arXiv:2309.03115

\bibitem[{Benjamini \& Hochberg(1995)}]{Benjamini_Hochberg_1995}
Benjamini, Y. \& Hochberg, Y. 1995, Journal of the Royal Statistical Society:
  Series B (Methodological), 57, 289

\bibitem[{{Best} \& {Heckman}(2012)}]{BestHeckman:2012}
{Best}, P.~N. \& {Heckman}, T.~M. 2012, \mnras, 421, 1569

\bibitem[{{Blandford} \& {Payne}(1982)}]{Blandford_Payne}
{Blandford}, R.~D. \& {Payne}, D.~G. 1982, \mnras, 199, 883

\bibitem[{{Blandford} \& {Znajek}(1977)}]{Blandford_Znajek}
{Blandford}, R.~D. \& {Znajek}, R.~L. 1977, \mnras, 179, 433

\bibitem[{{Bondi}(1952)}]{Bondi:1952}
{Bondi}, H. 1952, \mnras, 112, 195

\bibitem[{{B{\"o}ttcher} {et~al.}(2013){B{\"o}ttcher}, {Reimer}, {Sweeney}, \&
  {Prakash}}]{Boettcher:2013}
{B{\"o}ttcher}, M., {Reimer}, A., {Sweeney}, K., \& {Prakash}, A. 2013, \apj,
  768, 54

\bibitem[{{Buson} {et~al.}(2023){Buson}, {Tramacere}, {Oswald}, {Barbano},
  {Fichet de Clairfontaine}, {Pfeiffer}, {Azzollini}, {Baghmanyan}, \&
  {Ajello}}]{Buson:2023}
{Buson}, S., {Tramacere}, A., {Oswald}, L., {et~al.} 2023, arXiv e-prints,
  arXiv:2305.11263

\bibitem[{{Buson} {et~al.}(2022{\natexlab{a}}){Buson}, {Tramacere}, {Pfeiffer},
  {Oswald}, {de Menezes}, {Azzollini}, \& {Ajello}}]{Buson:2022}
{Buson}, S., {Tramacere}, A., {Pfeiffer}, L., {et~al.} 2022{\natexlab{a}},
  \apjl, 933, L43

\bibitem[{{Buson} {et~al.}(2022{\natexlab{b}}){Buson}, {Tramacere}, {Pfeiffer},
  {Oswald}, {de Menezes}, {Azzollini}, \& {Ajello}}]{Buson_erratum}
{Buson}, S., {Tramacere}, A., {Pfeiffer}, L., {et~al.} 2022{\natexlab{b}},
  \apjl, 934, L38

\bibitem[{Cao \& Jiang(1999)}]{Cao:1999}
Cao, X. \& Jiang, D.~R. 1999, Monthly Notices of the Royal Astronomical
  Society, 307, 802

\bibitem[{{Carnall}(2017)}]{SpectRes}
{Carnall}, A.~C. 2017, arXiv e-prints, arXiv:1705.05165

\bibitem[{{Celotti} {et~al.}(1997){Celotti}, {Padovani}, \&
  {Ghisellini}}]{Celotti:1997}
{Celotti}, A., {Padovani}, P., \& {Ghisellini}, G. 1997, \mnras, 286, 415

\bibitem[{{Cepa} {et~al.}(2003){Cepa}, {Aguiar-Gonzalez}, {Bland-Hawthorn},
  {Castaneda}, {Cobos}, {Correa}, {Espejo}, {Fragoso-Lopez}, {Fuentes},
  {Gigante}, {Gonzalez}, {Gonzalez-Escalera}, {Gonzalez-Serrano},
  {Joven-Alvarez}, {Lopez-Ruiz}, {Militello}, {Cano}, {Perez}, {Perez},
  {Rasilla}, {Sanchez}, \& {Tejada}}]{osiris}
{Cepa}, J., {Aguiar-Gonzalez}, M., {Bland-Hawthorn}, J., {et~al.} 2003, in
  Society of Photo-Optical Instrumentation Engineers (SPIE) Conference Series,
  Vol. 4841, Instrument Design and Performance for Optical/Infrared
  Ground-based Telescopes, ed. M.~{Iye} \& A.~F.~M. {Moorwood}, 1739--1749

\bibitem[{Cerruti(2020)}]{Cerruti:2020}
Cerruti, M. 2020, Galaxies, 8

\bibitem[{{Chen} {et~al.}(2021){Chen}, {Gu}, {Fan}, {Zhou}, {Yuan}, {Gu}, {Wu},
  {Xiong}, {Guo}, {Ding}, \& {Yu}}]{Chen:2021}
{Chen}, Y., {Gu}, Q., {Fan}, J., {et~al.} 2021, \apj, 913, 93

\bibitem[{{Condon} {et~al.}(1998){Condon}, {Cotton}, {Greisen}, {Yin},
  {Perley}, {Taylor}, \& {Broderick}}]{NVSS}
{Condon}, J.~J., {Cotton}, W.~D., {Greisen}, E.~W., {et~al.} 1998, \aj, 115,
  1693

\bibitem[{{Croom} {et~al.}(2004){Croom}, {Smith}, {Boyle}, {Shanks}, {Miller},
  {Outram}, \& {Loaring}}]{2QZ:2004}
{Croom}, S.~M., {Smith}, R.~J., {Boyle}, B.~J., {et~al.} 2004, \mnras, 349,
  1397

\bibitem[{{Decarli} {et~al.}(2011){Decarli}, {Dotti}, \&
  {Treves}}]{Decarli:2011}
{Decarli}, R., {Dotti}, M., \& {Treves}, A. 2011, \mnras, 413, 39

\bibitem[{{Decarli} {et~al.}(2008){Decarli}, {Labita}, {Treves}, \&
  {Falomo}}]{Decarli:2008}
{Decarli}, R., {Labita}, M., {Treves}, A., \& {Falomo}, R. 2008, \mnras, 387,
  1237

\bibitem[{{Deconto-Machado} {et~al.}(2023){Deconto-Machado}, {del Olmo Orozco},
  {Marziani}, {Perea}, \& {Stirpe}}]{DecontoMachado:2023}
{Deconto-Machado}, A., {del Olmo Orozco}, A., {Marziani}, P., {Perea}, J., \&
  {Stirpe}, G.~M. 2023, \aap, 669, A83

\bibitem[{{Dermer} {et~al.}(2014){Dermer}, {Murase}, \& {Inoue}}]{Dermer:2014}
{Dermer}, C.~D., {Murase}, K., \& {Inoue}, Y. 2014, Journal of High Energy
  Astrophysics, 3, 29

\bibitem[{{Drinkwater} {et~al.}(1997){Drinkwater}, {Webster}, {Francis},
  {Condon}, {Ellison}, {Jauncey}, {Lovell}, {Peterson}, \&
  {Savage}}]{Drinkwater:1997}
{Drinkwater}, M.~J., {Webster}, R.~L., {Francis}, P.~J., {et~al.} 1997, \mnras,
  284, 85

\bibitem[{{Du} {et~al.}(2013){Du}, {Bai}, \& {Xie}}]{Du:2013}
{Du}, L.~M., {Bai}, J.~M., \& {Xie}, Z.~H. 2013, \na, 18, 1

\bibitem[{{Dwelly} {et~al.}(2017){Dwelly}, {Salvato}, {Merloni}, {Brusa},
  {Buchner}, {Anderson}, {Boller}, {Brandt}, {Budav{\'a}ri}, {Clerc}, {Coffey},
  {Del Moro}, {Georgakakis}, {Green}, {Jin}, {Menzel}, {Myers}, {Nandra},
  {Nichol}, {Ridl}, {Schwope}, \& {Simm}}]{Dwelly:2017}
{Dwelly}, T., {Salvato}, M., {Merloni}, A., {et~al.} 2017, \mnras, 469, 1065

\bibitem[{Eddington(1926)}]{Eddington:1926}
Eddington, A.~S. 1926, The internal constitution of the stars (Cambridge
  University Press)

\bibitem[{{Fan} {et~al.}(2016){Fan}, {Yang}, {Liu}, {Luo}, {Lin}, {Yuan},
  {Xiao}, {Zhou}, {Hua}, \& {Pei}}]{Fan:2016}
{Fan}, J.~H., {Yang}, J.~H., {Liu}, Y., {et~al.} 2016, \apjs, 226, 20

\bibitem[{{Feigelson} \& {Nelson}(1985)}]{statistical_methods_I}
{Feigelson}, E.~D. \& {Nelson}, P.~I. 1985, \apj, 293, 192

\bibitem[{{Fichet de Clairfontaine} {et~al.}(2023){Fichet de Clairfontaine},
  {Buson}, {Pfeiffer}, {Marchesi}, {Azzollini}, {Baghmanyan}, {Tramacere},
  {Barbano}, \& {Oswald}}]{FichetDC:2023}
{Fichet de Clairfontaine}, G., {Buson}, S., {Pfeiffer}, L., {et~al.} 2023,
  \apjl, 958, L2

\bibitem[{{Fischer} {et~al.}(1998){Fischer}, {Hasinger}, {Schwope}, {Brunner},
  {Boller}, {Tr{\"u}mper}, {Voges}, \& {Neizvestnyj}}]{Fischer:1998}
{Fischer}, J.~U., {Hasinger}, G., {Schwope}, A.~D., {et~al.} 1998,
  Astronomische Nachrichten, 319, 347

\bibitem[{{Fossati} {et~al.}(1998){Fossati}, {Maraschi}, {Celotti}, {Comastri},
  \& {Ghisellini}}]{Fossati:1998}
{Fossati}, G., {Maraschi}, L., {Celotti}, A., {Comastri}, A., \& {Ghisellini},
  G. 1998, \mnras, 299, 433

\bibitem[{{Francis} {et~al.}(1991){Francis}, {Hewett}, {Foltz}, {Chaffee},
  {Weymann}, \& {Morris}}]{Francis:1991}
{Francis}, P.~J., {Hewett}, P.~C., {Foltz}, C.~B., {et~al.} 1991, \apj, 373,
  465

\bibitem[{{Freudling} {et~al.}(2013){Freudling}, {Romaniello}, {Bramich},
  {Ballester}, {Forchi}, {Garc{\'{\i}}a-Dabl{\'o}}, {Moehler}, \&
  {Neeser}}]{EsoReflex}
{Freudling}, W., {Romaniello}, M., {Bramich}, D.~M., {et~al.} 2013, \aap, 559,
  A96

\bibitem[{{Ghisellini} \& {Celotti}(2001)}]{Ghisellini_Celotti:2001}
{Ghisellini}, G. \& {Celotti}, A. 2001, \aap, 379, L1

\bibitem[{{Ghisellini} {et~al.}(2009){Ghisellini}, {Maraschi}, \&
  {Tavecchio}}]{Ghisellini:2009}
{Ghisellini}, G., {Maraschi}, L., \& {Tavecchio}, F. 2009, \mnras, 396, L105

\bibitem[{{Ghisellini} {et~al.}(2017){Ghisellini}, {Righi}, {Costamante}, \&
  {Tavecchio}}]{Ghisellini:2017}
{Ghisellini}, G., {Righi}, C., {Costamante}, L., \& {Tavecchio}, F. 2017,
  \mnras, 469, 255

\bibitem[{{Ghisellini} {et~al.}(2011{\natexlab{a}}){Ghisellini}, {Tagliaferri},
  {Foschini}, {Ghirlanda}, {Tavecchio}, {Della Ceca}, {Haardt}, {Volonteri}, \&
  {Gehrels}}]{Ghisellini2011_highred}
{Ghisellini}, G., {Tagliaferri}, G., {Foschini}, L., {et~al.}
  2011{\natexlab{a}}, \mnras, 411, 901

\bibitem[{{Ghisellini} \& {Tavecchio}(2008)}]{Ghisellini:2008}
{Ghisellini}, G. \& {Tavecchio}, F. 2008, \mnras, 387, 1669

\bibitem[{{Ghisellini} {et~al.}(2011{\natexlab{b}}){Ghisellini}, {Tavecchio},
  {Foschini}, \& {Ghirlanda}}]{Ghisellini:2011}
{Ghisellini}, G., {Tavecchio}, F., {Foschini}, L., \& {Ghirlanda}, G.
  2011{\natexlab{b}}, \mnras, 414, 2674

\bibitem[{{Ghisellini} {et~al.}(2010){Ghisellini}, {Tavecchio}, {Foschini},
  {Ghirlanda}, {Maraschi}, \& {Celotti}}]{Ghisellini:2010}
{Ghisellini}, G., {Tavecchio}, F., {Foschini}, L., {et~al.} 2010, \mnras, 402,
  497

\bibitem[{{Ghisellini} {et~al.}(2012){Ghisellini}, {Tavecchio}, {Foschini},
  {Sbarrato}, {Ghirlanda}, \& {Maraschi}}]{Ghisellini:2012}
{Ghisellini}, G., {Tavecchio}, F., {Foschini}, L., {et~al.} 2012, \mnras, 425,
  1371

\bibitem[{{Ghisellini} {et~al.}(2014){Ghisellini}, {Tavecchio}, {Maraschi},
  {Celotti}, \& {Sbarrato}}]{Ghisellini:2014}
{Ghisellini}, G., {Tavecchio}, F., {Maraschi}, L., {Celotti}, A., \&
  {Sbarrato}, T. 2014, \nat, 515, 376

\bibitem[{{Gimeno} {et~al.}(2016){Gimeno}, {Roth}, {Chiboucas}, {Hibon},
  {Boucher}, {White}, {Rippa}, {Labrie}, {Turner}, {Hanna}, {Lazo},
  {P{\'e}rez}, {Rogers}, {Rojas}, {Placco}, \& {Murowinski}}]{GMOS:2016}
{Gimeno}, G., {Roth}, K., {Chiboucas}, K., {et~al.} 2016, in Society of
  Photo-Optical Instrumentation Engineers (SPIE) Conference Series, Vol. 9908,
  Ground-based and Airborne Instrumentation for Astronomy VI, ed. C.~J.
  {Evans}, L.~{Simard}, \& H.~{Takami}, 99082S

\bibitem[{{Giommi} {et~al.}(2013){Giommi}, {Padovani}, \&
  {Polenta}}]{Giommi:2013}
{Giommi}, P., {Padovani}, P., \& {Polenta}, G. 2013, \mnras, 431, 1914

\bibitem[{{Giommi} {et~al.}(2012){Giommi}, {Padovani}, {Polenta}, {Turriziani},
  {D'Elia}, \& {Piranomonte}}]{Giommi:2012}
{Giommi}, P., {Padovani}, P., {Polenta}, G., {et~al.} 2012, \mnras, 420, 2899

\bibitem[{{Graham}(2007)}]{Graham:2007}
{Graham}, A.~W. 2007, \mnras, 379, 711

\bibitem[{{Healey} {et~al.}(2008){Healey}, {Romani}, {Cotter}, {Michelson},
  {Schlafly}, {Readhead}, {Giommi}, {Chaty}, {Grenier}, \&
  {Weintraub}}]{CGRaBS}
{Healey}, S.~E., {Romani}, R.~W., {Cotter}, G., {et~al.} 2008, \apjs, 175, 97

\bibitem[{{Healey} {et~al.}(2007){Healey}, {Romani}, {Taylor}, {Sadler},
  {Ricci}, {Murphy}, {Ulvestad}, \& {Winn}}]{Crates:2007}
{Healey}, S.~E., {Romani}, R.~W., {Taylor}, G.~B., {et~al.} 2007, \apjs, 171,
  61

\bibitem[{{Heckman} \& {Best}(2014)}]{Heckman:2014}
{Heckman}, T.~M. \& {Best}, P.~N. 2014, \araa, 52, 589

\bibitem[{{Hewett} \& {Wild}(2010)}]{Hewett:2010}
{Hewett}, P.~C. \& {Wild}, V. 2010, \mnras, 405, 2302

\bibitem[{Homan {et~al.}(2021)Homan, Cohen, Hovatta, Kellermann, Kovalev,
  Lister, Popkov, Pushkarev, Ros, \& Savolainen}]{Homan_2021}
Homan, D.~C., Cohen, M.~H., Hovatta, T., {et~al.} 2021, The Astrophysical
  Journal, 923, 67

\bibitem[{{Hovatta} {et~al.}(2021){Hovatta}, {Lindfors}, {Kiehlmann},
  {Max-Moerbeck}, {Hodges}, {Liodakis}, {L{\"a}hteem{\"a}ki}, {Pearson},
  {Readhead}, {Reeves}, {Suutarinen}, {Tammi}, \& {Tornikoski}}]{Hovatta:2021}
{Hovatta}, T., {Lindfors}, E., {Kiehlmann}, S., {et~al.} 2021, \aap, 650, A83

\bibitem[{{IceCube Collaboration}(2013)}]{icecube2013}
{IceCube Collaboration}. 2013, {Science}, 342, 1242856

\bibitem[{{IceCube Collaboration} {et~al.}(2018){IceCube Collaboration},
  {Aartsen}, {Ackermann}, {Adams}, {Aguilar}, {Ahlers}, {Ahrens}, {Al Samarai},
  {Altmann}, {Andeen}, {Anderson}, {Ansseau}, {Anton}, {Arg{\"u}elles},
  {Auffenberg}, {Axani}, {Bagherpour}, {Bai}, {Barron}, {Barwick}, {Baum},
  {Bay}, {Beatty}, {Becker Tjus}, {Becker}, {BenZvi}, {Berley}, {Bernardini},
  {Besson}, {Binder}, {Bindig}, {Blaufuss}, {Blot}, {Bohm}, {B{\"o}rner},
  {Bos}, {B{\"o}ser}, {Botner}, {Bourbeau}, {Bourbeau}, {Bradascio}, {Braun},
  {Brenzke}, {Bretz}, {Bron}, {Brostean-Kaiser}, {Burgman}, {Busse}, {Carver},
  {Cheung}, {Chirkin}, {Christov}, {Clark}, {Classen}, {Coenders}, {Collin},
  {Conrad}, {Coppin}, {Correa}, {Cowen}, {Cross}, {Dave}, {Day}, {de
  Andr{\'e}}, {De Clercq}, {DeLaunay}, {Dembinski}, {De Ridder}, {Desiati}, {de
  Vries}, {de Wasseige}, {de With}, {DeYoung}, {D{\'\i}az-V{\'e}lez}, {di
  Lorenzo}, {Dujmovic}, {Dumm}, {Dunkman}, {Dvorak}, {Eberhardt}, {Ehrhardt},
  {Eichmann}, {Eller}, {Evenson}, {Fahey}, {Fazely}, {Felde}, {Filimonov},
  {Finley}, {Flis}, {Franckowiak}, {Friedman}, {Fritz}, {Gaisser}, {Gallagher},
  {Gerhardt}, {Ghorbani}, {Glauch}, {Gl{\"u}senkamp}, {Goldschmidt},
  {Gonzalez}, {Grant}, {Griffith}, {Haack}, {Hallgren}, {Halzen}, {Hanson},
  {Hebecker}, {Heereman}, {Helbing}, {Hellauer}, {Hickford}, {Hignight},
  {Hill}, {Hoffman}, {Hoffmann}, {Hoinka}, {Hokanson-Fasig}, {Hoshina},
  {Huang}, {Huber}, {Hultqvist}, {H{\"u}nnefeld}, {Hussain}, {In}, {Iovine},
  {Ishihara}, {Jacobi}, {Japaridze}, {Jeong}, {Jero}, {Jones}, {Kalaczynski},
  {Kang}, {Kappes}, {Kappesser}, {Karg}, {Karle}, {Katz}, {Kauer}, {Keivani},
  {Kelley}, {Kheirandish}, {Kim}, {Kim}, {Kintscher}, {Kiryluk}, {Kittler},
  {Klein}, {Koirala}, {Kolanoski}, {K{\"o}pke}, {Kopper}, {Kopper},
  {Koschinsky}, {Koskinen}, {Kowalski}, {Krings}, {Kroll}, {Kr{\"u}ckl},
  {Kunwar}, {Kurahashi}, {Kuwabara}, {Kyriacou}, {Labare}, {Lanfranchi},
  {Larson}, {Lauber}, {Leonard}, {Lesiak-Bzdak}, {Leuermann}, {Liu}, {Lozano
  Mariscal}, {Lu}, {L{\"u}nemann}, {Luszczak}, {Madsen}, {Maggi}, {Mahn},
  {Mancina}, {Maruyama}, {Mase}, {Maunu}, {Meagher}, {Medici}, {Meier},
  {Menne}, {Merino}, {Meures}, {Miarecki}, {Micallef}, {Moment{\'e}},
  {Montaruli}, {Moore}, {Morse}, {Moulai}, {Nahnhauer}, {Nakarmi}, {Naumann},
  {Neer}, {Niederhausen}, {Nowicki}, {Nygren}, {Obertacke Pollmann}, {Olivas},
  {O'Murchadha}, {O'Sullivan}, {Palczewski}, {Pandya}, {Pankova}, {Peiffer},
  {Pepper}, {P{\'e}rez de los Heros}, {Pieloth}, {Pinat}, {Plum}, {Price},
  {Przybylski}, {Raab}, {R{\"a}del}, {Rameez}, {Rauch}, {Rawlins}, {Rea},
  {Reimann}, {Relethford}, {Relich}, {Resconi}, {Rhode}, {Richman},
  {Robertson}, {Rongen}, {Rott}, {Ruhe}, {Ryckbosch}, {Rysewyk}, {Safa},
  {S{\"a}lzer}, {Sanchez Herrera}, {Sandrock}, {Sandroos}, {Santander},
  {Sarkar}, {Sarkar}, {Satalecka}, {Schlunder}, {Schmidt}, {Schneider},
  {Schoenen}, {Sch{\"o}neberg}, {Schumacher}, {Sclafani}, {Seckel},
  {Seunarine}, {Soedingrekso}, {Soldin}, {Song}, {Spiczak}, {Spiering},
  {Stachurska}, {Stamatikos}, {Stanev}, {Stasik}, {Stein}, {Stettner},
  {Steuer}, {Stezelberger}, {Stokstad}, {St{\"o}{\ss}l}, {Strotjohann},
  {Stuttard}, {Sullivan}, {Sutherland}, {Taboada}, {Tatar}, {Tenholt},
  {Ter-Antonyan}, {Terliuk}, {Tilav}, {Toale}, {Tobin}, {Toennis}, {Toscano},
  {Tosi}, {Tselengidou}, {Tung}, {Turcati}, {Turley}, {Ty}, {Unger}, {Usner},
  {Vandenbroucke}, {Van Driessche}, {van Eijk}, {van Eijndhoven}, {Vanheule},
  {van Santen}, {Vogel}, {Vraeghe}, {Walck}, {Wallace}, {Wallraff}, {Wandler},
  {Wandkowsky}, {Waza}, {Weaver}, {Weiss}, {Wendt}, {Werthebach}, {Westerhoff},
  {Whelan}, {Whitehorn}, {Wiebe}, {Wiebusch}, {Wille}, {Williams}, {Wills},
  {Wolf}, {Wood}, {Wood}, {Woschnagg}, {Xu}, {Xu}, {Xu}, {Yanez}, {Yodh},
  {Yoshida}, {Yuan}, {Fermi-LAT Collaboration}, {Abdollahi}, {Ajello},
  {Angioni}, {Baldini}, {Ballet}, {Barbiellini}, {Bastieri}, {Bechtol},
  {Bellazzini}, {Berenji}, {Bissaldi}, {Blandford}, {Bonino}, {Bottacini},
  {Bregeon}, {Bruel}, {Buehler}, {Burnett}, {Burns}, {Buson}, {Cameron},
  {Caputo}, {Caraveo}, {Cavazzuti}, {Charles}, {Chen}, {Cheung}, {Chiang},
  {Chiaro}, {Ciprini}, {Cohen-Tanugi}, {Conrad}, {Costantin}, {Cutini},
  {D'Ammando}, {de Palma}, {Digel}, {Di Lalla}, {Di Mauro}, {Di Venere},
  {Dom{\'\i}nguez}, {Favuzzi}, {Franckowiak}, {Fukazawa}, {Funk}, {Fusco},
  {Gargano}, {Gasparrini}, {Giglietto}, {Giomi}, {Giommi}, {Giordano},
  {Giroletti}, {Glanzman}, {Green}, {Grenier}, {Grondin}, {Guiriec}, {Harding},
  {Hayashida}, {Hays}, {Hewitt}, {Horan}, {J{\'o}hannesson}, {Kadler},
  {Kensei}, {Kocevski}, {Krauss}, {Kreter}, {Kuss}, {La Mura}, {Larsson},
  {Latronico}, {Lemoine-Goumard}, {Li}, {Longo}, {Loparco}, {Lovellette},
  {Lubrano}, {Magill}, {Maldera}, {Malyshev}, {Manfreda}, {Mazziotta},
  {McEnery}, {Meyer}, {Michelson}, {Mizuno}, {Monzani}, {Morselli},
  {Moskalenko}, {Negro}, {Nuss}, {Ojha}, {Omodei}, {Orienti}, {Orlando},
  {Palatiello}, {Paliya}, {Perkins}, {Persic}, {Pesce-Rollins}, {Piron},
  {Porter}, {Principe}, {Rain{\`o}}, {Rando}, {Rani}, {Razzano}, {Razzaque},
  {Reimer}, {Reimer}, {Renault-Tinacci}, {Ritz}, {Rochester}, {Saz Parkinson},
  {Sgr{\`o}}, {Siskind}, {Spandre}, {Spinelli}, {Suson}, {Tajima}, {Takahashi},
  {Tanaka}, {Thayer}, {Thompson}, {Tibaldo}, {Torres}, {Torresi}, {Tosti},
  {Troja}, {Valverde}, {Vianello}, {Vogel}, {Wood}, {Wood}, {Zaharijas}, {MAGIC
  Collaboration}, {Ahnen}, {Ansoldi}, {Antonelli}, {Arcaro}, {Baack},
  {Babi{\'c}}, {Banerjee}, {Bangale}, {Barres de Almeida}, {Barrio}, {Becerra
  Gonz{\'a}lez}, {Bednarek}, {Bernardini}, {Berti}, {Bhattacharyya}, {Biland},
  {Blanch}, {Bonnoli}, {Carosi}, {Carosi}, {Ceribella}, {Chatterjee}, {Colak},
  {Colin}, {Colombo}, {Contreras}, {Cortina}, {Covino}, {Cumani}, {Da Vela},
  {Dazzi}, {De Angelis}, {De Lotto}, {Delfino}, {Delgado}, {Di Pierro},
  {Dom{\'\i}nguez}, {Dominis Prester}, {Dorner}, {Doro}, {Einecke},
  {Elsaesser}, {Fallah Ramazani}, {Fern{\'a}ndez-Barral}, {Fidalgo}, {Foffano},
  {Pfrang}, {Fonseca}, {Font}, {Franceschini}, {Fruck}, {Galindo}, {Gallozzi},
  {Garc{\'\i}a L{\'o}pez}, {Garczarczyk}, {Gaug}, {Giammaria}, {Godinovi{\'c}},
  {Gora}, {Guberman}, {Hadasch}, {Hahn}, {Hassan}, {Hayashida}, {Herrera},
  {Hose}, {Hrupec}, {Inoue}, {Ishio}, {Konno}, {Kubo}, {Kushida}, {Lelas},
  {Lindfors}, {Lombardi}, {Longo}, {L{\'o}pez}, {Maggio}, {Majumdar},
  {Makariev}, {Maneva}, {Manganaro}, {Mannheim}, {Maraschi}, {Mariotti},
  {Mart{\'\i}nez}, {Masuda}, {Mazin}, {Minev}, {M}, {Mirzoyan}, {Moralejo},
  {Moreno}, {Moretti}, {Nagayoshi}, {Neustroev}, {Niedzwiecki}, {Nievas
  Rosillo}, {Nigro}, {Nilsson}, {Ninci}, {Nishijima}, {Noda}, {Nogu{\'e}s},
  {Paiano}, {Palacio}, {Paneque}, {Paoletti}, {Paredes}, {Pedaletti},
  {Peresano}, {Persic}, {Prada Moroni}, {Prandini}, {Puljak}, {Rodriguez
  Garcia}, {Reichardt}, {Rhode}, {Rib{\'o}}, {Rico}, {Righi}, {Rugliancich},
  {Saito}, {Satalecka}, {Schweizer}, {Sitarek}, {{\v{S}}nidaric
  {\textasciiacute}}, {Sobczynska}, {Stamerra}, {Strzys}, {Suri{\'c}},
  {Takahashi}, {Tavecchio}, {Temnikov}, {Terzi{\'c}}, {Teshima},
  {Torres-Alb{\`a}}, {Treves}, {Tsujimoto}, {Vanzo}, {Vazquez Acosta}, {Vovk},
  {Ward}, {Will}, {S}, {Zaric {\textasciiacute}}, {AGILE Team}, {Lucarelli},
  {Tavani}, {Piano}, {Donnarumma}, {Pittori}, {Verrecchia}, {Barbiellini},
  {Bulgarelli}, {Caraveo}, {Cattaneo}, {Colafrancesco}, {Costa}, {Di Cocco},
  {Ferrari}, {Gianotti}, {Giuliani}, {Lipari}, {Mereghetti}, {Morselli},
  {Pacciani}, {Paoletti}, {Parmiggiani}, {Pellizzoni}, {Picozza}, {Pilia},
  {Rappoldi}, {Trois}, {Vercellone}, {Vittorini}, {ASAS-SN Team}, {Stanek},
  {Franckowiak}, {Kochanek}, {Beacom}, {Thompson}, {Holoien}, {Dong}, {Prieto},
  {Shappee}, {Holmbo}, {HAWC Collaboration}, {Abeysekara}, {Albert}, {Alfaro},
  {Alvarez}, {Arceo}, {Arteaga-Vel{\'a}zquez}, {Avila Rojas}, {Ayala Solares},
  {Becerril}, {Belmont-Moreno}, {Bernal}, {Caballero-Mora}, {Capistr{\'a}n},
  {Carrami{\~n}ana}, {Casanova}, {Castillo}, {Cotti}, {Cotzomi}, {Couti{\~n}o
  de Le{\'o}n}, {De Le{\'o}n}, {De la Fuente}, {Diaz Hernandez}, {Dichiara},
  {Dingus}, {DuVernois}, {D{\'\i}az-V{\'e}lez}, {Ellsworth}, {Engel},
  {Fiorino}, {Fleischhack}, {Fraija}, {Garc{\'\i}a-Gonz{\'a}lez}, {Garfias},
  {Gonz{\'a}lez Mu{\~n}oz}, {Gonz{\'a}lez}, {Goodman}, {Hampel-Arias},
  {Harding}, {Hernandez}, {Hona}, {Hueyotl-Zahuantitla}, {Hui},
  {H{\"u}ntemeyer}, {Iriarte}, {Jardin-Blicq}, {Joshi}, {Kaufmann}, {Kunde},
  {Lara}, {Lauer}, {Lee}, {Lennarz}, {Le{\'o}n Vargas}, {Linnemann},
  {Longinotti}, {Luis-Raya}, {Luna-Garc{\'\i}a}, {Malone}, {Marinelli},
  {Martinez}, {Martinez-Castellanos}, {Mart{\'\i}nez-Castro},
  {Mart{\'\i}nez-Huerta}, {Matthews}, {Miranda-Romagnoli}, {Moreno},
  {Mostaf{\'a}}, {Nayerhoda}, {Nellen}, {Newbold}, {Nisa}, {Noriega-Papaqui},
  {Pelayo}, {Pretz}, {P{\'e}rez-P{\'e}rez}, {Ren}, {Rho}, {Rivi{\`e}re},
  {Rosa-Gonz{\'a}lez}, {Rosenberg}, {Ruiz-Velasco}, {Ruiz-Velasco}, {Salesa
  Greus}, {Sandoval}, {Schneider}, {Schoorlemmer}, {Sinnis}, {Smith},
  {Springer}, {Surajbali}, {Tibolla}, {Tollefson}, {Torres}, {Villase{\~n}or},
  {Weisgarber}, {Werner}, {Yapici}, {Gaurang}, {Zepeda}, {Zhou}, {{\'A}lvarez},
  {H.~E.~S.~S. Collaboration}, {Abdalla}, {Ang{\"u}ner}, {Armand}, {Backes},
  {Becherini}, {Berge}, {B{\"o}ttcher}, {Boisson}, {Bolmont}, {Bonnefoy},
  {Bordas}, {Brun}, {B{\"u}chele}, {Bulik}, {Caroff}, {Carosi}, {Casanova},
  {Cerruti}, {Chakraborty}, {Chandra}, {Chen}, {Colafrancesco}, {Davids},
  {Deil}, {Devin}, {Djannati-Ata{\"\i}}, {Egberts}, {Emery}, {Eschbach},
  {Fiasson}, {Fontaine}, {Funk}, {F{\"u}{\ss}ling}, {Gallant}, {Gat{\'e}},
  {Giavitto}, {Glawion}, {Glicenstein}, {Gottschall}, {Grondin}, {Haupt},
  {Henri}, {Hinton}, {Hoischen}, {Holch}, {Huber}, {Jamrozy}, {Jankowsky},
  {Jankowsky}, {Jouvin}, {Jung-Richardt}, {Kerszberg}, {Kh{\'e}lifi}, {King},
  {Klepser}, {Kluz {\textasciiacute}niak}, {Komin}, {Kraus}, {Lefaucheur},
  {Lemi{\`e}re}, {Lemoine-Goumard}, {Lenain}, {Leser}, {Lohse},
  {L{\'o}pez-Coto}, {Lorentz}, {Lypova}, {Marandon}, {Guillem
  Mart{\'\i}-Devesa}, {Maurin}, {Mitchell}, {Moderski}, {Mohamed}, {Mohrmann},
  {Moulin}, {Murach}, {de Naurois}, {Niederwanger}, {Niemiec}, {Oakes},
  {O'Brien}, {Ohm}, {Ostrowski}, {Oya}, {Panter}, {Parsons}, {Perennes},
  {Piel}, {Pita}, {Poireau}, {Priyana Noel}, {Prokoph}, {P{\"u}hlhofer},
  {Quirrenbach}, {Raab}, {Rauth}, {Renaud}, {Rieger}, {Rinchiuso}, {Romoli},
  {Rowell}, {Rudak}, {Sasaki}, {Sanchez}, {Schlickeiser}, {Sch{\"u}ssler},
  {Schulz}, {Schwanke}, {Seglar-Arroyo}, {Shafi}, {Simoni}, {Sol}, {Stegmann},
  {Steppa}, {Tavernier}, {Taylor}, {Tiziani}, {Trichard}, {Tsirou}, {van
  Eldik}, {van Rensburg}, {van Soelen}, {Veh}, {Vincent}, {Voisin}, {Wagner},
  {Wagner}, {Wierzcholska}, {Zanin}, {Zdziarski}, {Zech}, {Ziegler}, {Zorn},
  {{\.Z}ywucka}, {INTEGRAL Team}, {Savchenko}, {Ferrigno}, {Bazzano}, {Diehl},
  {Kuulkers}, {Laurent}, {Mereghetti}, {Natalucci}, {Panessa}, {Rodi},
  {Ubertini}, {Kanata}, Teams, {Morokuma}, {Ohta}, {Tanaka}, {Mori},
  {Yamanaka}, {Kawabata}, {Utsumi}, {Nakaoka}, {Kawabata}, {Nagashima},
  {Yoshida}, {Matsuoka}, {Itoh}, {Kapteyn Team}, {Keel}, {Liverpool Telescope
  Team}, {Copperwheat}, {Steele}, {Swift/NuSTAR Team}, {Cenko}, {Cowen},
  {DeLaunay}, {Evans}, {Fox}, {Keivani}, {Kennea}, {Marshall}, {Osborne},
  {Santander}, {Tohuvavohu}, {Turley}, {VERITAS Collaboration}, {Abeysekara},
  {Archer}, {Benbow}, {Bird}, {Brill}, {Brose}, {Buchovecky}, {Buckley},
  {Bugaev}, {Christiansen}, {Connolly}, {Cui}, {Daniel}, {Errando}, {Falcone},
  {Feng}, {Finley}, {Fortson}, {Furniss}, {Gueta}, {H{\"u}tten}, {Hervet},
  {Hughes}, {Humensky}, {Johnson}, {Kaaret}, {Kar}, {Kelley-Hoskins},
  {Kertzman}, {Kieda}, {Krause}, {Krennrich}, {Kumar}, {Lang}, {Lin}, {Maier},
  {McArthur}, {Moriarty}, {Mukherjee}, {Nieto}, {O'Brien}, {Ong}, {Otte},
  {Park}, {Petrashyk}, {Pohl}, {Popkow}, {Pueschel}, {Quinn}, {Ragan},
  {Reynolds}, {Richards}, {Roache}, {Rulten}, {Sadeh}, {Santander}, {Scott},
  {Sembroski}, {Shahinyan}, {Sushch}, {Tr{\'e}panier}, {Tyler}, {Vassiliev},
  {Wakely}, {Weinstein}, {Wells}, {Wilcox}, {Wilhelm}, {Williams}, {Zitzer},
  {VLA/B Team}, {Tetarenko}, {Kimball}, {Miller-Jones}, \&
  {Sivakoff}}]{icecube2018}
{IceCube Collaboration}, {Aartsen}, M.~G., {Ackermann}, M., {et~al.} 2018,
  Science, 361, eaat1378

\bibitem[{{Inskip} {et~al.}(2010){Inskip}, {Tadhunter}, {Morganti}, {Holt},
  {Ramos Almeida}, \& {Dicken}}]{Inskip:2010}
{Inskip}, K.~J., {Tadhunter}, C.~N., {Morganti}, R., {et~al.} 2010, \mnras,
  407, 1739

\bibitem[{{Isobe} {et~al.}(1986){Isobe}, {Feigelson}, \&
  {Nelson}}]{statistical_methods_II}
{Isobe}, T., {Feigelson}, E.~D., \& {Nelson}, P.~I. 1986, \apj, 306, 490

\bibitem[{{Jones} {et~al.}(2009){Jones}, {Read}, {Saunders}, {Colless},
  {Jarrett}, {Parker}, {Fairall}, {Mauch}, {Sadler}, {Watson}, {Burton},
  {Campbell}, {Cass}, {Croom}, {Dawe}, {Fiegert}, {Frankcombe}, {Hartley},
  {Huchra}, {James}, {Kirby}, {Lahav}, {Lucey}, {Mamon}, {Moore}, {Peterson},
  {Prior}, {Proust}, {Russell}, {Safouris}, {Wakamatsu}, {Westra}, \&
  {Williams}}]{6dFGS:2009}
{Jones}, D.~H., {Read}, M.~A., {Saunders}, W., {et~al.} 2009, \mnras, 399, 683

\bibitem[{{Jones} {et~al.}(2004){Jones}, {Saunders}, {Colless}, {Read},
  {Parker}, {Watson}, {Campbell}, {Burkey}, {Mauch}, {Moore}, {Hartley},
  {Cass}, {James}, {Russell}, {Fiegert}, {Dawe}, {Huchra}, {Jarrett}, {Lahav},
  {Lucey}, {Mamon}, {Proust}, {Sadler}, \& {Wakamatsu}}]{6dFGS:2004}
{Jones}, D.~H., {Saunders}, W., {Colless}, M., {et~al.} 2004, \mnras, 355, 747

\bibitem[{{Kalfountzou} {et~al.}(2012){Kalfountzou}, {Jarvis}, {Bonfield}, \&
  {Hardcastle}}]{Kalfountzou:2012}
{Kalfountzou}, E., {Jarvis}, M.~J., {Bonfield}, D.~G., \& {Hardcastle}, M.~J.
  2012, \mnras, 427, 2401

\bibitem[{{Kausch} {et~al.}(2015){Kausch}, {Noll}, {Smette}, {Kimeswenger},
  {Barden}, {Szyszka}, {Jones}, {Sana}, {Horst}, \& {Kerber}}]{Molecfit2}
{Kausch}, W., {Noll}, S., {Smette}, A., {et~al.} 2015, \aap, 576, A78

\bibitem[{{Koss} {et~al.}(2017){Koss}, {Trakhtenbrot}, {Ricci}, {Lamperti},
  {Oh}, {Berney}, {Schawinski}, {Balokovi{\'c}}, {Baronchelli}, {Crenshaw},
  {Fischer}, {Gehrels}, {Harrison}, {Hashimoto}, {Hogg}, {Ichikawa}, {Masetti},
  {Mushotzky}, {Sartori}, {Stern}, {Treister}, {Ueda}, {Veilleux}, \&
  {Winter}}]{Koss:2017}
{Koss}, M., {Trakhtenbrot}, B., {Ricci}, C., {et~al.} 2017, \apj, 850, 74

\bibitem[{{Kouch} {et~al.}(2024){Kouch}, {Lindfors}, {Hovatta}, {Liodakis},
  {Koljonen}, {Nilsson}, {Kiehlmann}, {Max-Moerbeck}, {Readhead}, {Reeves},
  {Pearson}, {Jormanainen}, {Ramazani}, \& {Graham}}]{Kouch:2024}
{Kouch}, P.~M., {Lindfors}, E., {Hovatta}, T., {et~al.} 2024, \aap, 690, A111

\bibitem[{{Kovalev} {et~al.}(2023){Kovalev}, {Plavin}, {Pushkarev}, \&
  {Troitsky}}]{Kovalev_2023}
{Kovalev}, Y.~Y., {Plavin}, A.~V., {Pushkarev}, A.~B., \& {Troitsky}, S.~V.
  2023, Galaxies, 11, 84

\bibitem[{{Krau{\ss}} {et~al.}(2018){Krau{\ss}}, {Deoskar}, {Baxter}, {Kadler},
  {Kreter}, {Langejahn}, {Mannheim}, {Polko}, {Wang}, \& {Wilms}}]{Krauss:2018}
{Krau{\ss}}, F., {Deoskar}, K., {Baxter}, C., {et~al.} 2018, \aap, 620, A174

\bibitem[{{Kun} {et~al.}(2022){Kun}, {Bartos}, {Becker Tjus}, {Biermann},
  {Franckowiak}, \& {Halzen}}]{Kun:2022}
{Kun}, E., {Bartos}, I., {Becker Tjus}, J., {et~al.} 2022, \apj, 934, 180

\bibitem[{{Labrie} {et~al.}(2019){Labrie}, {Anderson}, {C{\'a}rdenes},
  {Simpson}, \& {Turner}}]{DRAGONS}
{Labrie}, K., {Anderson}, K., {C{\'a}rdenes}, R., {Simpson}, C., \& {Turner},
  J. E.~H. 2019, in Astronomical Society of the Pacific Conference Series, Vol.
  523, Astronomical Data Analysis Software and Systems XXVII, ed. P.~J.
  {Teuben}, M.~W. {Pound}, B.~A. {Thomas}, \& E.~M. {Warner}, 321

\bibitem[{{Lainez} {et~al.}(2024){Lainez}, {Dominguez}, {Paliya},
  {Alvarez-Crespo}, {Ajello}, {Finke}, {Nievas-Rosillo}, {Contreras}, \&
  {Desai}}]{Lainez_2024}
{Lainez}, M., {Dominguez}, A., {Paliya}, V.~S., {et~al.} 2024, in 38th
  International Cosmic Ray Conference, 558

\bibitem[{{Landoni} {et~al.}(2020){Landoni}, {Falomo}, {Paiano}, \&
  {Treves}}]{ZBLLAC}
{Landoni}, M., {Falomo}, R., {Paiano}, S., \& {Treves}, A. 2020, \apjs, 250, 37

\bibitem[{Lang(1974)}]{Lang:1974}
Lang, K. 1974, Astrophysical Formulae: Volume I \& Volume II: Radiation, Gas
  Processes and High Energy Astrophysics/Space, Time, Matter and Cosmology
  (Springer Science \& Business Media)

\bibitem[{{Lavalley} {et~al.}(1992){Lavalley}, {Isobe}, \& {Feigelson}}]{ASURV}
{Lavalley}, M., {Isobe}, T., \& {Feigelson}, E. 1992, in Astronomical Society
  of the Pacific Conference Series, Vol.~25, Astronomical Data Analysis
  Software and Systems I, ed. D.~M. {Worrall}, C.~{Biemesderfer}, \&
  J.~{Barnes}, 245

\bibitem[{{Le{\'o}n-Tavares} {et~al.}(2011){Le{\'o}n-Tavares}, {Valtaoja},
  {Chavushyan}, {Tornikoski}, {A{\~n}orve}, {Nieppola}, \&
  {L{\"a}hteenm{\"a}ki}}]{LeonTavares:2011}
{Le{\'o}n-Tavares}, J., {Valtaoja}, E., {Chavushyan}, V.~H., {et~al.} 2011,
  \mnras, 411, 1127

\bibitem[{{Longair}(1994)}]{Longair:book}
{Longair}, M.~S. 1994, {High energy astrophysics. Vol.2: Stars, the galaxy and
  the interstellar medium}, Vol.~2 (Cambridge University Press)

\bibitem[{{Mannheim}(1993)}]{Mannheim:1993}
{Mannheim}, K. 1993, \aap, 269, 67

\bibitem[{{Maraschi} \& {Tavecchio}(2003)}]{Maraschi:2003}
{Maraschi}, L. \& {Tavecchio}, F. 2003, \apj, 593, 667

\bibitem[{{Massaro} {et~al.}(2014){Massaro}, {Maselli}, {Leto}, {Marchegiani},
  {Perri}, {Giommi}, \& {Piranomonte}}]{BZCat}
{Massaro}, E., {Maselli}, A., {Leto}, C., {et~al.} 2014, {Multifrequency
  Catalogue of Blazars - 5th Edition}

\bibitem[{{Massaro} {et~al.}(2011){Massaro}, {D'Abrusco}, {Ajello}, {Grindlay},
  \& {Smith}}]{Massaro:2011}
{Massaro}, F., {D'Abrusco}, R., {Ajello}, M., {Grindlay}, J.~E., \& {Smith},
  H.~A. 2011, \apjl, 740, L48

\bibitem[{{Massaro} {et~al.}(2015){Massaro}, {Landoni}, {D'Abrusco},
  {Milisavljevic}, {Paggi}, {Masetti}, {Smith}, \& {Tosti}}]{Massaro:2015}
{Massaro}, F., {Landoni}, M., {D'Abrusco}, R., {et~al.} 2015, \aap, 575, A124

\bibitem[{{McLure} \& {Dunlop}(2004)}]{McLure:2004}
{McLure}, R.~J. \& {Dunlop}, J.~S. 2004, in Multiwavelength AGN Surveys, ed.
  R.~{M{\'u}jica} \& R.~{Maiolino}, 389--392

\bibitem[{{Mingo} {et~al.}(2014){Mingo}, {Hardcastle}, {Croston}, {Dicken},
  {Evans}, {Morganti}, \& {Tadhunter}}]{Mingo:2014}
{Mingo}, B., {Hardcastle}, M.~J., {Croston}, J.~H., {et~al.} 2014, \mnras, 440,
  269

\bibitem[{{Morris} {et~al.}(1991){Morris}, {Stocke}, {Gioia}, {Schild},
  {Wolter}, {Maccacaro}, \& {della Ceca}}]{Morris:1991}
{Morris}, S.~L., {Stocke}, J.~T., {Gioia}, I.~M., {et~al.} 1991, \apj, 380, 49

\bibitem[{{M{\"u}cke} {et~al.}(2003){M{\"u}cke}, {Protheroe}, {Engel},
  {Rachen}, \& {Stanev}}]{Mucke:2003}
{M{\"u}cke}, A., {Protheroe}, R.~J., {Engel}, R., {Rachen}, J.~P., \& {Stanev},
  T. 2003, Astroparticle Physics, 18, 593

\bibitem[{{Murowinski} {et~al.}(1998){Murowinski}, {Bond}, {Crampton},
  {Davidge}, {Fletcher}, {Leckie}, {Morbey}, {Roberts}, {Saddlemyer},
  {Sebesta}, {Stilburn}, {Szeto}, {Allington-Smith}, {Content}, {Davies},
  {Dodsworth}, {Haynes}, {Robinson}, {Robertson}, {Webster}, {Lee}, {Beard},
  {Dickson}, {Kelly}, {Bennet}, {Ellis}, {Hastings}, \& {Williams}}]{GMOS:1998}
{Murowinski}, R.~G., {Bond}, T., {Crampton}, D., {et~al.} 1998, in Society of
  Photo-Optical Instrumentation Engineers (SPIE) Conference Series, Vol. 3355,
  Optical Astronomical Instrumentation, ed. S.~{D'Odorico}, 188--195

\bibitem[{{Nolan} {et~al.}(2012){Nolan}, {Abdo}, {Ackermann}, {Ajello},
  {Allafort}, {Antolini}, {Atwood}, {Axelsson}, {Baldini}, {Ballet},
  {Barbiellini}, {Bastieri}, {Bechtol}, {Belfiore}, {Bellazzini}, {Berenji},
  {Bignami}, {Blandford}, {Bloom}, {Bonamente}, {Bonnell}, {Borgland},
  {Bottacini}, {Bouvier}, {Brandt}, {Bregeon}, {Brigida}, {Bruel}, {Buehler},
  {Burnett}, {Buson}, {Caliandro}, {Cameron}, {Campana}, {Ca{\~n}adas},
  {Cannon}, {Caraveo}, {Casandjian}, {Cavazzuti}, {Ceccanti}, {Cecchi},
  {{\c{C}}elik}, {Charles}, {Chekhtman}, {Cheung}, {Chiang}, {Chipaux},
  {Ciprini}, {Claus}, {Cohen-Tanugi}, {Cominsky}, {Conrad}, {Corbet}, {Cutini},
  {D'Ammando}, {Davis}, {de Angelis}, {DeCesar}, {DeKlotz}, {De Luca}, {den
  Hartog}, {de Palma}, {Dermer}, {Digel}, {Silva}, {Drell}, {Drlica-Wagner},
  {Dubois}, {Dumora}, {Enoto}, {Escande}, {Fabiani}, {Falletti}, {Favuzzi},
  {Fegan}, {Ferrara}, {Focke}, {Fortin}, {Frailis}, {Fukazawa}, {Funk},
  {Fusco}, {Gargano}, {Gasparrini}, {Gehrels}, {Germani}, {Giebels},
  {Giglietto}, {Giommi}, {Giordano}, {Giroletti}, {Glanzman}, {Godfrey},
  {Grenier}, {Grondin}, {Grove}, {Guillemot}, {Guiriec}, {Gustafsson},
  {Hadasch}, {Hanabata}, {Harding}, {Hayashida}, {Hays}, {Hill}, {Horan},
  {Hou}, {Hughes}, {Iafrate}, {Itoh}, {J{\'o}hannesson}, {Johnson}, {Johnson},
  {Johnson}, {Johnson}, {Kamae}, {Katagiri}, {Kataoka}, {Katsuta}, {Kawai},
  {Kerr}, {Kn{\"o}dlseder}, {Kocevski}, {Kuss}, {Lande}, {Landriu},
  {Latronico}, {Lemoine-Goumard}, {Lionetto}, {Llena Garde}, {Longo},
  {Loparco}, {Lott}, {Lovellette}, {Lubrano}, {Madejski}, {Marelli}, {Massaro},
  {Mazziotta}, {McConville}, {McEnery}, {Mehault}, {Michelson}, {Minuti},
  {Mitthumsiri}, {Mizuno}, {Moiseev}, {Mongelli}, {Monte}, {Monzani},
  {Morselli}, {Moskalenko}, {Murgia}, {Nakamori}, {Naumann-Godo}, {Norris},
  {Nuss}, {Nymark}, {Ohno}, {Ohsugi}, {Okumura}, {Omodei}, {Orlando}, {Ormes},
  {Ozaki}, {Paneque}, {Panetta}, {Parent}, {Perkins}, {Pesce-Rollins},
  {Pierbattista}, {Pinchera}, {Piron}, {Pivato}, {Porter}, {Racusin},
  {Rain{\`o}}, {Rando}, {Razzano}, {Razzaque}, {Reimer}, {Reimer}, {Reposeur},
  {Ritz}, {Rochester}, {Romani}, {Roth}, {Rousseau}, {Ryde}, {Sadrozinski},
  {Salvetti}, {Sanchez}, {Saz Parkinson}, {Sbarra}, {Scargle}, {Schalk},
  {Sgr{\`o}}, {Shaw}, {Shrader}, {Siskind}, {Smith}, {Spandre}, {Spinelli},
  {Stephens}, {Strickman}, {Suson}, {Tajima}, {Takahashi}, {Takahashi},
  {Tanaka}, {Thayer}, {Thayer}, {Thompson}, {Tibaldo}, {Tibolla}, {Tinebra},
  {Tinivella}, {Torres}, {Tosti}, {Troja}, {Uchiyama}, {Vandenbroucke}, {Van
  Etten}, {Van Klaveren}, {Vasileiou}, {Vianello}, {Vitale}, {Waite},
  {Wallace}, {Wang}, {Werner}, {Winer}, {Wood}, {Wood}, {Wood}, {Yang}, \&
  {Zimmer}}]{2FGL}
{Nolan}, P.~L., {Abdo}, A.~A., {Ackermann}, M., {et~al.} 2012, \apjs, 199, 31

\bibitem[{{Osmer} {et~al.}(1994){Osmer}, {Porter}, \& {Green}}]{Osmer:1994}
{Osmer}, P.~S., {Porter}, A.~C., \& {Green}, R.~F. 1994, \apj, 436, 678

\bibitem[{{Osterbrock} \& Ferland(2006)}]{Osterbrock}
{Osterbrock}, D.~E. \& Ferland, G.~J. 2006, {Astrophysics Of Gas Nebulae and
  Active Galactic Nuclei} (University science books)

\bibitem[{{Padovani}(1992)}]{Padovani:1992}
{Padovani}, P. 1992, \mnras, 257, 404

\bibitem[{{Padovani} {et~al.}(2022{\natexlab{a}}){Padovani}, {Boccardi},
  {Falomo}, \& {Giommi}}]{Padovani_PKS:2022}
{Padovani}, P., {Boccardi}, B., {Falomo}, R., \& {Giommi}, P.
  2022{\natexlab{a}}, \mnras, 511, 4697

\bibitem[{{Padovani} {et~al.}(2022{\natexlab{b}}){Padovani}, {Giommi},
  {Falomo}, {Oikonomou}, {Petropoulou}, {Glauch}, {Resconi}, {Treves}, \&
  {Paiano}}]{Padovani:2022}
{Padovani}, P., {Giommi}, P., {Falomo}, R., {et~al.} 2022{\natexlab{b}},
  \mnras, 510, 2671

\bibitem[{{Padovani} {et~al.}(2012){Padovani}, {Giommi}, \&
  {Rau}}]{Padovani:2012}
{Padovani}, P., {Giommi}, P., \& {Rau}, A. 2012, \mnras, 422, L48

\bibitem[{{Padovani} {et~al.}(2019){Padovani}, {Oikonomou}, {Petropoulou},
  {Giommi}, \& {Resconi}}]{Padovani:2019}
{Padovani}, P., {Oikonomou}, F., {Petropoulou}, M., {Giommi}, P., \& {Resconi},
  E. 2019, \mnras, 484, L104

\bibitem[{{Padovani} {et~al.}(2016){Padovani}, {Resconi}, {Giommi}, {Arsioli},
  \& {Chang}}]{Padovani:2016}
{Padovani}, P., {Resconi}, E., {Giommi}, P., {Arsioli}, B., \& {Chang}, Y.~L.
  2016, \mnras, 457, 3582

\bibitem[{{Paiano} {et~al.}(2023){Paiano}, {Falomo}, {Treves}, {Padovani},
  {Giommi}, {Scarpa}, {Bisogni}, \& {Marini}}]{Paiano:2023}
{Paiano}, S., {Falomo}, R., {Treves}, A., {et~al.} 2023, \mnras, 521, 2270

\bibitem[{{Paiano} {et~al.}(2018){Paiano}, {Falomo}, {Treves}, \&
  {Scarpa}}]{Paiano:2018}
{Paiano}, S., {Falomo}, R., {Treves}, A., \& {Scarpa}, R. 2018, \apjl, 854, L32

\bibitem[{{Paiano} {et~al.}(2020){Paiano}, {Falomo}, {Treves}, \&
  {Scarpa}}]{Paiano:2020}
{Paiano}, S., {Falomo}, R., {Treves}, A., \& {Scarpa}, R. 2020, \mnras, 497, 94

\bibitem[{{Paiano} {et~al.}(2017){Paiano}, {Landoni}, {Falomo}, {Treves},
  {Scarpa}, \& {Righi}}]{Paiano:2017}
{Paiano}, S., {Landoni}, M., {Falomo}, R., {et~al.} 2017, \apj, 837, 144

\bibitem[{{Paliya} {et~al.}(2021){Paliya}, {Dom{\'\i}nguez}, {Ajello},
  {Olmo-Garc{\'\i}a}, \& {Hartmann}}]{Paliya:2021}
{Paliya}, V.~S., {Dom{\'\i}nguez}, A., {Ajello}, M., {Olmo-Garc{\'\i}a}, A., \&
  {Hartmann}, D. 2021, \apjs, 253, 46

\bibitem[{{Paliya} {et~al.}(2017){Paliya}, {Marcotulli}, {Ajello}, {Joshi},
  {Sahayanathan}, {Rao}, \& {Hartmann}}]{Paliya:2017}
{Paliya}, V.~S., {Marcotulli}, L., {Ajello}, M., {et~al.} 2017, \apj, 851, 33

\bibitem[{{Park} \& {Trippe}(2017)}]{Park:2017}
{Park}, J. \& {Trippe}, S. 2017, \apj, 834, 157

\bibitem[{{Pe{\~n}a-Herazo} {et~al.}(2021){Pe{\~n}a-Herazo}, {Massaro}, {Gu},
  {Paggi}, {Landoni}, {D'Abrusco}, {Ricci}, {Masetti}, \&
  {Chavushyan}}]{PenaHerazo_changinglook}
{Pe{\~n}a-Herazo}, H.~A., {Massaro}, F., {Gu}, M., {et~al.} 2021, \aj, 161, 196

\bibitem[{Pfeiffer(2022)}]{Pfeiffer_BT}
Pfeiffer, L. 2022, Master thesis, {Julius-Maximilians-Universität Würzburg}

\bibitem[{Pfeiffer(2024)}]{Pfeiffer_MT}
Pfeiffer, L. 2024, Master thesis, {Julius-Maximilians-Universität Würzburg}

\bibitem[{{Pfleiderer} \& {Krommidas}(1982)}]{statistical_methods_IV}
{Pfleiderer}, J. \& {Krommidas}, P. 1982, \mnras, 198, 281

\bibitem[{{Piranomonte} {et~al.}(2007){Piranomonte}, {Perri}, {Giommi},
  {Landt}, \& {Padovani}}]{Piranomonte:2007}
{Piranomonte}, S., {Perri}, M., {Giommi}, P., {Landt}, H., \& {Padovani}, P.
  2007, \aap, 470, 787

\bibitem[{{Plavin} {et~al.}(2020){Plavin}, {Kovalev}, {Kovalev}, \&
  {Troitsky}}]{Plavin:2020}
{Plavin}, A., {Kovalev}, Y.~Y., {Kovalev}, Y.~A., \& {Troitsky}, S. 2020, \apj,
  894, 101

\bibitem[{{Plavin} {et~al.}(2021){Plavin}, {Kovalev}, {Kovalev}, \&
  {Troitsky}}]{Plavin_2021}
{Plavin}, A.~V., {Kovalev}, Y.~Y., {Kovalev}, Y.~A., \& {Troitsky}, S.~V. 2021,
  \apj, 908, 157

\bibitem[{{Plavin} {et~al.}(2023){Plavin}, {Kovalev}, {Kovalev}, \&
  {Troitsky}}]{Plavin:2023}
{Plavin}, A.~V., {Kovalev}, Y.~Y., {Kovalev}, Y.~A., \& {Troitsky}, S.~V. 2023,
  \mnras, 523, 1799

\bibitem[{{Plotkin} {et~al.}(2011){Plotkin}, {Markoff}, {Trager}, \&
  {Anderson}}]{Plotkin:2011}
{Plotkin}, R.~M., {Markoff}, S., {Trager}, S.~C., \& {Anderson}, S.~F. 2011,
  \mnras, 413, 805

\bibitem[{{Prochaska} {et~al.}(2020){Prochaska}, {Hennawi}, {Westfall},
  {Cooke}, {Wang}, {Hsyu}, {Davies}, {Farina}, \& {Pelliccia}}]{Pypeit_1}
{Prochaska}, J., {Hennawi}, J., {Westfall}, K., {et~al.} 2020, The Journal of
  Open Source Software, 5, 2308

\bibitem[{{Punsly} \& {Zhang}(2011)}]{Punsly:2011}
{Punsly}, B. \& {Zhang}, S. 2011, \mnras, 412, L123

\bibitem[{{Ramos Almeida} {et~al.}(2013){Ramos Almeida}, {Bessiere},
  {Tadhunter}, {Inskip}, {Morganti}, {Dicken}, {Gonz{\'a}lez-Serrano}, \&
  {Holt}}]{Ramos:2013}
{Ramos Almeida}, C., {Bessiere}, P.~S., {Tadhunter}, C.~N., {et~al.} 2013,
  \mnras, 436, 997

\bibitem[{{Rector} {et~al.}(2000){Rector}, {Stocke}, {Perlman}, {Morris}, \&
  {Gioia}}]{Rector:2000}
{Rector}, T.~A., {Stocke}, J.~T., {Perlman}, E.~S., {Morris}, S.~L., \&
  {Gioia}, I.~M. 2000, \aj, 120, 1626

\bibitem[{{Ren} {et~al.}(2014){Ren}, {Rebassa-Mansergas}, {Luo}, {Zhao},
  {Xiang}, {Liu}, {Zhao}, {Jin}, \& {Zhang}}]{Ren:2014}
{Ren}, J.~J., {Rebassa-Mansergas}, A., {Luo}, A.~L., {et~al.} 2014, \aap, 570,
  A107

\bibitem[{{Ruan} {et~al.}(2014){Ruan}, {Anderson}, {Plotkin}, {Brandt},
  {Burnett}, {Myers}, \& {Schneider}}]{Ruan:2014}
{Ruan}, J.~J., {Anderson}, S.~F., {Plotkin}, R.~M., {et~al.} 2014, \apj, 797,
  19

\bibitem[{Sanchez~Zaballa {et~al.}(2025)Sanchez~Zaballa, Buson, Marchesi,
  Tombesi, Dauser, Wilms, \& Azzollini}]{SanchezZaballa:2025}
Sanchez~Zaballa, J.~M., Buson, S., Marchesi, S., {et~al.} 2025, The
  Astrophysical Journal, 988, 120

\bibitem[{{Santoro} {et~al.}(2020){Santoro}, {Tadhunter}, {Baron}, {Morganti},
  \& {Holt}}]{Santoro:2020}
{Santoro}, F., {Tadhunter}, C., {Baron}, D., {Morganti}, R., \& {Holt}, J.
  2020, \aap, 644, A54

\bibitem[{{Sbarrato} {et~al.}(2012){Sbarrato}, {Ghisellini}, {Maraschi}, \&
  {Colpi}}]{Sbarrato:2012}
{Sbarrato}, T., {Ghisellini}, G., {Maraschi}, L., \& {Colpi}, M. 2012, \mnras,
  421, 1764

\bibitem[{{Sbarrato} {et~al.}(2014){Sbarrato}, {Padovani}, \&
  {Ghisellini}}]{Sbarrato:2014}
{Sbarrato}, T., {Padovani}, P., \& {Ghisellini}, G. 2014, \mnras, 445, 81

\bibitem[{{Sbarufatti} {et~al.}(2009){Sbarufatti}, {Ciprini}, {Kotilainen},
  {Decarli}, {Treves}, {Veronesi}, \& {Falomo}}]{Sbaruffati_2009}
{Sbarufatti}, B., {Ciprini}, S., {Kotilainen}, J., {et~al.} 2009, \aj, 137, 337

\bibitem[{{Shakura} \& {Sunyaev}(1973)}]{Shakura:1973}
{Shakura}, N.~I. \& {Sunyaev}, R.~A. 1973, \aap, 24, 337

\bibitem[{{Shaw} {et~al.}(2012){Shaw}, {Romani}, {Cotter}, {Healey},
  {Michelson}, {Readhead}, {Richards}, {Max-Moerbeck}, {King}, \&
  {Potter}}]{Shaw:2012}
{Shaw}, M.~S., {Romani}, R.~W., {Cotter}, G., {et~al.} 2012, \apj, 748, 49

\bibitem[{{Shaw} {et~al.}(2013){Shaw}, {Romani}, {Cotter}, {Healey},
  {Michelson}, {Readhead}, {Richards}, {Max-Moerbeck}, {King}, \&
  {Potter}}]{Shaw:2013}
{Shaw}, M.~S., {Romani}, R.~W., {Cotter}, G., {et~al.} 2013, \apj, 764, 135

\bibitem[{{Shen} \& {Liu}(2012)}]{Shen:2012}
{Shen}, Y. \& {Liu}, X. 2012, \apj, 753, 125

\bibitem[{{Shen} {et~al.}(2011){Shen}, {Richards}, {Strauss}, {Hall},
  {Schneider}, {Snedden}, {Bizyaev}, {Brewington}, {Malanushenko},
  {Malanushenko}, {Oravetz}, {Pan}, \& {Simmons}}]{Shen:2011}
{Shen}, Y., {Richards}, G.~T., {Strauss}, M.~A., {et~al.} 2011, \apjs, 194, 45

\bibitem[{{Smette} {et~al.}(2015){Smette}, {Sana}, {Noll}, {Horst}, {Kausch},
  {Kimeswenger}, {Barden}, {Szyszka}, {Jones}, {Gallenne}, {Vinther},
  {Ballester}, \& {Taylor}}]{Molecfit1}
{Smette}, A., {Sana}, H., {Noll}, S., {et~al.} 2015, \aap, 576, A77

\bibitem[{{Smith} {et~al.}(1998){Smith}, {Boyle}, {Shanks}, {Croom}, {Miller},
  \& {Read}}]{2QZ:1998}
{Smith}, R.~J., {Boyle}, B.~J., {Shanks}, T., {et~al.} 1998, in New Horizons
  from Multi-Wavelength Sky Surveys, ed. B.~J. {McLean}, D.~A. {Golombek},
  J.~J.~E. {Hayes}, \& H.~E. {Payne}, Vol. 179, 348

\bibitem[{{Song} {et~al.}(2012){Song}, {Luo}, {Comte}, {Bai}, {Zhang}, {Du},
  {Zhang}, {Chen}, {Zuo}, \& {Zhao}}]{Song:2012}
{Song}, Y.-H., {Luo}, A.~L., {Comte}, G., {et~al.} 2012, Research in Astronomy
  and Astrophysics, 12, 453

\bibitem[{{Stathopoulos} {et~al.}(2022){Stathopoulos}, {Petropoulou}, {Giommi},
  {Vasilopoulos}, {Padovani}, \& {Mastichiadis}}]{Stathopoulos:2022}
{Stathopoulos}, S.~I., {Petropoulou}, M., {Giommi}, P., {et~al.} 2022, \mnras,
  510, 4063

\bibitem[{{Titov} {et~al.}(2013){Titov}, {Stanford}, {Johnston}, {Pursimo},
  {Hunstead}, {Jauncey}, {Maslennikov}, \& {Boldycheva}}]{Titov:2013}
{Titov}, O., {Stanford}, L.~M., {Johnston}, H.~M., {et~al.} 2013, \aj, 146, 10

\bibitem[{{Tody}(1986)}]{Tody:1986}
{Tody}, D. 1986, in Society of Photo-Optical Instrumentation Engineers (SPIE)
  Conference Series, Vol. 627, Instrumentation in astronomy VI, ed. D.~L.
  {Crawford}, 733

\bibitem[{{Tody}(1993)}]{Tody:1993}
{Tody}, D. 1993, in Astronomical Society of the Pacific Conference Series,
  Vol.~52, Astronomical Data Analysis Software and Systems II, ed. R.~J.
  {Hanisch}, R.~J.~V. {Brissenden}, \& J.~{Barnes}, 173

\bibitem[{{Torrealba} {et~al.}(2012){Torrealba}, {Chavushyan},
  {Cruz-Gonz{\'a}lez}, {Arshakian}, {Bertone}, \&
  {Rosa-Gonz{\'a}lez}}]{Torrealba_2012}
{Torrealba}, J., {Chavushyan}, V., {Cruz-Gonz{\'a}lez}, I., {et~al.} 2012,
  \rmxaa, 48, 9

\bibitem[{{Truebenbach} \& {Darling}(2017)}]{Truebenbach:2017}
{Truebenbach}, A.~E. \& {Darling}, J. 2017, \apjs, 233, 3

\bibitem[{{Urry}(2004)}]{Urry:2004}
{Urry}, C. 2004, in Astronomical Society of the Pacific Conference Series, Vol.
  311, AGN Physics with the Sloan Digital Sky Survey, ed. G.~T. {Richards} \&
  P.~B. {Hall}, 49

\bibitem[{{Urry} \& {Padovani}(1995)}]{Urry:1995}
{Urry}, C.~M. \& {Padovani}, P. 1995, \pasp, 107, 803

\bibitem[{{Vermeulen} {et~al.}(1995){Vermeulen}, {Ogle}, {Tran}, {Browne},
  {Cohen}, {Readhead}, {Taylor}, \& {Goodrich}}]{Vermeulen:1995}
{Vermeulen}, R.~C., {Ogle}, P.~M., {Tran}, H.~D., {et~al.} 1995, \apjl, 452, L5

\bibitem[{{Vernet} {et~al.}(2011){Vernet}, {Dekker}, {D'Odorico}, {Kaper},
  {Kjaergaard}, {Hammer}, {Randich}, {Zerbi}, {Groot}, {Hjorth}, {Guinouard},
  {Navarro}, {Adolfse}, {Albers}, {Amans}, {Andersen}, {Andersen}, {Binetruy},
  {Bristow}, {Castillo}, {Chemla}, {Christensen}, {Conconi}, {Conzelmann},
  {Dam}, {de Caprio}, {de Ugarte Postigo}, {Delabre}, {di Marcantonio},
  {Downing}, {Elswijk}, {Finger}, {Fischer}, {Flores}, {Fran{\c{c}}ois},
  {Goldoni}, {Guglielmi}, {Haigron}, {Hanenburg}, {Hendriks}, {Horrobin},
  {Horville}, {Jessen}, {Kerber}, {Kern}, {Kiekebusch}, {Kleszcz}, {Klougart},
  {Kragt}, {Larsen}, {Lizon}, {Lucuix}, {Mainieri}, {Manuputy}, {Martayan},
  {Mason}, {Mazzoleni}, {Michaelsen}, {Modigliani}, {Moehler}, {M{\o}ller},
  {Norup S{\o}rensen}, {N{\o}rregaard}, {P{\'e}roux}, {Patat}, {Pena}, {Pragt},
  {Reinero}, {Rigal}, {Riva}, {Roelfsema}, {Royer}, {Sacco}, {Santin},
  {Schoenmaker}, {Spano}, {Sweers}, {Ter Horst}, {Tintori}, {Tromp}, {van
  Dael}, {van der Vliet}, {Venema}, {Vidali}, {Vinther}, {Vola}, {Winters},
  {Wistisen}, {Wulterkens}, \& {Zacchei}}]{XShooter}
{Vernet}, J., {Dekker}, H., {D'Odorico}, S., {et~al.} 2011, \aap, 536, A105

\bibitem[{{Vestergaard} \& {Osmer}(2009)}]{Vestergaard:2009}
{Vestergaard}, M. \& {Osmer}, P.~S. 2009, \apj, 699, 800

\bibitem[{{Vestergaard} \& {Peterson}(2006)}]{Vestergaard:2006}
{Vestergaard}, M. \& {Peterson}, B.~M. 2006, \apj, 641, 689

\bibitem[{{Wang} {et~al.}(2002){Wang}, {Staubert}, \& {Ho}}]{Wang:2002}
{Wang}, J.-M., {Staubert}, R., \& {Ho}, L.~C. 2002, \apj, 579, 554

\bibitem[{{White} {et~al.}(1997){White}, {Becker}, {Helfand}, \&
  {Gregg}}]{FIRST}
{White}, R.~L., {Becker}, R.~H., {Helfand}, D.~J., \& {Gregg}, M.~D. 1997,
  \apj, 475, 479

\bibitem[{{Xiao} {et~al.}(2022){Xiao}, {Fan}, {Ouyang}, {Hu}, {Chen}, {Fu}, \&
  {Zhang}}]{Xiao:2022}
{Xiao}, H., {Fan}, J., {Ouyang}, Z., {et~al.} 2022, \apj, 936, 146

\bibitem[{{Xie} \& {Li}(2012)}]{Xie:2012}
{Xie}, Z.~H. \& {Li}, Z. 2012, \pasj, 64, 33

\bibitem[{{Xiong} {et~al.}(2015){Xiong}, {Zhang}, {Bai}, \&
  {Zhang}}]{Xiong:2015}
{Xiong}, D., {Zhang}, X., {Bai}, J., \& {Zhang}, H. 2015, \mnras, 450, 3568

\bibitem[{{Zhao} {et~al.}(2012){Zhao}, {Zhao}, {Chu}, {Jing}, \&
  {Deng}}]{LAMOST:2012}
{Zhao}, G., {Zhao}, Y., {Chu}, Y., {Jing}, Y., \& {Deng}, L. 2012, arXiv
  e-prints, arXiv:1206.3569

\bibitem[{{Zhou} {et~al.}(2021){Zhou}, {Kamionkowski}, \& {Liang}}]{Zhou:2021}
{Zhou}, B., {Kamionkowski}, M., \& {Liang}, Y.-f. 2021, \prd, 103, 123018

\end{thebibliography}

\newpage

\begin{appendix}
\section{The target sample properties}
\label{append: tables}

Table \ref{tab: associations} lists the 5BZCat sources proposed as associated with neutrino hotspots. The first four columns report the name, coordinates, and L value of the neutrino hotspots (for more details see \origin, \originII). The following two columns list the associated $5$BZCat blazars with the corresponding redshift information. For the redshift, we report the values collected by inspecting the most recent literature and cross-checking with the optical spectrum available from this work. The seventh column reports the blazar \emph{Fermi}-LAT counterpart from the Third Data Release of the Fourth \emph{Fermi} Large Area Telescope (LAT) AGN Catalog \citep[$4$LAC-DR$3$, ][]{4LAC-DR3}. Finally, the last column lists the reference to the optical spectrum used for the object. Eight blazars of our sample of interest have already been included in other works as promising candidates for the production of high-energy IceCube neutrinos \citep[][see Appendix \ref{sec:append dataset} for more details on the individual objects]{icecube2018, Krauss:2018, Plavin:2020, Hovatta:2021, Padovani_PKS:2022, Stathopoulos:2022, Plavin:2023}. They are highlighted with the $\diamond$ symbol in Table \ref{tab: associations}. 

Table \ref{table:results} lists the physical properties of the candidate neutrino emitter-blazars, estimated as explained in Section \ref{sec: analysis}.

Table \ref{table: AD test} summarizes the main findings of the statistical analysis (Section \ref{subsec: statistical analysis}). The first column reports the analyzed quantity, while the second, third, and fourth columns detail the tested samples, including the number of measurements and the number of upper limits either excluded or included in the analysis. The fifth and sixth columns present the resulting pre-trial pvalues from the A-D and Peto logrank tests, respectively. The $\ast$ symbol indicates the statistically significant ($\gtrsim3\sigma$) post-trial pvalues (see also Appendix \ref{sec: append survival analysis}).

\bgroup
\def\arraystretch{0.93}%
\begin{table*}[!h]
\caption{List of candidate PeVatron blazars--neutrino hotspot associations from \origin\, and \originII.
}
\label{tab: associations}
\centering
\begin{threeparttable}[t]
\resizebox{0.8\textwidth}{!}{\begin{tabular}{ccclcccc}        
\hline
IceCube hotspot & $\alpha_{\rm hs} \left[ ^{\circ} \right]$ & $\delta_{\rm hs} \left[ ^{\circ} \right]$ & L & 5BZCat blazar & Redshift & \emph{Fermi}-LAT counterpart & Optical spectroscopy reference \\
\hline
IC~J2242$-$0540  & $340.75$  &  $-5.68$  &  $4.012$  & 5BZB~J2243$-$0609            &  $0.30$$^a$  &  $-$                 &   VLT X$-$Shooter $^{A}$            \\
IC~J0359$-$0746  & $59.85$   &  $-7.78$  &  $5.565$  & 5BZQ~J0357$-$0751            &  $1.05$      &  $-$                 &   $6$dFGS-DR$3$                     \\
IC~J0256$-$2146  & $44.12$   &  $-21.78$ &  $4.873$  & 5BZQ~J0256$-$2137            &  $1.47$      &  $-$                 &   VLT X$-$Shooter $^{A}$            \\
IC~J2037$-$2216  & $309.38$  &  $-22.27$ &  $4.664$  & 5BZQ~J2036$-$2146            &  $2.299$     &  $-$                 &   Gemini GMOS$-$S $^{A}$            \\
IC~J0630$-$2353  & $97.56$   &  $-23.89$ &  $4.420$  & 5BZB~J0630$-$2406            &  $1.239<{\rm z}<1.33^b$    &  4FGL~J0630.9$-$2406 &   \citet{Shaw:2013}                 \\
IC~J0359$-$2551  & $59.94$   &  $-25.86$ &  $4.356$  & 5BZB~J0359$-$2615            &  $1.47^c$    &  4FGL~J0359.4$-$2616 &   Gemini GMOS$-$S $^{A}$            \\
IC~J0145$-$3154  & $26.28$   &  $-31.91$ &  $4.937$  & 5BZU~J0143$-$3200            &  $0.375$     &  4FGL~J0143.5$-$3156 &   $2$QZ                             \\
IC~J2001$-$3314  & $300.41$  &  $-33.24$ &  $4.905$  & 5BZQ~J2003$-$3251            &  $3.773$     &  $-$                 &   \citet{Paliya:2017}               \\
IC~J2304$-$3614  & $346.03$  &  $-36.24$ &  $4.025$  & 5BZQ~J2304$-$3625            &  $0.962$     &  $-$                 &   Gemini GMOS$-$S $^{A}$            \\
IC~J1818$-$6315  & $274.50$  &  $-63.26$ &  $4.030$  & 5BZU~J1819$-$6345            &  $0.063$     &  $-$                 &   \citet{Santoro:2020}              \\
\hline
IC~J0241$-$0214  &	$40.43$	 &   $-2.24$ &  $3.827$  & 5BZQ~J0239$-$0234            &  $1.11413$  &  $-$                 &   SDSS$-$DR16                       \\
IC~J0243$+$0009  &	$40.78$	 &   $0.15$	 &  $6.752$  & 5BZB~J0243$+$0046            &  $0.409$    &  4FGL~J0242.9$+$0045 &   SDSS$-$DR16                       \\
IC~J0146$+$0027  &	$26.72$	 &   $0.45$	 &  $3.020$  & 5BZB~J0148$+$0129            &  $0.94206$  &  4FGL~J0148.6$+$0127 &   SDSS$-$DR17                       \\
IC~J0310$+$0205  &	$47.64$	 &   $2.09$	 &  $3.028$  & 5BZQ~J0312$+$0133            &  $0.664$    &  4FGL~J0312.8$+$0134 &   \citet{Paliya:2017}               \\
IC~J0419$+$0241  &	$64.86$	 &   $2.69$	 &  $3.155$  & 5BZQ~J0422$+$0219            &  $2.277$    &  4FGL~J0422.8$+$0225 &   \citet{Paliya:2017}               \\
IC~J0509$+$0531  &	$77.34$	 &   $5.53$	 &  $4.125$  & 5BZB~J0509$+$0541 $\diamond$ &  $0.3365$   &  4FGL~J0509.4$+$0542 &   \citet{Padovani:2019}             \\
IC~J0359$+$0549  &	$59.77$	 &   $5.83$	 &  $3.413$  & 5BZQ~J0400$+$0550            &  $0.758$    &  $-$                 &   LAMOST-DR$5$                      \\
IC~J0316$+$0625  &	$49.22$	 &   $6.43$	 &  $3.334$  & 5BZB~J0314$+$0619            &  $0.62$     &  4FGL~J0314.3$+$0620 &   LAMOST-DR$5$                      \\
IC~J2241$+$0720  &	$340.31$ &   $7.33$	 &  $3.265$  & 5BZQ~J2238$+$0724 $\diamond$ &  $1.011$    &  $-$                 &   GTC OSIRIS $^{A}$                 \\
IC~J0505$+$1247  &	$76.29$	 &   $12.79$ &  $5.192$	 & 5BZB~J0502$+$1338 $\diamond$ &  $0.35^d$   &  4FGL~J0502.5$+$1340 &   APO DIS                           \\
IC~J1420$+$1305  &	$215.16$ &   $13.09$ &  $3.444$	 & 5BZQ~J1420$+$1205            &  $4.026$    &  $-$                 &   \citet{Shen:2011}                 \\
IC~J0031$+$1524  &	$7.91$	 &   $15.40$ &  $3.667$	 & 5BZB~J0035$+$1515            &  $0.57^e$   &  4FGL~J0035.2$+$1514 &   GTC OSIRIS $^{A}$                 \\
IC~J1138$+$1822  &	$174.73$ &   $18.37$ &  $3.379$	 & 5BZQ~J1143$+$1843            &  $0.37435$  &  $-$                 &   \citet{Shen:2011}                 \\
IC~J1550$+$1840  &	$237.66$ &   $18.68$ &  $3.291$	 & 5BZB~J1546$+$1817 $\diamond$ & $3.61156^f$ &  4FGL~J1546.5$+$1816 &   SDSS$-$DR17                       \\
IC~J1151$+$2309  &	$177.89$ &   $23.16$ &  $4.317$	 & 5BZB~J1150$+$2417            & $0.209^d$   &  4FGL~J1150.4$+$2418 &   \citet{Sbaruffati_2009}           \\
IC~J1427$+$2348  &	$216.91$ &   $23.81$ &  $4.179$	 & 5BZB~J1427$+$2348 $\diamond$ & $0.604$     &  4FGL~J1427.0$+$2348 &   \citet{Padovani_PKS:2022}         \\
IC~J0222$+$2407  &	$35.51$  &	$24.13$	 &  $3.107$	 & 5BZU~J0220$+$2509            & $0.48$      &  $-$                 &   GTC OSIRIS $^{A}$                 \\
IC~J0123$+$2536  &	$20.92$  &	$25.61$	 &  $3.772$	 & 5BZQ~J0122$+$2502            & $2.025$     &  $-$                 &   SDSS$-$DR17                       \\
IC~J1124$+$2626  &	$171.21$ &	$26.44$	 &  $3.440$	 & 5BZQ~J1125$+$2610 $\diamond$ & $2.35005$   &  4FGL~J1125.2$+$2557 &   SDSS$-$DR17                       \\
IC~J1044$+$2716  &	$161.19$ &	$27.28$	 &  $3.175$	 & 5BZQ~J1047$+$2635            & $2.563636$  &  $-$                 &   SDSS$-$DR17                       \\
IC~J0737$+$2817  &	$114.43$ &	$28.29$	 &  $3.352$	 & 5BZB~J0737$+$2846            & $0.272$     &  $-$                 &   \citet{Xiong:2015}                \\
IC~J0134$+$3102  &	$23.73$	 &	$31.04$	 &  $3.167$	 & 5BZQ~J0137$+$3122            & $1.73063$   &  $-$                 &   SDSS$-$DR17                       \\
IC~J1712$+$3237  &	$258.05$ &	$32.62$	 &  $3.088$	 & 5BZU~J1706$+$3214            & $1.06929$   &  $-$                 &   \citet{Shen:2011}                 \\
IC~J0205$+$3435  &	$31.46$	 &	$34.59$	 &	$3.133$	 & 5BZB~J0208$+$3523            & $0.318$     &  4FGL~J0208.6$+$3523 &   \citet{Morris:1991}               \\
IC~J0959$+$3710  &	$149.94$ &	$37.17$	 &	$3.166$	 & 5BZB~J1004$+$3752            & $0.44$      &  $-$                 &   \citet{Xiong:2015}                \\
IC~J1535$+$3806  &	$233.79$ &	$38.11$	 &	$3.073$	 & 5BZU~J1536$+$3742            & $0.679$     &  $-$                 &   SDSS$-$DR17                       \\
IC~J1619$+$3840  &	$244.86$ &	$38.68$	 &	$3.734$	 & 5BZQ~J1617$+$3801            & $1.60867$   &  $-$                 &   \citet{Shen:2011}                 \\
IC~J1209$+$3938  &	$182.46$ &	$39.64$	 &	$3.915$	 & 5BZB~J1210$+$3929 $\diamond$ & $0.616$     &  4FGL~J1210.3$+$3928 &   \citet{Rector:2000}               \\
IC~J1244$+$4037  &	$191.07$ &	$40.62$	 &	$3.644$	 & 5BZQ~J1243$+$4043            & $1.518$     &  $-$                 &   \citet{Titov:2013}                \\
IC~J1200$+$4212  &	$180.18$ &	$42.21$	 &	$4.791$	 & 5BZG~J1156$+$4238            & $0.17162$   &  $-$                 &   \citet{Rector:2000}               \\
IC~J1126$+$4248  &	$171.57$ &	$42.81$	 &	$3.332$	 & 5BZB~J1122$+$4316            & $0.43516$   &  $-$                 &   SDSS$-$DR17                       \\
IC~J1741$+$4336  &	$265.45$ &	$43.61$	 &	$3.133$	 & 5BZQ~J1740$+$4348            & $2.246$     &  $-$                 &   GTC OSIRIS $^{A}$                 \\
IC~J2249$+$4436  &	$342.33$ &	$44.60$	 &	$3.214$	 & 5BZB~J2247$+$4413            & $>0.31^g$   &  4FGL~J2247.8$+$4413 &   GTC OSIRIS $^{A}$                 \\
IC~J1522$+$4558  &	$230.74$ &	$45.98$	 &	$3.490$	 & 5BZB~J1523$+$4606            & $--^g$      &  $-$                 &   SDSS$-$DR$13$                     \\
IC~J0805$+$5005  &	$121.33$ &	$50.09$	 &	$4.005$	 & 5BZQ~J0808$+$4950            & $1.4344$    &  4FGL~J0808.5$+$4950 &   \citet{Shen:2011}                 \\
IC~J1327$+$5028  &	$201.86$ &	$50.48$	 &	$3.638$	 & 5BZQ~J1327$+$5008            & $1.01191^i$ &  4FGL~1328.9$+$5019  &   \citet{Shen:2011}                 \\
IC~J1643$+$5052  &	$250.93$ &	$50.87$	 &	$3.397$	 & 5BZU~J1647$+$4950            & $0.0475$    &  4FGL~J1647.5$+$4950 &   \citet{Paliya:2021}               \\
IC~J0512$+$5701  &	$78.12$	 &	$57.02$	 &	$3.671$	 & 5BZQ~J0514$+$5602             & $2.19$      &  $-$                 &   GTC OSIRIS $^{A}$                 \\
IC~J0546$+$5843  &	$86.54$	 &	$58.73$	 &	$3.234$	 & 5BZB~J0540$+$5823 $\diamond$  & $>0.1$      &  4FGL~J0540.5$+$5823 &   \citet{Paiano:2020}               \\
IC~J1855$+$6530  &	$283.87$ &	$65.51$	 &	$3.041$	 & 5BZB~J1848$+$6537             & $0.364^l$   &  4FGL~J1848.5$+$6537 &   GTC OSIRIS $^{A}$                 \\
IC~J1937$+$7330  &	$294.50$ &	$73.50$	 &	$3.548$	 & 5BZQ~J1927$+$7358             & $0.302$     &  $-$                 &   \citet{Park:2017}                 \\
IC~J1355$+$7743  &	$208.88$ &	$77.73$	 &	$3.052$	 & 5BZQ~J1357$+$7643             & $1.585$     &  4FGL~J1358.1$+$7642 &   \citet{Shaw:2012}                 \\
\hline
\end{tabular}
}
\tablefoot{The first four columns report the information about the neutrino hotspots: the name, the equatorial coordinates (J$2000$) and the $L$ value. The following three columns report the following information for the candidate 5BZCat blazar counterpart: the redshift, the blazar \emph{Fermi}-LAT counterpart in the $4$LAC-DR$3$ catalog, and the optical spectrum used in this work (see also Appendices \ref{sec:append dataset} and \ref{sec:spectra}). We collected the redshifts by consulting the most recent literature, and cross-checking the literature value with our analysis of the optical spectrum, when available. Sources proposed as candidate neutrino emitters in other works are highlighted with a $\diamond$ symbol.}
% \begin{tablenotes}
%     \begin{minipage}{0.84\linewidth}
%     \vspace{0.1cm}
% 	\small 
% 	\item
\small{
    \tablefoottext{a}{Redshift from \citet{Crates:2007}.}
    \tablefoottext{b}{Lower limit on the redshift from \citet{Shaw:2013}, upper limit from \citet{Lainez_2024}.}
    \tablefoottext{c}{Redshift from \citet{Drinkwater:1997}.}
    \tablefoottext{d}{Redshift from \citet{Truebenbach:2017}.}
    \tablefoottext{e}{Estimated in this work.}
    \tablefoottext{f}{Redshift from SDSS-DR17.}
    \tablefoottext{g}{Redshift from \citet{Shaw:2013}.}
    \tablefoottext{h}{Tentative estimation ${\rm z} = 1.012$ in \citet{Dwelly:2017}, based on a featureless SDSS$-$DR$13$ spectrum.}
    \tablefoottext{i}{Redshift from \citet{Hewett:2010}.}
    \tablefoottext{l}{Redshift from \citet{Piranomonte:2007}.}
    \tablefoottext{A}{From this work.}}
%     \end{minipage}
% \end{tablenotes}
\end{threeparttable}
\end{table*}
\egroup

\begin{table*}[h]
\centering
\caption{Physical properties of the candidate PeVatron blazar sample estimated in this work.}
\label{table:results}
\resizebox{0.8\textwidth}{!}{\begin{tabular}{ccccccccc}
\toprule
 Blazar & $\rm{M}_{\rm{BH}} \left[ \rm{M}_{\odot} \right]$   & $\accretion$   & $\gratio$   & $\Pradio \left[ \rm{W}\cdot\rm{Hz}^{-1} \right]$   & $\rm{r}_{\rm{BLR}} \left[\rm{cm}\right]$   & $\rm{r}_{\rm{DT}} \left[\rm{cm}\right]$   & $\rm{L}_{\rm{bol}} \left[\rm{erg}\cdot\rm{s}^{-1}\right]$   \\
\midrule
 5BZB~J2243$-$0609 & $1.72\times10^{7}$ & $3.56\times10^{-4}$ & $<5.48\times10^{-2}$ & $3.22\times10^{25}$ & $8.41\times10^{15}$  & $2.10\times10^{17}$ & $1.34\times10^{45}$ \\
 5BZQ~J0357$-$0751 & $3.03\times10^{10}$  & $2.10\times10^{-3}$ & $<6.43\times10^{-4}$ & $2.16\times10^{27}$ & $8.57\times10^{17}$  & $2.14\times10^{19}$ & $1.33\times10^{47}$ \\
 5BZQ~J0256$-$2137 & $2.04\times10^{9}$ & $3.65\times10^{-3}$ & $<2.21\times10^{-2}$ & $5.17\times10^{27}$ & $2.93\times10^{17}$  & $7.32\times10^{18}$ & $2.11\times10^{46}$ \\
 5BZQ~J2036$-$2146 & $<3.72\times10^{9}$  & $<5.71\times10^{-4}$   & $<1.49\times10^{-2}$  & $8.86\times10^{27}$ & $<1.56\times10^{17}$ & $<3.91\times10^{18}$   & $<1.32\times10^{46}$   \\
 5BZB~J0630$-$2406 & $<3.79\times10^{9}$  & $<5.79\times10^{-4}$   & $\sim1.02\times10^{0}$   & $9.64\times10^{26}$ & $<2.65\times10^{17}$ & $<6.62\times10^{18}$   & $<2.05\times10^{46}$   \\
 5BZB~J0359$-$2615 & $<3.65\times10^{8}$ & $<1.38\times10^{-4}$ & $<1.22\times10^{0}$   & $1.12\times10^{28}$ & $<2.41\times10^{16}$  & $<6.02\times10^{17}$ & $<5.45\times10^{45}$ \\
 5BZU~J0143$-$3200 & $3.36\times10^{10}$  & $2.47\times10^{-3}$ & $1.88\times10^{-4}$  & $3.79\times10^{25}$ & $9.78\times10^{17}$  & $2.44\times10^{19}$ & $9.25\times10^{46}$ \\
 5BZQ~J2003$-$3251 & $3.98\times10^{9}$ & $1.07\times10^{-1}$ & $<1.13\times10^{-1}$ & $6.27\times10^{28}$ & $2.22\times10^{18}$  & $5.55\times10^{19}$ & $9.86\times10^{47}$ \\
 5BZQ~J2304$-$3625 & $<5.81\times10^{7}$ & $<4.76\times10^{-5}$ & $<2.68\times10^{-1}$ & $1.25\times10^{27}$ & $<5.65\times10^{15}$  & $<1.41\times10^{17}$ & $<1.06\times10^{45}$ \\
 5BZU~J1819$-$6345 & $6.80\times10^{8}$ & $1.87\times10^{-5}$ & $<1.82\times10^{-5}$  & $1.91\times10^{26}$ & $1.21\times10^{16}$  & $3.04\times10^{17}$ & $1.09\times10^{44}$ \\
 \midrule
 5BZQ~J0239$-$0234 & $3.04\times10^{9}$ & $1.88\times10^{-3}$ & $<7.46\times10^{-3}$ & $2.15\times10^{27}$ & $2.57\times10^{17}$  & $6.41\times10^{18}$ & $2.58\times10^{46}$ \\
 5BZB~J0243$+$0046 & $<9.69\times10^{7}$  & $<7.04\times10^{-5}$   & $<1.13\times10^{-1}$  & $1.84\times10^{25}$ & $<8.88\times10^{15}$ & $<2.20\times10^{17}$   & $<1.99\times10^{45}$   \\
 5BZB~J0148$+$0129 & $<7.96\times10^{8}$ & $<2.47\times10^{-4}$ & $<1.85\times10^{-1}$  & $4.11\times10^{26}$ & $<4.76\times10^{16}$  & $<1.19\times10^{18}$ & $<5.79\times10^{45}$ \\
 5BZQ~J0312$+$0133 & $5.37\times10^{8}$ & $4.05\times10^{-3}$ & $6.03\times10^{-1}$  & $9.00\times10^{26}$ & $1.57\times10^{17}$  & $3.96\times10^{18}$ & $5.02\times10^{45}$ \\
 5BZQ~J0422$+$0219 & $7.94\times10^{8}$ & $4.34\times10^{-2}$ & $1.32\times10^{1}$   & $2.81\times10^{28}$ & $6.29\times10^{17}$  & $1.58\times10^{19}$ & $7.96\times10^{46}$ \\
 5BZB~J0509$+$0541 & $3.00\times10^{8}$ & $7.50\times10^{-4}$ & $7.50\times10^{-1}$  & $2.07\times10^{26}$ & $5.48\times10^{16}$  & $1.37\times10^{18}$ & $1.70\times10^{45}$ \\
 5BZQ~J0400$+$0550 & $2.83\times10^{8}$ & $1.51\times10^{-2}$ & $<3.07\times10^{-2}$ & $1.32\times10^{27}$ & $2.21\times10^{17}$  & $5.54\times10^{18}$ & $1.67\times10^{46}$ \\
 5BZB~J0314$+$0619 & $<5.86\times10^{8}$  & $<2.01\times10^{-4}$   & $<6.49\times10^{-1}$  & $4.83\times10^{25}$ & $<3.69\times10^{16}$ & $<9.23\times10^{17}$   & $<4.95\times10^{45}$   \\
 5BZQ~J2238$+$0724 & $5.89\times10^{8}$  & $<1.97\times10^{-2}$   & $3.02\times10^{-2}$  & $6.62\times10^{26}$ & $3.66\times10^{17}$ & $9.15\times10^{18}$   & $2.73\times10^{46}$   \\
 5BZB~J0502$+$1338 & $1.94\times10^{8}$ & $1.19\times10^{-4}$ & $2.62\times10^{-1}$  & $2.31\times10^{26}$ & $1.63\times10^{16}$  & $4.08\times10^{17}$ & $2.08\times10^{44}$ \\
 5BZQ~J1420$+$1205 & $1.92\times10^{9}$ & $3.69\times10^{-2}$ & $<2.73\times10^{-1}$ & $1.51\times10^{28}$ & $9.05\times10^{17}$  & $2.26\times10^{19}$ & $1.11\times10^{47}$ \\
 5BZB~J0035$+$1515 & $7.25\times10^{8}$  & $1.79\times10^{-4}$   & $1.82\times10^{-1}$  & $2.24\times10^{25}$ & $3.87\times10^{16}$ & $9.69\times10^{17}$   & $8.07\times10^{45}$   \\
 5BZQ~J1143$+$1843 & $6.22\times10^{8}$ & $1.29\times10^{-2}$ & $<2.55\times10^{-3}$ & $2.83\times10^{25}$ & $3.04\times10^{17}$  & $7.61\times10^{18}$ & $1.23\times10^{46}$ \\
 5BZB~J1546$+$1817 & $<1.39\times10^{9}$  & $<9.19\times10^{-4}$   & $<4.47\times10^{-1}$  & $6.19\times10^{27}$ & $<1.22\times10^{17}$ & $<3.04\times10^{18}$   & $<2.97\times10^{45}$   \\
 5BZB~J1150$+$2417 & $<4.75\times10^{8}$  & $<1.61\times10^{-4}$   & $<4.27\times10^{-2}$  & $1.04\times10^{26}$ & $<2.97\times10^{16}$ & $<7.44\times10^{17}$   & $<4.59\times10^{45}$   \\
 5BZB~J1427$+$2348 & $8.00\times10^{8}$ & $1.00\times10^{-3}$ & $3.10\times10^{0}$   & $6.71\times10^{26}$ & $1.00\times10^{17}$  & $2.50\times10^{18}$ & $3.00\times10^{45}$ \\
 5BZU~J0220$+$2509 & $8.60\times10^{7}$  & $2.70\times10^{-3}$   & $<3.34\times10^{-2}$  & $1.31\times10^{25}$ & $5.18\times10^{16}$ & $1.29\times10^{18}$   & $4.12\times10^{45}$   \\
 5BZQ~J0122$+$2502 & $1.87\times10^{9}$ & $2.95\times10^{-2}$ & $<5.32\times10^{-2}$ & $2.22\times10^{28}$ & $7.88\times10^{17}$  & $1.97\times10^{19}$ & $7.03\times10^{46}$ \\
 5BZQ~J1125$+$2610 & $3.38\times10^{9}$ & $9.87\times10^{-3}$ & $7.86\times10^{-1}$ & $4.12\times10^{28}$ & $6.04\times10^{17}$  & $1.51\times10^{19}$ & $8.43\times10^{46}$ \\
 5BZQ~J1047$+$2635 & $4.37\times10^{9}$ & $2.80\times10^{-3}$ & $<4.05\times10^{-2}$ & $2.27\times10^{27}$ & $3.66\times10^{17}$  & $9.15\times10^{18}$ & $3.49\times10^{46}$ \\
 5BZB~J0737$+$2846 & $<1.26\times10^{8}$  & $<5.35\times10^{-4}$   & $<6.01\times10^{-3}$  & $3.78\times10^{24}$ & $<2.79\times10^{16}$ & $<6.97\times10^{17}$   & $<1.55\times10^{44}$   \\
 5BZQ~J0137$+$3122 & $1.11\times10^{8}$ & $9.15\times10^{-3}$ & $<6.04\times10^{-1}$ & $8.79\times10^{27}$ & $1.09\times10^{17}$  & $2.72\times10^{18}$ & $5.17\times10^{45}$ \\
 5BZU~J1706$+$3214 & $2.87\times10^{9}$ & $1.02\times10^{-2}$ & $<7.13\times10^{-3}$ & $8.50\times10^{26}$ & $5.82\times10^{17}$  & $1.45\times10^{19}$ & $1.37\times10^{47}$ \\
 5BZB~J0208$+$3523 & $<5.07\times10^{8}$  & $<4.60\times10^{-5}$   & $<9.13\times10^{-2}$  & $1.69\times10^{24}$ & $<1.64\times10^{16}$ & $<4.10\times10^{17}$   & $<5.38\times10^{43}$   \\
 5BZB~J1004$+$3752 & $<3.89\times10^{8}$ & $<4.54\times10^{-4}$ & $<5.97\times10^{-3}$ & $3.56\times10^{25}$ & $<4.52\times10^{16}$  & $<1.13\times10^{18}$ & $<4.08\times10^{44}$ \\
 5BZU~J1536$+$3742 & $<5.54\times10^{7}$ & $<6.51\times10^{-5}$ & $<1.20\times10^{-1}$ & $9.57\times10^{25}$ & $<6.45\times10^{15}$  & $<1.61\times10^{17}$ & $1.11\times10^{46}$ \\
 5BZQ~J1617$+$3801 & $2.69\times10^{9}$ & $3.12\times10^{-3}$ & $<1.95\times10^{-2}$ & $3.17\times10^{26}$ & $3.11\times10^{17}$  & $7.77\times10^{18}$ & $4.68\times10^{46}$ \\
 5BZB~J1210$+$3929 & $<1.74\times10^{9}$ & $<8.11\times10^{-6}$ & $<2.10\times10^{-2}$  & $3.13\times10^{25}$ & $<1.28\times10^{16}$  & <$3.20\times10^{17}$ & $<2.97\times10^{45}$ \\
 5BZQ~J1243$+$4043  & $3.34\times10^{8}$ & $1.30\times10^{-3}$ & $<1.46\times10^{-1}$ & $2.91\times10^{27}$ & $7.08\times10^{16}$  & $1.77\times10^{18}$ & $4.48\times10^{45}$ \\
 5BZG~J1156$+$4238 & $<1.74\times10^{8}$ & $<2.85\times10^{-5}$ & $<1.54\times10^{-3}$ & $1.17\times10^{24}$ & $<7.57\times10^{15}$  & $<1.89\times10^{17}$ & $<7.31\times10^{45}$ \\
 5BZB~J1122$+$4316 & $<1.59\times10^{8}$  & $<9.23\times10^{-5}$   & $<1.43\times10^{-2}$  & $2.13\times10^{24}$ & $<2.30\times10^{16}$ & $<3.25\times10^{17}$   & $<2.41\times10^{44}$   \\
 5BZQ~J1740$+$4348 & $1.76\times10^{9}$  & $1.18\times10^{-2}$   & $<7.30\times10^{-2}$  & $4.45\times10^{27}$ & $4.90\times10^{17}$ & $1.22\times10^{19}$   & $4.80\times10^{46}$   \\
 5BZB~J2247$+$4413 & $<1.65\times10^{8}$  & $<9.90\times10^{-5}$   & $<1.49\times10^{-1}$  & $2.26\times10^{25}$ & $<1.37\times10^{16}$ & $<3.43\times10^{17}$   & $<2.99\times10^{45}$   \\
 5BZB~J1523$+$4606 & $<8.40\times10^{7}$  & $<5.60\times10^{-5}$   & $<2.13\times10^{-1}$  & $4.46\times10^{25}$ & $<7.37\times10^{15}$ & $<1.84\times10^{17}$   & $<2.36\times10^{45}$   \\
 5BZQ~J0808$+$4950 & $6.95\times10^{8}$ & $1.56\times10^{-2}$ & $2.24\times10^{0}$   & $1.48\times10^{28}$ & $3.54\times10^{17}$  & $8.86\times10^{18}$ & $3.82\times10^{46}$ \\
 5BZQ~J1327$+$5008 & $2.87\times10^{9}$ & $1.83\times10^{-3}$ & $1.03\times10^{-1}$ & $1.64\times10^{27}$ & $2.46\times10^{17}$  & $6.16\times10^{18}$ & $4.74\times10^{45}$ \\
 5BZU~J1647$+$4950 & $1.44\times10^{7}$ & $1.12\times10^{-3}$ & $8.13\times10^{-2}$  & $9.57\times10^{23}$ & $1.36\times10^{16}$  & $3.41\times10^{17}$ & $3.72\times10^{43}$ \\
 5BZQ~J0514$+$5602 & $8.06\times10^{8}$  & $3.73\times10^{-3}$   & $<1.49\times10^{-1}$  & $1.22\times10^{28}$ & $1.86\times10^{17}$ & $4.66\times10^{18}$   & $6.94\times10^{45}$   \\
 5BZB~J0540$+$5823 & $<7.14\times10^{7}$  & $<3.25\times10^{-5}$   & $<2.80\times10^{-2}$  & $7.79\times10^{23}$ & $<5.18\times10^{15}$ & $<1.29\times10^{17}$   & $<3.64\times10^{44}$   \\
 5BZB~J1848$+$6537 & $<6.68\times10^{7}$  & $<5.55\times10^{-5}$   & $1.23\times10^{0}$  & $3.71\times10^{24}$ & $6.54\times10^{15}$ & $1.64\times10^{17}$   & $<1.69\times10^{45}$   \\
 5BZQ~J1927$+$7358 & $3.71\times10^{8}$ & $2.00\times10^{-2}$ & $<2.59\times10^{-3}$ & $1.19\times10^{27}$ & $2.93\times10^{17}$  & $7.32\times10^{18}$ & $1.88\times10^{46}$ \\
 5BZQ~J1357$+$7643 & $1.83\times10^{9}$ & $7.54\times10^{-3}$ & $3.81\times10^{0}$   & $1.10\times10^{28}$ & $1.26\times10^{17}$  & $3.15\times10^{18}$ & $6.59\times10^{45}$ \\
\bottomrule
\end{tabular}}
\end{table*}

\begin{table*}[]
\centering
\caption{Results of the statistical analysis.}
\label{table: AD test}
\resizebox{0.8\linewidth}{!}{
\begin{tabular}{clcccl}
\toprule
Quantity & Compared samples & \# Measurements & \# ULs & \thead{A-D pvalue \\ pre-trial} & \thead{Peto logrank pvalue \\ pre-trial} \\
\midrule
$\accretion$                                     & Our sample vs. P$21$        & $32$ vs. $658$  &  $20$ vs. $336$   &   $7.00\times10^{-4}$  &   $3.10\times10^{-3}$  \\
                                                                & Our sample vs. P$17$        & $32$ vs. $47$   &   $20$ vs. $-$    &   $2.01\times10^{-1}$ & $6.54\times10^{-1}$           \\
                                                                & Our sample vs. S$12$        & $32$ vs. $78$   &   $20$ vs. $83$   &   $8.09\times10^{-1}$ & $1.00\times10^{0}$           \\
\hline
$L_{\gamma} \left[ {\rm erg} \cdot {\rm s}^{-1} \right]$        & our sample vs. P$21$        & $24$ vs. $1000$  & $28$ vs. $-$ &   $8.36\times10^{-1}$ &   $1.47\times10^{-2}$                  \\
                                                                & our sample vs. P$17$        & $24$ vs. $307$  & $28$ vs. $191$    &   $1.00\times10^{-4}$  &   $6.04\times10^{-4}$           \\
                                                                & our sample vs. S$12$        & $24$ vs. $100$  & $28$ vs. $61$ &   $5.51\times10^{-2}$ & $4.13\times10^{-1}$                           \\
                                                                & our sample vs. BZCat        & $24$ vs. $1111$ & $28$ vs. $1666$   &   $1.44\times10^{-1}$ &   $1.72\times10^{-2}$        \\
                                                                & our sample vs. $4$LAC-DR$3$ & $24$ vs. $1846$ & $28$ vs. $-$  &   $7.34\times10^{-1}$ &   $5.32\times10^{-2}$  \\
\hline
$\Pradio \left[ {\rm W} \cdot {\rm Hz}^{-1} \right]$ & our sample vs. P$21$        & $52$ vs. $914$  & $-$ vs. $-$   &   $8.50\times10^{-2}$ &   $3.57\times10^{-2}$             \\
                                                                & our sample vs. P$17$        & $52$ vs. $416$  & $-$ vs. $-$   &   $2.10\times10^{-2}$ &   $1.03\times10^{-2}$             \\
                                                                & our sample vs. S$12$        & $52$ vs. $74$   & $-$ vs. $-$   &   $1.00\times10^{-4}$   &   $1.53\times10^{-4}\,\,\,\ast$                        \\
                                                                & our sample vs. BZCat        & $52$ vs. $2747$ & $-$ vs. $-$   &   $3.44\times10^{-1}$ &   $2.33\times10^{-1}$     \\
\hline
$z$                                                             & our sample vs. P$21$        & $49$ vs. $1000$ & $3$ vs. $-$   &   $2.15\times10^{-2}$ &   $2.79\times10^{-2}$              \\
                                                                & our sample vs. P$17$        & $49$ vs. $498$  & $3$ vs. $-$   &   $1.25\times10^{-1}$  &   $1.83\times10^{-1}$     \\
                                                                & our sample vs. S$12$        & $49$ vs. $162$  & $3$ vs. $-$   &   $1.00\times10^{-4}$  &   $1.52\times10^{-5}\,\,\,\ast$         \\
                                                                & our sample vs. BZCat        & $49$ vs. $2753$ & $3$ vs. $-$   &   $6.82\times10^{-1}$ &   $9.42\times10^{-1}$              \\
                                                                & our sample vs. $4$LAC-DR$3$ & $49$ vs. $1822$ & $3$ vs. $-$   &   $5.50\times10^{-3}$ &  $1.03\times10^{-2}$            \\
\hline
$\gratio$                                                       & our sample vs. P$21$        & $14$ vs. $658$  & $38$ vs. $342$    &   $7.95\times10^{-2}$ &  $5.94\times10^{-1}$          \\
                                                                & our sample vs. P$17$        & $14$ vs. $37$   & $38$ vs. $466$    &   $1.04\times10^{-3}$ &  $1.41\times10^{-2}$  \\
                                                                & our sample vs. S$12$        & $14$ vs. $78$   & $38$ vs. $62$     &   $3.12\times10^{-1}$ &  $5.77\times10^{-1}$        \\
\hline
$L_{\rm disk} \left[ {\rm erg} \cdot {\rm s}^{-1} \right]$      & our sample vs. P$21$        & $32$ vs. $658$  & $20$ vs. $342$    &   $9.19\times10^{-2}$ &   $6.06\times10^{-1}$           \\
                                                                & our sample vs. P$17$        & $32$ vs. $47$   & $20$ vs. $-$  &   $6.90\times10^{-3}$ &  $1.32\times10^{-2}$  \\
                                                                & our sample vs. S$12$        & $32$ vs. $78$   & $20$ vs. $83$ &   $1.51\times10^{-1}$ &  $2.59\times10^{-1}$               \\
\hline
$\mbh \left[ \Msun \right]$                           & our sample vs. P$21$        & $32$ vs. $658$  & $20$ vs. $342$    &   $1.08\times10^{-2}$ &   $4.66\times10^{-4}$  \\
                                                                & our sample vs. P$17$        & $32$ vs. $47$   & $20$ vs. $-$  &   $5.00\times10^{-4}$  &   $1.70\times10^{-6}\,\,\,\ast$                             \\
                                                                & our sample vs. S$12$        & $32$ vs. $78$   & $20$ vs. $83$ &   $2.31\times10^{-2}$ &   $3.48\times10^{-1}$               \\
\hline
$r_{\rm BLR} \left[ {\rm cm} \right]$                           & our sample vs. P$21$        & $32$ vs. $658$  & $20$ vs. $342$    &   $1.02\times10^{-1}$ &  $5.68\times10^{-1}$            \\
                                                                & our sample vs. P$17$        & $32$ vs. $47$   & $20$ vs. $-$  &   $7.67\times10^{-3}$ &   $1.14\times10^{-2}$  \\
                                                                & our sample vs. S$12$        & $32$ vs. $78$   & $20$ vs. $84$ &   $1.41\times10^{-1}$ &  $2.36\times10^{-1}$                \\
\hline
$r_{\rm DT} \left[ {\rm cm} \right]$                            & our sample vs. P$21$        & $32$ vs. $658$  & $20$ vs. $342$    &   $9.75\times10^{-2}$ &  $5.72\times10^{-1}$             \\
                                                                & our sample vs. P$17$        & $32$ vs. $47$   & $20$ vs. $-$  &   $7.76\times10^{-3}$ &    $1.18\times10^{-2}$  \\
                                                                & our sample vs. S$12$        & $32$ vs. $78$   & $20$ vs. $84$ &   $1.64\times10^{-1}$ &  $2.46\times10^{-1}$                 \\
\bottomrule
\end{tabular}}
\tablefoot{Results of the Anderson-Darling and Peto logrank tests (see also Appendix \ref{sec: append survival analysis}) performed on the candidate PeVatron blazars and the reference samples of \citet{Paliya:2021, Paliya:2017, Sbarrato:2012, BZCat, 4LAC-DR3}. The first column lists the physical property on which the test was applied. The second, third and fourth report the compared samples, with the corresponding number of measurements and the number of upper limits (ULs) excluded from the A-D and included in the Peto logrank. In the fifth and sixth columns, the pre-trial A-D and Peto logrank pvalues, respectively. The three cases in which the Peto logrank pvalue is considered statistically significant with the Benjamini-Hochberg method are highlighted with a $\ast$ symbol (see text for further details).}
\end{table*}

%%%%%%%%%%%%%%%%%%%%%%%%%%%%%%%%%%%%%%%%%%%%%%
\section{Instruments and datasets}
\label{sec:append dataset}

 In this section, we provide additional information about the data collected via optical spectroscopy. Table \ref{table:spectra} lists the instrument, observation date, and total exposure time for each spectrum used in the analysis. For two of the candidate PeVatron blazars located in the southern celestial hemisphere, a spectrum was already available in the Third Data Release of the Six-degree Field Galaxy Survey \citep[$6$dFGS-DR$3$, ][]{6dFGS:2004, 6dFGS:2009} and $2$dF QSO Redshift Survey \citep[$2$QZ; ][]{2QZ:1998, 2QZ:2004}. The literature provides useful information about two other southern objects. We observed the remaining five sources with the X-Shooter \citep{XShooter} spectrograph on the European Southern Observatory (ESO) Very Large Telescope (VLT) and the Gemini South Multi-Object Spectrographs \citep[GMOS, ][]{GMOS:1998, GMOS:2016}. In the northern hemisphere, $20$ out of $42$ objects have a publicly available optical spectrum in the Data Release $17$ of the Sloan Digital Sky Survey \citep[SDSS-DR$17$, ][]{SDSS17}, the ZBLLAC database \citep{ZBLLAC} and the Fifth Data Release of the Large Sky Area Multi-Object Fiber Spectroscopic Telescope \citep[LAMOST-DR$5$, ][]{LAMOST:2012}. The LAMOST catalog provides spectra with relative flux calibration. This may affect the accuracy of the absolute flux level but does not dramatically impact the spectroscopic determinations \citep{Ren:2014, Song:2012}. The literature provides all the desired properties for $15$ objects. Five other blazars appear in several published papers, with only information about the redshift and no public spectrum. In addition, we observed seven of the northern objects with the GTC OSIRIS spectrograph. As mentioned in Section \ref{subsec:quantities}, to estimate the accretion properties, we only used the permitted lines coming from the broad line region (\Ha\ $\lambda 6563$ \AA, \Hb\ $\lambda 4861$ \AA, \CIV\ $\lambda 1549$ \AA, \MgII\ $\lambda 2798$ \AA) either through detection or limit estimation. In this work, other BLR lines and the oxygen transitions originated in the NLR are analyzed to check the redshift value and the intensity of external radiation fields.

 %%%%%%%%%%%%%%%%%%%%%%%
\subsection{Very Large Telescope}
\label{subsec:VLT}

We observed 5BZB~J2243$-$0609 with the European Southern Observatory (ESO) Very Large Telescope at Paranal Observatory (VLT) X-Shooter spectrograph in the ESO-VLT-U$3$ telescope on September $23$rd, $2022$ (MJD $59845$), program id $109$.$22$WZ.$001$ in the three arms of the instrument at a maximum airmass of $1.5$ and seeing of $2.0"$. The median signal-to-noise ratio per order and the reference spectral resolving power were, respectively, $9.56$ and $5573$ in the NIR (${\rm slit width} = 0.9"$), $13.84$ and $8935$ in the VIS (${\rm slit width} = 0.9"$), $8.00$ and $5453$ in the UVB (${\rm slit} = 0.9"$). Consistently with the redshift ${\rm z} = 0.30$, we detected  \Ha, \Hb, \MgII, and \OII\ emission lines (rest-frame ${\rm EW} = 7.43, 0.91, 4.18, 2.34$, respectively).

Instead, we observed 5BZQ~J0256$-$2137 on July $26$th, $2022$ (MJD $59786$), program id $109$.$22$WZ.$001$, at a maximum airmass of $1.8$ and seeing of $2.0"$.
The median signal-to-noise ratio per order and the reference spectral resolving power were, respectively, $2.86$ and $5573$ in the NIR (${\rm slit width} = 0.9"$), $12.72$ and $8935$ in the VIS (${\rm slit width} = 0.9"$), $18.69$ and $5453$ in the UVB (${\rm slit} = 1.0"$). In agreement with the redshift ${\rm z} = 1.47$, we identified the emission lines of \Ha, \Hb, \CIV, \MgII, and \OIII\ (rest-frame ${\rm EW} = 210.53, 3.68, 15.51, 14.57, 25.92$, respectively).

For \Gsource, the spectrum was obtained by \citet{Shaw:2013} with the Focal Reducer and low dispersion spectrograph \citep[FORS$2$; ][]{FORS}. This appears featureless in terms of emission, but the presence of absorption lines (Fe~\MakeUppercase{\romannumeral2} $2344.2$, $2374.4$, $2382.7$, $2586.6$, $2600.1$ \AA) due to intervening systems (either a galaxy or low-excitation clouds in the halo of the host galaxy) places a firm lower limit on the redshift $z > 1.239$. Also, the absence of the hydrogen Ly$\alpha$ line in absorption places a statistically based upper limit $z_{\rm max} < 1.91$, using the constraints imposed by the signal-to-noise ratio of the spectrum. For our analysis, we adopted the procedure described in Section \ref{subsec:quantities} to estimate upper limits.

%%%%%%%%%%%%%%%%%%%%%%%
\subsection{Gemini South}
We observed three sources using the R$150$ grism of the Gemini Multi-Object Spectrographs of Gemini South \citep[GMOS-S, ][]{GMOS:2016} within the program GS-$2022$B-DD-$201$. We set the central wavelength on the grism to $950\,{\rm nm}$ and used a $1.0"$ slit width. The grating has a resolution ${\rm R} = 631$, and dispersion of $0.193\, {\rm nm} \cdot {\rm pixel}^{-1}$. 

Two spectra of 5BZQ~J2036$-$2146 in four exposures of $300\,{\rm s}$ each were taken on October $1$st, $2022$ (MJD $59853$) and four exposures of $800\,{\rm s}$ on November 18th 2022 (MJD $59901$). The mean airmass for the observations was $1.02$ and $1.33$, respectively.

For 5BZB~J0359$-$2615, we acquired two spectra: four exposures of $300\, {\rm s}$ each on October $1$st, 2022 (MJD $59853$) and four exposures of $t_{\rm exp} = 1.8\, {\rm ks}$ on November $20$th $2022$ (MJD $59903$). The mean airmass was $1.17$ and $1.61$, respectively.

Finally, for 5BZQ~J2304$-$3625, six exposures of $300\, {\rm s}$ each were taken on November $16$th, 2022 (MJD $59899$), two exposures of $1.35\, {\rm ks}$ on November 19th 2022 (MJD $59902$) and four exposures (two of $1.35\, {\rm ks}$, and two of $300\, {\rm s}$) on November 23rd 2022 (MJD $59906$). The mean airmass was $1.02$, $1.10$ and $1.24$, respectively.

Instrumental issues on the CCD of the spectrograph contaminated the acquisition, making part of the spectra unusable. Careful adjustments had to be applied during the reduction procedure and the analysis had to be limited to the wavelength range of $4900$ to $10100$ \AA. Due to this contamination, we made no tentative identification of emission features and placed upper limits on the lines instead.

%%%%%%%%%%%%%%%%%%%%%%%
\subsection{Gran Telescopio Canarias}
We observed seven sources with the R$1000$B Optical System for Imaging and low-Intermediate-Resolution Integrated Spectroscopy of the Gran Telescopio Canarias \citep[GTC OSIRIS, ][]{osiris}. 

We observed \RXJ\ on November $5$th, $2023$ (MJD $60253$) with two exposures of $1\, {\rm ks}$ each. The mean airmass was $1.09$ and $1.12$. We detected \MgII and \OII (with rest-frame ${\rm EW} = 0.63$ \AA\, and $1.04$ \AA, respectively), which lead to the redshift estimation ${\rm z} = 0.57$.

We observed 5BZQ~J2238$+$0724 on November $11$th, $2023$ (MJD $60252$) with two exposures of $700\, {\rm s}$ each. The mean airmass was $1.08$ and $1.07$. We detected C~\MakeUppercase{\romannumeral3}], \MgII, and \OII\ (rest-frame ${\rm EW} = 19.72, 40.01, 0.68$, respectively), which lead to the confirmation of the redshift ${\rm z} = 1.011$.

We observed 5BZU~J0220$+$2509 on November $4$th, $2023$ (MJD $60252$) with two exposures of $500\, {\rm s}$ each. The mean airmass was $1.01$ and $1.02$. We detected \Hb, \MgII, \OII, and \OIII\ (rest-frame ${\rm EW} = 8.82, 121.08, 3.52, 25.52$, respectively), which lead to the confirmation of the redshift ${\rm z} = 0.48$.

The new GTC spectrum of 5BZQ~J1740$+$4348 was acquired on March $19$th, $2024$ (MJD $60388$) with a single exposure of $1.5\,{\rm ks}$ at airmass $1.31$. We detected Ly$\alpha$, \CIV, He~\MakeUppercase{\romannumeral2}, and C~\MakeUppercase{\romannumeral3}] (rest-frame ${\rm EW} = 112.75, 37.92, 1.89, 18.92$, respectively), confirming the redshift ${\rm z} = 2.246$.

We observed 5BZB~J2247$+$4413 on November $3$rd, $2023$ (MJD $60251$) with two exposures of $700\,{\rm s}$ each, airmass $1.04$. The result is a featureless spectrum, on which we placed limits.

We performed two acquisitions of $1\,{\rm ks}$ each for 5BZQ~J0514$+$5602 on November $5$th, $2023$ (MJD $60253$), at airmass $1.19$ and $1.17$. The detection of Ly$\alpha$, \CIV, and C~\MakeUppercase{\romannumeral3}] emission lines (rest-frame ${\rm EW} = 41.13, 114.33, 16.77$, respectively) allowed us to confirm the redshift value ${\rm z} = 2.19$.

Finally, we observed 5BZB~J1848$+$6537 on April $5$th, $2024$ (MJD $60405$) with two exposures of $1\,{\rm ks}$ each, at airmass $1.39$ and $1.36$. No emission line appears in the resulting spectrum, therefore we estimated upper limits on the properties of this blazar.

%%%%%%%%%%%%%%%%%%%%%%%
\subsection{Properties from the literature}
\label{subsec:literature}
Some blazars of our sample already appeared in previous works. Here we list the information collected from the literature. 

For 5BZQ~J2003$-$3251, we used the average of the published physical properties \citep{Xiong:2015, Cao:1999, Osmer:1994, Paliya:2017}. This blazar was also observed in \citet{DecontoMachado:2023}.

For 5BZU~J1819$-$6345, the paper of \citet{Mingo:2014} provided the black hole mass and Eddington luminosity as measured from the $K_{\rm s}$ and r magnitudes \citep{Inskip:2010, Ramos:2013} using the $\mbh - L_{\rm K}$ and $\mbh - L_{\rm r}$ correlations \citep{Graham:2007}. They concluded that the spectra showed clear traces of an optical disk and dust lane. 

We took the values of black hole mass and disk luminosity for 5BZQ~J0239$-$0234, 5BZQ~J0312$+$0133, 5BZQ~J0422$+$0219, and 5BZQ~J0808$+$4950 from \citet{Paliya:2017}. 5BZQ~J0422$+$0219 appeared also in \citet{Ghisellini2011_highred}, in which the authors perform the SED leptonic modeling and report the resulting physical parameters. 5BZQ~J0808$+$4950 was also included in the samples of \citet{Shen:2011} and \citet{Paliya:2021}.

The blazar 5BZQ~J2238$+$0724 was identified as a good candidate for IceCube high-energy neutrinos production in \citet{Hovatta:2021, Plavin:2023}.

The value of the redshift for 5BZB~J0502$+$1338 was estimated in \cite{Truebenbach:2017}. \citet{Plavin:2020, Plavin:2023, Hovatta:2021} identified this blazar as a promising candidate for IceCube high-energy neutrino emission.

The blazars 5BZQ~J1420$+$1205 and 5BZU~J1706$+$3214 were included in the sample of \citet{Shen:2011}, in which they provided the values of the redshift, the bolometric luminosity, and the black hole mass.

The literature provided several different attempts to inspect \RXJ, first classified as a BL Lac \citep{Fischer:1998}: the SDSS spectrum analyzed in \citet{Shaw:2013} appeared featureless as the follow-up observation performed with the $10$ m Gran Telescopio Canarias by \citet{Paiano:2017}. Based on the signal-to-noise ratio and the minimum expected equivalent width, they placed a lower limit of ${\rm z} > 0.55$ on the redshift, consistent with our estimation.

The SDSS$-$DR$17$ optical spectra of 5BZQ~J1143$+$1843, 5BZQ~J1617$+$3801, 5BZQ~J1327$+$5008 were also analyzed in \citet{Shen:2011}.

For \Lsource, we used the redshift estimate of \citet{Truebenbach:2017} and the VLT$-$FORS$1$ spectrum \citep{Sbaruffati_2009}. The source was also included in the sample studied in \citet{Ghisellini:2011}, in which they reported no detection of emission lines but listed the physical properties used as parameters for the theoretical SED modeling. The SDSS$-$DR6 spectrum of this blazar was also used in \citet{Sbarrato:2012} to place upper limits.

The blazar \PKS\ (5BZB~J1427$+$2348) was already identified as a promising candidate neutrino emitter \citep{IceCube10y:2020}, and its properties extensively studied in the literature \citep{Paiano:2017, Padovani_PKS:2022}.

The two objects 5BZB~J0737$+$2846 and 5BZB~J1004$+$3752 were studied in \citet{LeonTavares:2011} and \citet{Xiong:2015}, which provided the values of the black hole mass as obtained from stellar velocity dispersion estimations and the jet kinetic power. We assumed the value of $P_{\rm jet}$ as an upper limit for the luminosity of the accretion disk, and derived the other physical quantities of our interest consequently.

The properties of 5BZB~J0208$+$3523 were also available from the literature \citep{Du:2013, Wang:2002, Morris:1991}. The spectrum was used for the redshift estimation and employed for placing $3\sigma$ limits on the lines' luminosities.

To study 5BZB~J1210$+$3929 and 5BZG~J1156$+$4238, we took the limits on the luminosity of emission lines from \citet{Rector:2000} and the black hole mass from \citet{Wang:2002}. From these, we estimated the other quantities of our interest. \citet{Stathopoulos:2022} identified 5BZB~J1210$+$3929 as a promising candidate for neutrinos emission during X-ray flares detected by the X-ray Telescope of the Neil Gehrels Swift Observatory (Swift XRT).

The value of the redshift of the object \changlook\ was estimated by the analysis of \citet{Titov:2013} on a spectrum taken with the Alhambra Faint Object Spectrograph and Camera (ALFOSC) on the $2.5$ m Nordic Optical Telescope (NOT), which we used in our analysis. This blazar, along with 5BZU~J0143$-$3200, 5BZQ~J1125$+$2610 and 5BZQ~J1047$+$2635, were also studied in the survey of \citet{PenaHerazo_changinglook} on changing-look blazars. 5BZQ~J1125$+$2610 was reported as a good candidate for IceCube high-energy neutrinos emission also in \citet{Hovatta:2021, Plavin:2023}. 

Two featureless optical spectra of 5BZB~J2247$+$4413 were taken with the Double Spectrograph on the $200"$ Hale Telescope at Mt. Palomar and the $4$ m Mayall telescope at the Kitt Peak National Observatory (KPNO) and analyzed by \citet{Shaw:2013} and \citet{Massaro:2015}, however, they are not publicly available.

The object 5BZU~J1647$+$4950 was included in the analysis of \citet{Paliya:2021, Xiao:2022}.

The position of 5BZB~J0540$+$5823 was already identified as consistent with an IceCube event \citep{Padovani:2016}. We used the GTC OSIRIS spectrum of \citet{Paiano:2020} to place limits.

The redshift of 5BZB~J1848$+$6537 was determined in the work of \citet{Piranomonte:2007}, based the identification of galaxy absorption features typical of ellipticals.

The spectra of 5BZQ~J1927$+$7358 were broadly studied in the literature \citep{Torrealba_2012, Park:2017, Koss:2017}. It appeared also in the discussion of \citet{Xie:2012} about the accretion disk-jet connection when exploring the correlation between the observed BLR flux and the radio, near-IR, and X-ray fluxes.

For 5BZQ~J1357$+$7643, the Palomar $200"$ Double Spectrograph (DBSP) spectrum was analyzed in the work of \citet{Shaw:2012}.
The object was also included in the samples of \citet{Paliya:2017} and \citet{Chen:2021}.

\bgroup
\def\arraystretch{1}%
\begin{table}
    \centering
    \caption{Details about the spectral dataset.}
    \label{table:spectra}
    \resizebox{0.8\columnwidth}{!}{\begin{tabular}{cccc}
    \toprule
    Blazar & Instrument & Observation date $\left[ {\rm MJD} \right]$ & $t_{\rm exp} \left[ {\rm ks} \right]$ \\
    \midrule 
    5BZB~J2243$-$0609   &  VLT X$-$Shooter        & $59845$         &  $11.73$    \\ 
    5BZQ~J0357$-$0751   &  AAO $6$dF              & $53291$         &  $1.80$     \\ 
    5BZQ~J0256$-$2137   &  VLT X$-$Shooter        & $59786$         &  $8.79$     \\
    5BZQ~J2036$-$2146   &  GMOS$-$S               & $59853, 59901$  &  $4.40$     \\ 
    5BZB~J0630$-$2406   &  VLT FORS$2$            & $55191$         &  $0.60$     \\ 
    5BZB~J0359$-$2615   &  GMOS$-$S               & $59853, 59903$  &  $8.40$     \\
    5BZU~J0143$-$3200   &  AAO $6$dF              & $52169$         &  $3.30$     \\
    5BZQ~J2238$+$0724   &  GTC OSIRIS             & $60252$         &  $1.40$     \\
    5BZQ~J2304$-$3625   &  GMOS$-$S               & $59899, 59902$  &  $7.80$     \\
    5BZQ~J0239$-$0234   &  APO BOSS               & $57011$         &  $4.50$     \\
    5BZB~J0243$+$0046   &  APO BOSS               & $52177$         &  $3.60$     \\
    5BZB~J0148$+$0129   &  APO BOSS               & $56900$         &  $5.40$     \\
    5BZB~J0509$+$0541   &  GTC OSIRIS             & $58080 - 58138$ &  $48.50$    \\
    5BZQ~J0400$+$0550   &  LAMOST LRS             & $57394$         &  $5.82$     \\
    5BZB~J0502$+$1338   &  APO BOSS               & $57450$         &  $0.30$     \\
    5BZQ~J1420$+$1205   &  APO BOSS               & $53885$         &  $2.74$     \\
    5BZB~J0035$+$1515   &  GTC OSIRIS             & $60253$         &  $2.00$     \\
    5BZQ~J1143$+$1843   &  APO BOSS               & $54179$         &  $3.51$     \\
    5BZB~J1546$+$1817   &  APO BOSS               & $55337$         &  $3.60$     \\
    5BZB~J1150$+$2417   &  VLT FORS$1$            & $54483$         &  $2.40$     \\
    5BZB~J1427$+$2348   &  GTC OSIRIS             & $57203$         &  $0.90$     \\
    5BZU~J0220$+$2509   &  GTC OSIRIS             & $60252$         &  $1.00$     \\
    5BZQ~J0122$+$2502   &  APO BOSS               & $57286$         &  $4.50$     \\
    5BZQ~J1125$+$2610   &  APO BOSS               & $56306$         &  $2.70$     \\
    5BZQ~J1047$+$2635   &  APO BOSS               & $56340$         &  $3.60$     \\
    5BZB~J0737$+$2846   &  APO BOSS               & $52232$         &  $3.20$     \\
    5BZQ~J0137$+$3122   &  APO BOSS               & $56567$         &  $2.70$     \\
    5BZU~J1706$+$3214   &  Shane KAST             & $50258$         &  $0.90$     \\
    5BZB~J1004$+$3752   &  APO BOSS               & $52993$         &  $2.70$     \\
    5BZU~J1536$+$3742   &  APO BOSS               & $52753$         &  $1.92$     \\
    5BZQ~J1617$+$3801   &  APO BOSS               & $52767$         &  $3.90$     \\
    5BZQ~J1243$+$4043   &  NOT ALFOSC             & $56035$         &  $2.40$     \\
    5BZB~J1122$+$4316   &  APO BOSS               & $56013$         &  $3.60$     \\
    5BZQ~J1740$+$4348   &  GTC OSIRIS             & $60388$         &  $1.50$     \\
    5BZB~J2247$+$4413   &  GTC OSIRIS             & $60252$         &  $1.40$     \\
    5BZQ~J1327$+$5008   &  APO BOSS               & $57423$         &  $3.60$     \\
    5BZQ~J0514$+$5602   &  GTC OSIRIS             & $60253$         &  $2.00$     \\
    5BZB~J0540$+$5823   &  GTC OSIRIS             & $58146$         &  $0.90$     \\
    5BZB~J1848$+$6537   &  GTC OSIRIS             & $60405$         &  $2.00$     \\
    \bottomrule
    \end{tabular}}
\end{table}
\egroup

%%%%%%%%%%%%%%%%%%%%%%%%%%%%%%%%%%%%%%%%%%%%%%
\section{Discussion about the upper limits}
\label{sec:append upper lim}
To exploit at best the information available from the optical spectroscopy, for featureless spectra we estimated limits on the physical quantities. Analogously, we placed limits on the $\gamma$-ray luminosity of sources without \emph{Fermi}-LAT detection. In Fig.\ref{fig:accretion regime no highlights}, we showed the trend of the $\accretion$ ratio as a function of $\gratio$ for inspecting the accretion regime. The approach is similar to \citet{Sbarrato:2012, Sbarrato:2014}, however, we used a different orientation of the arrows to indicate limits on both the optical and $\gamma$-rays. To explore the space of parameters for the $\accretion - \gratio$ plane when $L_{\rm BLR}$, $L_{\rm Edd}$ and $L_{\gamma}$ are upper limits, we chose the starting values of the three quantities and made them vary. Since $L_{\rm BLR}$ and $L_{\rm Edd}$ are always related through Eqs. \ref{eqn:bhmass} and \ref{eqn:lumblr}, in the simulation we used the coefficients corresponding to \MgII\ and constrained the range of variability. As shown in Fig.\ref{fig: limits}, the initial point (the red star) moved vertically only downward and horizontally in both directions. The area spanned as the quantities change is shown by the gray points, while the final values are indicated with the gray stars.

\begin{figure}
    \centering
    \includegraphics[width = 0.5\textwidth]{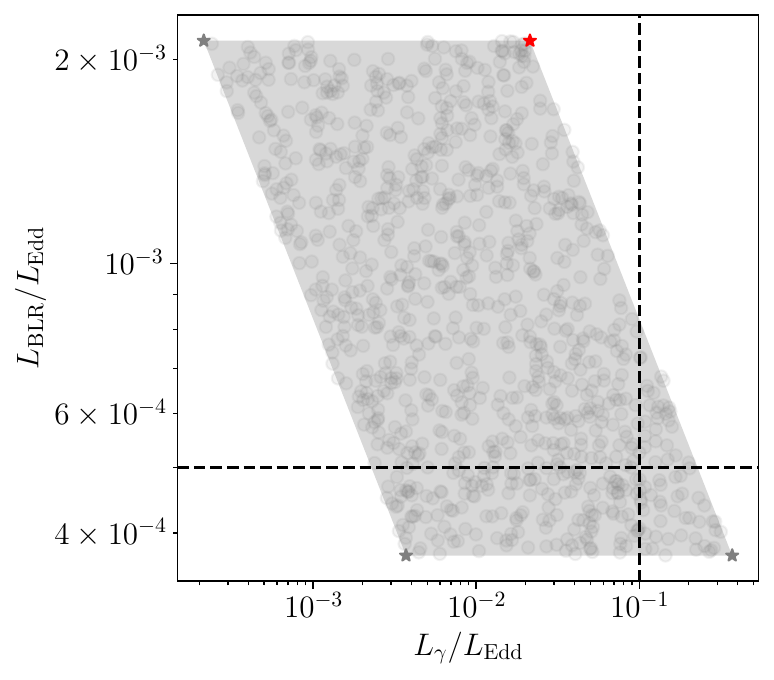}
    \caption{Trend of the accretion regime $\accretion$ as a function of $\gratio$ in the case of upper limits on both the optical and $\gamma$-rays. The red star is the chosen initial value, the gray points indicate the region spanned when the three quantities $L_{\rm BLR}$, $L_{\rm Edd}$, and $L_{\gamma}$ vary. The gray stars represent the final values of the parameter simulation. The dotted black lines represent the separation limits for the accretion efficiency, respectively $\accretion\sim5\times10^{-4}$ and $\gratio\sim0.1$ \citep{Ghisellini:2011, Sbarrato:2012}.  In this case, $L_{\rm BLR}$ varies between $\left[10^{44},10^{46}\right]\, {\rm erg} \cdot {\rm s}^{-1}$, $L_{\gamma}$ between $\left[10^{45}, 10^{47}\right]\, {\rm erg} \cdot {\rm s}^{-1}$.}
    \label{fig: limits}
\end{figure}

%%%%%%%%%%%%%%%%%%%%%%%%%%%%%%%%%%%%%%%%%%%%%%
\section{Evaluating the reliability of statistical methods for handling censored data}
\label{sec: append survival analysis}

In our study of PeVatron blazar candidates and comparison samples, we assessed the applicability of statistical methods that incorporate censored data, i.e., upper and/or lower limits (\saysay{left-} and \saysay{right-censored} data, respectively) into the analysis. Previous studies have utilized survival analysis techniques to handle such censored data effectively.
Specifically, the Kaplan–Meier estimator is employed to estimate the survival function of a given quantity in a sample, accommodating both uncensored and censored observations. To compare properties across different samples, tests such as the logrank and Peto logrank are utilized, which include weighting schemes to account for censored data points \citep{statistical_methods_I, ASURV, Padovani:1992, Homan_2021, Baldi_2022, Padovani:2022, Paiano:2023}.

We investigated both the logrank test and the Peto logrank test, the latter being reported to be less affected by differences in censoring patterns between samples \citep{statistical_methods_I, Padovani:2022, Paiano:2023}, and found that the results are highly sensitive to the rate of censored data and the sample size. To evaluate the suitability of these methodologies for our study, we conducted several analyses, especially covering cases where significant discrepancies were observed in Section \ref{sec:comparison}. Below, we summarize the findings most pertinent to our study.

\subsection*{Experiment setup and findings}
\label{subsec: app setup}
We utilized samples which list only measurements, such as the $5$BZCat redshift, with a total of $2753$ values, $4$LAC-DR$3$ $\gamma$-ray luminosity ($1846$), and P$17$ black hole mass ($47$).
We initially constructed comparison subsamples comprising only detections by randomly selecting values from the parent dataset. We generated multiple test subsamples with increasing sizes to evaluate the consistency of these subsamples with the original datasets. To do this, we employed the Kaplan–Meier estimator alongside the logrank and Peto logrank tests. As expected, these distributions remained compatible, regardless of the physical quantities tested, as illustrated in the left panels of Fig. \ref{fig: survival test}. In this way, we assessed and confirmed their comparability to the original datasets, and also validated the expected performance of both the logrank and Peto logrank tests.

Subsequently, we generated mock samples by introducing left-censored data. These mock samples were constructed to have the same size and distribution as the original dataset, with the only difference being that some measurements were flagged as ULs. This involved randomly selecting an increasing number of measurements and flagging them as ULs. We adopted the more extreme scenario of turning the measurements into upper limits, resulting in the lowest pvalues, as this represents the most conservative approach. We, then, applied survival analysis to compare these mock samples to the original sample, as shown in the right panels of the same Fig. \ref{fig: survival test}. Our findings revealed that the Peto logrank test is less biased by censoring patterns compared to the standard logrank test, as discussed in \cite{statistical_methods_I}. Consequently, we based our analysis on the Peto logrank test.
However, the resulting pvalues, plotted against the number of ULs, showed increasing discrepancies as the fraction of limits grows, highlighting potential limitations in the test's applicability. 

As an additional test, we evaluated the reliability of this method as a function of the fraction of upper limits relative to the total sample size, considering different sample sizes equal to  $20, 50, 100, 200$. From the parent sample, we extracted two subsamples of the same size. One was kept unmodified, i.e., its values are the same as the original sample and considered as measurements. In the second subsample, a progressively larger number of them were flagged as UL. In this way, the second subsample includes both measurements and limits, but the overall distribution is identical to the parent sample by construction.
We recorded the Peto logrank pvalues obtained over $100$ realizations for each scenario and estimated their median. The results obtained using the $5$BZCat redshift are shown in Fig. \ref{fig: bzcat same size}. The Figure illustrates the median Peto logrank pvalues plotted against the percentage of ULs relative to the total sample size. The horizontal dashed-dotted and dashed lines indicate the $3\sigma$ and $5\sigma$ levels, respectively. This analysis demonstrated that, as the proportion of ULs increases, the pvalues tend to decrease, indicating a higher likelihood of running into Type I errors. Another finding is that increasing the sample size, while maintaining the same percentage of censored data, increased the likelihood of false positives. This occurs because a larger sample inherently includes more censored data.

\subsection*{Applicability in the context of our study}
\label{subsec: app applicability}

In the analysis presented in Section \ref{sec:comparison}, three tests highlighted significant discrepancies $(\gtrsim3\sigma)$ with the Peto logrank approach. These are marked with the symbol $\ast$ in the rightmost column of Table \ref{table: AD test}. Two of these tests involved only measurements or a small amount of censored data. Specifically, these included the comparison with S$12$ of the redshift, where the total sample sizes were $49$ vs $162$, with $6\%$ censored data, and $\Pradio$, with total sample sizes of $52$ vs $74$. These cases are similar to the simulations presented in Figures \ref{fig: survival test} (bottom) and \ref{fig: bzcat same size} (orange, green, red), which demonstrated that when testing samples with little or no censored data, the Peto logrank test performs as expected and can therefore be regarded as reliable. They are also consistent with the corresponding A-D pvalues.

The third test compared the $\mbh$ distributions between the target sample and the P$17$ sample. The target sample consisted of $52$ observations, approximately $38\%$ of which were censored data, while the P$17$ sample comprised $47$ measurements with no censored data.
To assess the robustness of this result, we referred to simulations shown in Figures \ref{fig: survival test} (bottom) and \ref{fig: bzcat same size} (orange). These simulations with sample sizes and an UL fraction closest to those of the observed pvalue $1.70\times10^{-6}$ conservatively results in a pvalue of $>10^{-2}$. Thus, we expect the $38\%$ UL of the $\mbh$ test against P17 to have underestimated its intrinsic pvalue by a factor of $>10^{-2}$ in the worst-case scenario. In other words, the UL-corrected pvalue is expected to be $<1.70\times10^{-6}/0.01 = 1.70\times10^{-4}$, placing it at the third rank. Since the third rank Benjamini-Hochberg critical threshold is $\sim2.8\times10^{-4}$ and the UL-corrected pvalue is conservatively $<1.70\times10^{-4}$, we conclude that the $\mbh$ test using P17 is significant at the $\gtrsim 3\sigma$ level and may be attributable to an intrinsic difference between the tested samples.

\begin{figure}
\centering
\begin{subfigure}{\linewidth}
    \includegraphics[width=\linewidth,height=0.205\textheight]{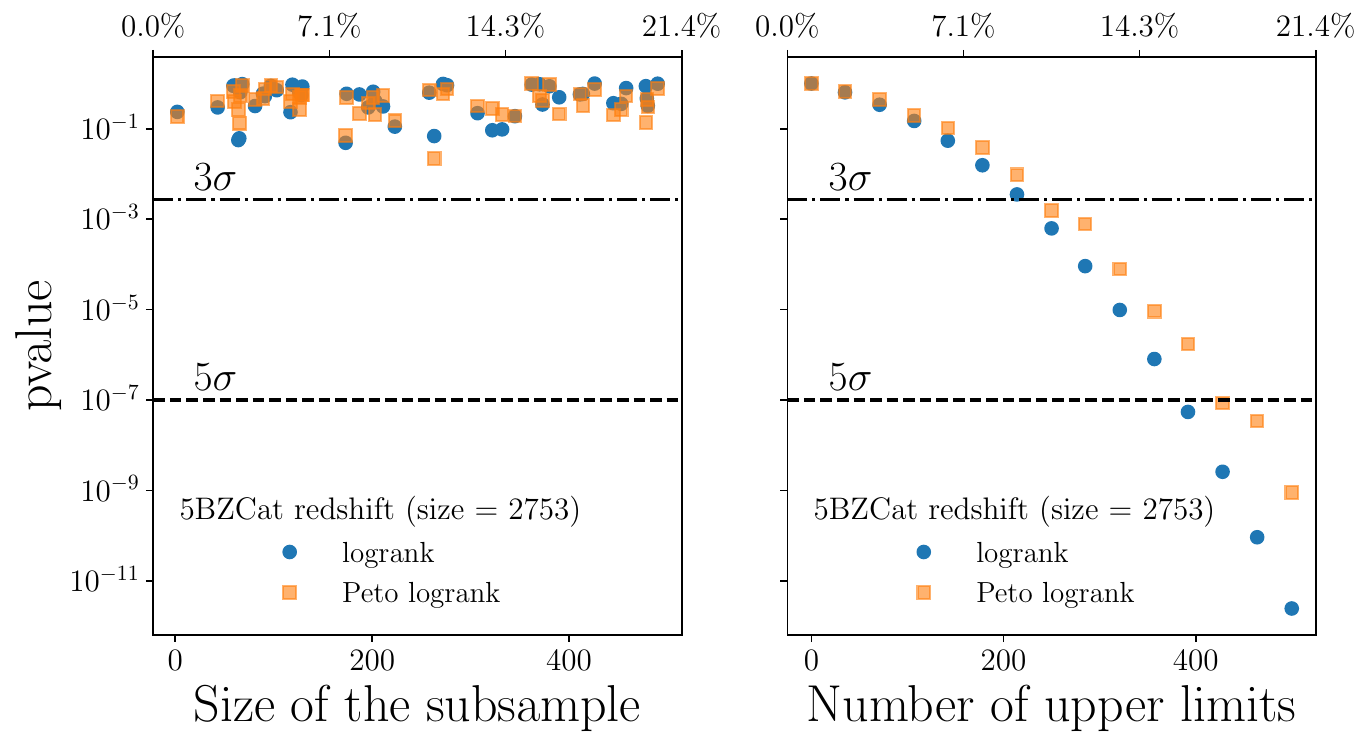}
\end{subfigure}
 \begin{subfigure}{\linewidth}
     \includegraphics[width=\linewidth,height=0.205\textheight]{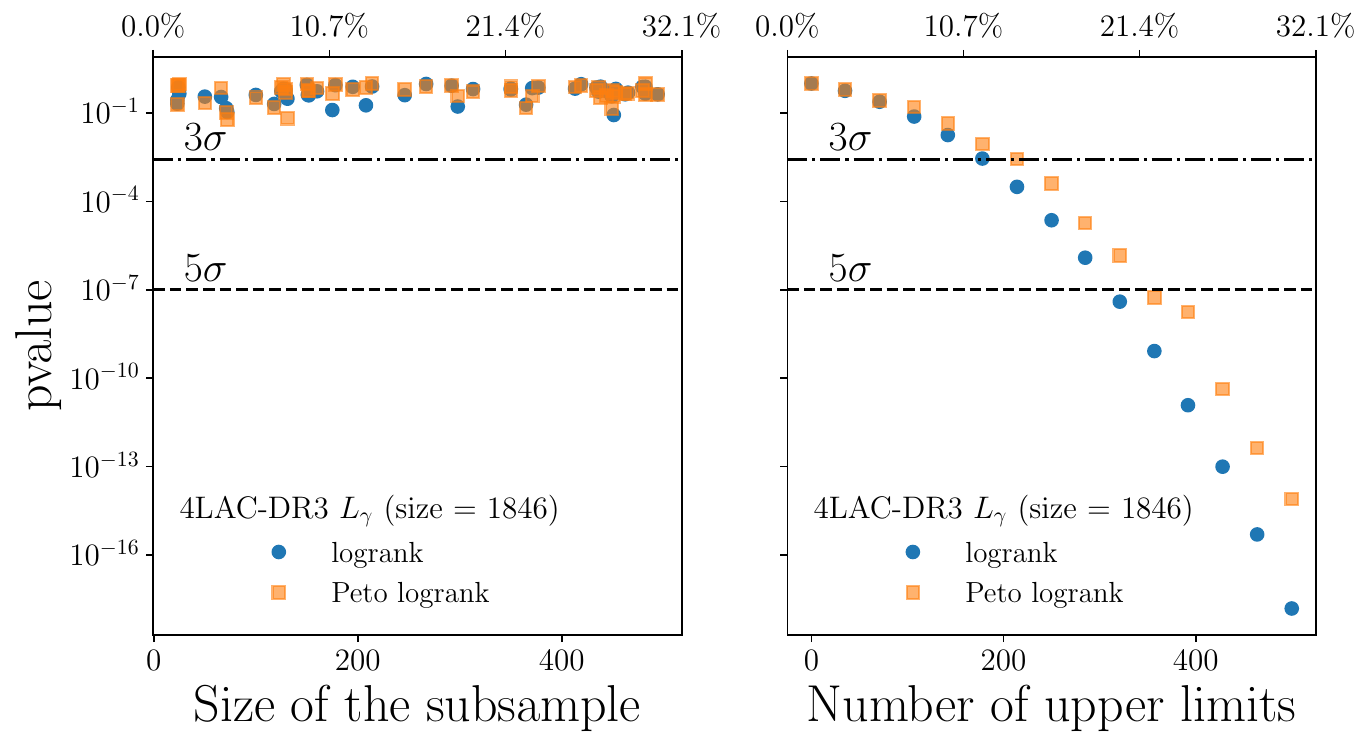}
 \end{subfigure}
 \begin{subfigure}{\linewidth}
    \includegraphics[width=0.97\linewidth,height=0.205\textheight]{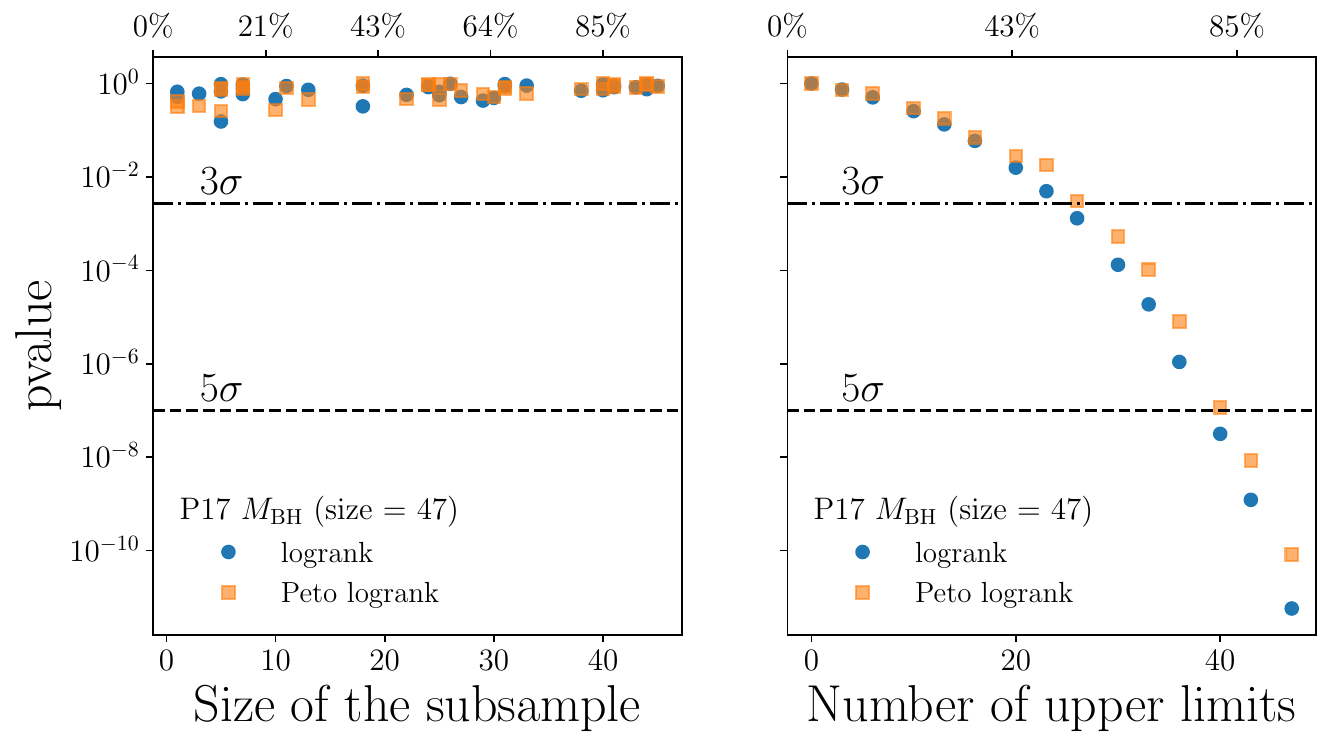}
\end{subfigure}
\caption{\footnotesize Redshift distribution of $5$BZCat (top), the $L_{\gamma}$ distribution of $4$LAC-DR$3$ (middle) and the $\mbh$ distribution of P$17$ (bottom) are used to access the reliability of the statistical approaches. The \textit{left panels} show the result of the logrank (blue points) and Peto logrank tests (orange squares) performed on the original population (only measurements) vs. comparison subsamples of progressively larger size, constructed by randomly selecting measured values from the parent dataset. The pvalue is shown as a function of the size of the selected subsample. The \textit{right panels} show the result of the logrank (blue points) and Peto logrank (orange squares) tests performed on the parent populations (only measurements) vs. mock samples of the same size as the parent population, where a progressively larger number of values is flagged as censored data. The pvalue is shown as a function of the number of limits. In both panels, the upper axis indicates the percentage with respect to the total size of the sample (see Appendix \ref{subsec: app setup}).}
    \label{fig: survival test}
\end{figure}

 \begin{figure}
    \centering
    \includegraphics[width = 0.5\textwidth]{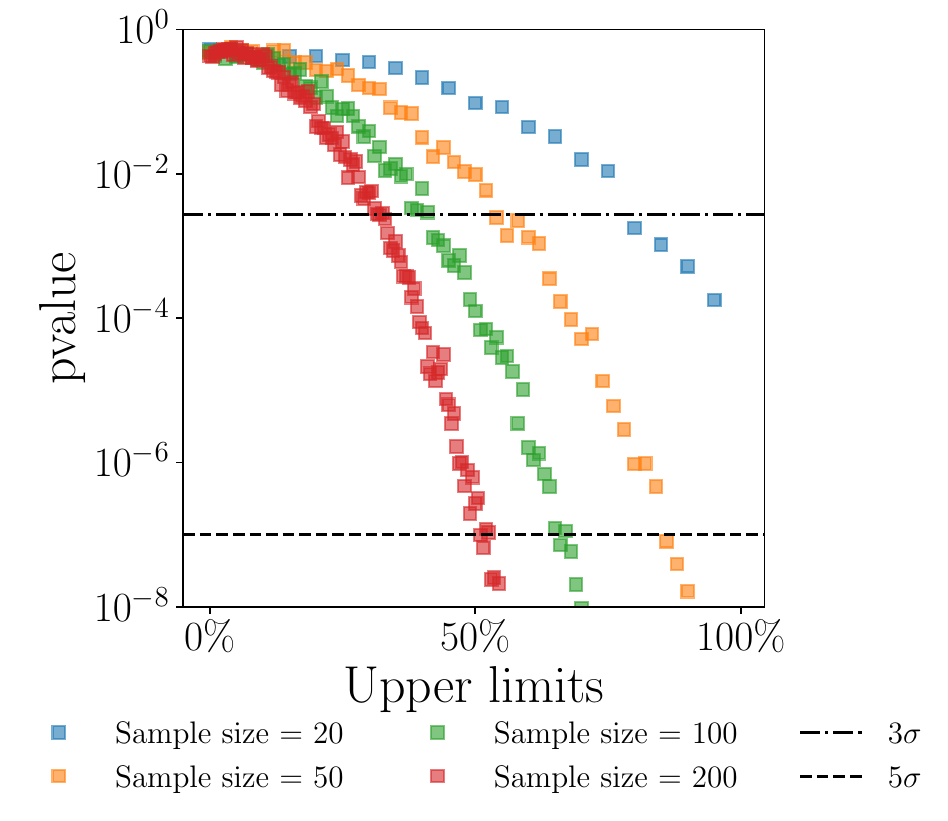}
    \caption{\footnotesize Peto logrank pvalues as a function of the number of upper limits (in percentage with respect to the total sample size). From the $5$BZCat redshift sample, two subsamples of equal size were extracted: the first is kept unmodified (only measurements), in the second a progressively larger number of measurements were flagged as UL. In this way, the second subsample includes both measurements and censored data. The Peto logrank test was applied to subsamples of the same size at each step. The results for subsample size $= 20$ in blue, $50$ in orange, $100$ in green, and $200$ in red are shown by plotting, for each case, the median of the pvalues obtained over $100$ realizations.}
    \label{fig: bzcat same size}
\end{figure}

 %%%%%%%%%%%%%%%%%%%%%%%%%%%%%%%%%%%%%%%%%%%%%%
 \section{Atlas of optical spectra}
 \label{sec:spectra}
In this section, we show the new optical spectra acquired for this work in  Fig.\ref{fig: spectra new} (see Table \ref{table:spectra} and Appendix \ref{sec:append dataset} for all the details, also for the spectra that were already available in the literature and/or public archives).

\begin{figure*}
    \centering
    \includegraphics[width = \textwidth]{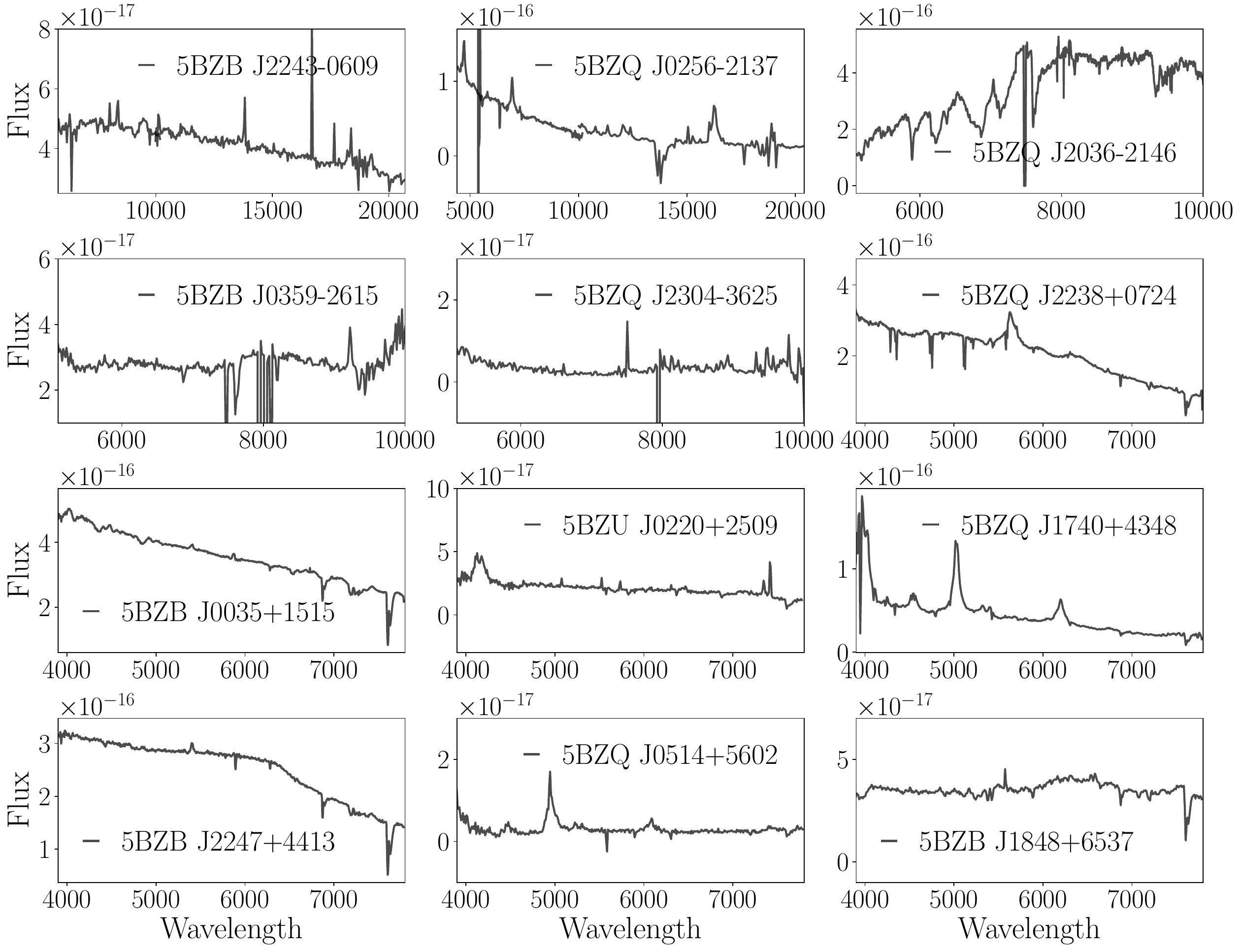}
    \caption{Newly acquired optical spectra for this work. The flux on the y-axis is in units $\left[ {\rm erg} \cdot {\rm cm}^{-2} \cdot {\rm s}^{-1} \cdot {\rm \AA}^{-1} \right]$, the wavelengths on the x-axis in $\left[ {\rm \AA} \right]$ (see Table \ref{table:spectra} and Appendix \ref{sec:append dataset} for all the details).}
    \label{fig: spectra new}
\end{figure*}

\end{appendix}

\end{document}